\documentclass[prd,twocolumn,showpacs,preprintnumbers,amsmath,amssymb]{revtex4}
\usepackage{graphicx}% Include figure files
\usepackage{dcolumn}% Align table columns on decimal point
\usepackage{bm}% bold math
\usepackage{longtable}
\usepackage{afterpage}
\def\lsim{\mathrel{\lower2.5pt\vbox{\lineskip=0pt\baselineskip=0pt
\hbox{$<$}\hbox{$\sim$}}}}
\def\gsim{\mathrel{\lower2.5pt\vbox{\lineskip=0pt\baselineskip=0pt
\hbox{$>$}\hbox{$\sim$}}}}

\newcommand{\imag}{{\mbox{Im}\,}}
\newcommand{\real}{{\mbox{Re}\,}}
\newcommand{\gev}{{\mbox{GeV}\,}}
\newcommand{\mev}{{\mbox{MeV}\,}}
\newcommand{\be}{\begin{equation}}
\newcommand{\ee}{\end{equation}}
\newcommand{\pepe}{\hbox{P.P.}}
\newcommand{\dd}{\hbox{d}\,}
\newcommand{\gammav}{ \Gamma }

\newcommand{\im}{\mbox{Im }}
\newcommand{\re}{\mbox{Re }}
\newcommand{\ra}{\rightarrow}

\newcommand{\mpi}{M_\pi}
\newcommand{\mps}{\mpi^2}
\newcommand{\inv}[1]{\frac{1}{#1}}

\begin{document}

%\preprint{hep ph/0203134}

\title{The pion-pion scattering amplitude. IV: \\
Improved analysis with once subtracted Roy-like equations up to 1100 MeV}

\author{R. Garc\'{\i}a-Mart\'{\i}n$^a$, R.~Kami\'nski$^b$,
  J. R. Pel\'aez$^a$, J. Ruiz de Elvira$^a$ and F. J. Yndur\'ain$^c$,\footnote{deceased}}

\affiliation{
$^a$Departamento de F{\'\i}sica Te{\'o}rica II,
 Universidad Complutense de Madrid, 28040   Madrid,\ Spain\\
$^b$Department of Theoretical Physics
Henryk Niewodnicza\'nski Institute of Nuclear Physics,
Polish Academy of Sciences,
31-342, 
Krak\'ow, Poland,\\
$^c$Departamento de F\'{\i}sica Te\'orica, C-XI
 Universidad Aut\'onoma de Madrid,
 Canto Blanco,
E-28049, Madrid, Spain.}

%\date{June 2007}% It is always \today, today,

\begin{abstract}
We improve our description of $\pi\pi$ scattering data by imposing
additional requirements to our previous fits,
in the form of once-subtracted Roy-like equations, 
while extending our analysis up to 1100 MeV. We provide 
simple and ready to use parametrizations of the amplitude.
In addition, we present a detailed description
and derivation of these  once-subtracted dispersion relations that,
in the 450 to 1100 MeV region,
provide an additional constraint which is much stronger 
 than
our previous requirements of Forward Dispersion Relations 
and standard Roy equations. 
The ensuing constrained amplitudes
describe the existing data with rather small uncertainties 
in the whole region from threshold
up to 1100 MeV, while satisfying very stringent dispersive constraints.
For the S0 wave, this requires
an improved matching of the low and high energy 
parametrizations. Also for this wave we have considered
the latest low energy $K_{\ell 4}$ decay results, including their isospin violation correction,
and we have removed some controversial 
data points. These changes on the data translate into better determinations of threshold and subthreshold
parameters which remove almost all disagreement
with previous Chiral Perturbation Theory and Roy equation calculations below 800 MeV. 
Finally, our results favor the dip structure of the 
S0 inelasticity around the controversial 1000 MeV region.
\end{abstract}

\maketitle

%\vspace{-.5cm}

\section{Introduction}

In a series of papers  \cite{Pelaez:2004vs,Kaminski:2006yv,Kaminski:2006qe}
that we will denote by PY05, KPY06 and KPY08, respectively, we have provided several sets
of precise phenomenological fits to $\pi\pi$ scattering data. 
The interest in a precise and model independent
description of the data available in this process is twofold:
On the one hand, it could be used at low energies
to extract information about the parameters of
Chiral Perturbation Theory (ChPT) \cite{Gasser:1983yg}, quark masses
 and the size of the chiral condensate, pionic atom decays or CP violation in the kaonic system. On the other hand,
in the intermediate energy region, it could provide model independent
information to identify the properties of hadronic resonances, 
particularly the scalar ones
which are related to the spontaneous chiral symmetry breaking of QCD and  
the possible existence of glueball states.

Pion-pion scattering is very special due to 
the strong constraints from  isospin, crossing and chiral symmetries, 
but mostly
from analyticity. The latter allows for a very rigorous
dispersive integral formalism that
relates the amplitude 
at any energy with an integral over the whole energy range, 
increasing the precision 
and providing information
on the amplitude even at energies where data are poor. 
Our aim is to provide reliable and model independent $\pi\pi$ scattering
amplitudes that describe data and are consistent, within uncertainties, with dispersion 
relations. Note that, since we would like to test ChPT,
we are not using it in our analysis, and that, in order to calculate 
dispersive integrals up to infinity, 
we have been using Regge parametrizations obtained 
from a fit to data on nucleon-nucleon, meson-nucleon and
pion-pion total cross sections \cite{Pelaez:2003ky}.
In this work we will further improve  our data analysis by imposing in the fits
an additional set of once-subtracted dispersion relations, that we will also 
derive and describe in detail, showing that they are much more precise in the 
intermediate energy region than
those we have used up to now.  
        
In general, on each paper of this series (or also in \cite{Yndurain:2007qm}),
we have first obtained a set of phenomenological 
``Unconstrained'' Fits to Data (UFD), which was fairly 
consistent with the dispersive requirements. Next, starting 
from that UFD set, we obtained
``Constrained'' Fits to Data (CFD) by imposing simultaneous fulfillment
of dispersion relations. These constrained fits not only describe data, but are 
remarkably consistent with the strong analyticity requirements. Furthermore,
the output of the dispersive integrals is model independent and very precise.

The constraints we imposed in the first two  papers of this series were
just a complete set of Forward Dispersion Relations (FDR), 
plus some crossing sum rules.
In the third paper, apart from including the most recent and reliable data 
up to that date on $K_{\ell 4}$ decays \cite{Rosselet:1976pu,Batley:2007zz},
we also imposed Roy Equations~\cite{Roy:1971tc},  because they constrain
the $t\neq0$ behavior of the amplitude,
while ensuring $s-t$ crossing symmetry. These equations, 
which had already been used in the 70's to analyze some of the existing data \cite{oldRoy},
as derived by S. M. Roy, have two subtractions and provide
a strong constraint in the low energy part of the partial waves. 
For this reason there has
been recently a considerable effort to analyze them in relation
with ChPT \cite{Bern}. They have also been recently 
used to eliminate \cite{Kaminski:2002pe}
the longstanding ambiguity about ``up'' or ``down''
type solutions of the S0 wave data analyses. 
Since Roy equations are written in terms of partial waves, they lead,
if supplemented with further theoretical input from ChPT \cite{Caprini:2005zr},
to precise predictions for resonance poles 
like the much debated $f_0(600)$.
Despite being listed with huge uncertainties in the Particle Data Book \cite{PDG},
several analyses using analytic methods or dispersive techniques 
with chiral constraints \cite{dispersiveanalysis,Yndurain:2007qm}, as well as those using Roy Eqs.
\cite{Caprini:2005zr}, are in 
fair agreement about its pole position, around $450-i250$~MeV. However, its nature remains 
controversial, since it might not be an ordinary meson \cite{Jaffe:1976ig}.
 A precise analysis of $\pi\pi$ scattering data may help clarifying the situation by
 studying the $f_0(600)$ parameters (like the coupling \cite{sigmacoupling}),
and the connection of the pole to QCD parameters \cite{milargen},
 although one have to bear in mind \cite{Achasov:2007fz} the difficulties  
to interpret the coupling in terms of simple intuitive models. Nevertheless,
let us remark that here we only aim at a precise description of data, which could later 
be used for those purposes among many others, 
but the interpretation of this resonance and the extension to the 
complex plane are beyond the scope of this work.

Back to Roy equations, when used only with data, as it is our case,
the S2 wave scattering length, which is very poorly known experimentally,
dominates completely the Roy equations uncertainties,
that become very large above roughly  450 MeV, for the
S0 and S2 waves. For that reason Roy equations do not provide a significant
additional constraint  for the amplitudes beyond that energy, once they are already constrained with FDR. 
In this work we will overcome that caveat with 
additional once-subtracted Roy-like equations
that have a much weaker dependence on scattering lengths. 
The fact that these additional equations have a much smaller uncertainty
above roughly 450 MeV will force us to refine the matching 
of our S0 wave parametrizations.

Let us remark, though, that our parametrizations
 are consistent with those in KPY08
within one standard deviation,
 with the only exception of the S0 wave. 
However,
the new central values satisfy Roy equations and the new once subtracted dispersion relations
better.
Moreover, we will now be able to extend the Roy equations analysis, 
both with one and two subtractions, up to 1115~MeV, 
instead of just the $K\bar{K}$ threshold.

Once again we remark that the functional form of the amplitude parametrizations
becomes irrelevant once the imaginary part of the amplitude
is used in the dispersive integrals, whose results are model independent.
In the understanding that
 running the dispersive representation could be tedious for the reader, 
we  provide results in terms 
of our simple and ready to use CFD parametrizations, which are very
good approximations to the dispersive result.

The plan of this work goes as follows: In Section \ref{sec:UFD} 
we very briefly comment
 on the simple unconstrained data fits (detailed in Appendix A) 
of all partial waves obtained in previous works. Only the S0 wave
is given in more detail in Section \ref{sec:S0wave} 
to introduce the new improvements. These are of two kinds:
On the one hand, the data has changed, since we are 
taking into account the final and more precise NA48/2 data \cite{ULTIMONA48}, 
including the threshold enhanced isospin violation correction to all $K_{\ell4}$ data,
and getting rid of the controversial $K\rightarrow2\pi$ datum.
On the other hand, we have improved our parametrization,  by imposing
a continuous derivative 
matching between the low and intermediate energy regions and allowing for more flexibility
in the parametrization
around the $f_0(980)$ region.

In Section \ref{sec:Dispersive}, after introducing 
FDRs and Roy equations very briefly, we present 
the once-subtracted
dispersion relations and compare their structure with the standard Roy equations.
Next, in section \ref{sec:CFD} 
we impose these new relations together with the constraints already used
in previous works (FDRs, sum rules, standard Roy equations...) to obtain the final 
representation for the amplitudes, i.e., the CFD set of amplitudes. 
In Sect.\ref{sec:thresholdand Adler} we study the threshold parameters and Adler zero determinations  stemming
from this constrained fit through the use of additional sum rules and dispersive integrals.
Then, in the discussion section, 
we compare these CFD with our previous results and other works in the literature, 
and we comment
on the  effect of considering different 
choices of data or parametrizations as a starting point to obtain our
final result. 
In particular, we show how our results favor a ``dip'' structure
in the S0 wave inelasticity right above 1000 MeV, which has been the subject of a longstanding controversy \cite{Au:1986vs}.
Finally, we present our conclusions.
In the Appendices we provide a list of all parametrizations 
and parameters of the UFD and CFD, as well as the detailed derivation 
of the once-subtracted relations together with all relevant integral kernels.
In appendix \ref{sec:ponderatedphases} we provide a table with the phase shifts in the elastic region, as obtained from the dispersive representation.

\section{The Unconstrained Fits to Data}
\label{sec:UFD}

\subsection{Our previous works}

To explain the motivation for further improvements in our previous amplitudes, 
we briefly describe next the results of the previous articles.
In particular, 

- In PY05~\cite{Pelaez:2004vs} we obtained simple and easy to use phenomenological
 parametrizations of $\pi\pi$ scattering data
whose consistency was checked by means of FDR
and several crossing sum rules. 
The P, S2, D0, D2, F, G0 and G2 partial waves were described by simple fits
to $\pi\pi$ scattering data up to 1.42 GeV. In the elastic regime, the
P wave was obtained from
a fit to the pion form factor. 
For the S0 wave, given the fact that there are several 
conflicting sets of data, we fitted first each set separately and
then performed another global fit only in the energy regions 
where different data
sets are consistent. 
Surprisingly, some of the most commonly used
 data sets failed to pass these consistency tests, although
the global fit was in fairly good agreement with FDR. Hence, it could
be used as a starting point for a constrained fit to data. 
This CFD was obtained
by imposing
FDR and crossing sum rules to be satisfied within errors, 
in the elastic regime and up to 925 MeV. 
As a result,  a precise 
description of the data
up to 925 MeV was obtained by means of a constrained 
fit, satisfying the 
FDR and sum rule requirements remarkably well. 

- In KPY06 \cite{Kaminski:2006yv} 
we refined our parametrizations above $K\bar{K}$ threshold, 
including more $\pi\pi$ data but, most importantly,
$\pi\pi\rightarrow K\bar{K}$ data in a coupled channel fit.
These reduced uncertainties forced us to slightly refine the UFD 
parametrizations of our D0, D2 and P waves between 1 and 1.42 GeV as well as
 the Regge parameters.
This led to a remarkable improvement in the consistency 
of the $\pi^0\pi^0$ FDR.

- In KPY08 \cite{Kaminski:2006qe} we also considered Roy equations \cite{Roy:1971tc}
 for our amplitudes 
below  $K\bar{K}$ threshold. The UFD fits, where we had previously
incorporated \cite{Yndurain:2007qm} the most reliable low energy
data from $K_{\ell 4}$ decays  to that date~\cite{Batley:2007zz}, 
satisfied Roy equations  fairly well and 
the agreement was remarkably good once they were imposed
into a new set of CFD. 

Since, in this work, we are going to consider
a set of dispersion relations {\it in addition} to the dispersive
constraints we have just described, our starting point will be
the UFD set already obtained in KPY08, 
that we describe only very briefly in the next subsections, but explain in detail in Appendix A.
The only exception will be the S0 wave, that we describe in Sect.~\ref{sec:S0wave}.
The reasons are the appearance of new data \cite{ULTIMONA48}, the existence of modifications on the
analysis of the old experimental results, and, in addition,  that we have found that
the new constraints are strong enough to require a better matching,
 with a continuous derivative, between the low and intermediate energy 
parametrizations.

\subsection{Notation}
\label{sec:notation}
For $\pi\pi\rightarrow\pi\pi$ scattering amplitudes of definite isospin $I$ in the $s$-channel, we write a partial
 wave decomposition as follows: 
\begin{eqnarray}
&&F^{(I)}(s,t)=\frac{8}{\pi} \sum_\ell (2\ell+1)P_\ell(\cos\theta) t^{(I)}_\ell(s),\\
&&t^{(I)}_\ell(s)=\frac{\sqrt{s}}{2 k}\hat f^{(I)}_\ell(s), \quad
\hat f^{(I)}_\ell(s)=\frac{\eta_\ell^{(I)}(s)e^{2i\delta_\ell^{(I)}(s)}-1}{2i},
\nonumber 
\end{eqnarray}
where $\delta_\ell^{(I)}(s)$ and $\eta_\ell^{(I)}(s)$ are the phase shift 
and inelasticity of the $I$, $\ell$ partial wave, $\ell$ is the angular momentum, 
and $k$ is the center of mass momentum.
In the elastic case, $\eta=1$ and 
\begin{equation}
\hat f^{(I)}_\ell(s)=\sin\delta_\ell^{(I)}(s)\,e^{i\delta_\ell^{(I)}(s)}.
\end{equation}
Note that $I=0,1,2$ and that
 whenever $I$ is even (odd) then $\ell$ is even (odd), and thus we will omit the isospin index for odd waves.
We may refer to partial waves either by their $I$, $\ell$ quantum numbers or by the
usual spectroscopic notation S0, S2, P, D0, D2, F, G0, G2, etc...

In addition, we recall the expressions for the so called
threshold parameters, which are the coefficients
of the amplitude expansion in powers of center of mass (CM) momenta around threshold:
\begin{eqnarray}
  \frac{s^{1/2}}{2 M_\pi k^{2\ell+1}}\real \hat{f}_\ell^{(I)}(s)\simeq
  a_\ell^{(I)}+b_\ell^{(I)} k^2+O(k^4).
  \label{eq:defthreshold}
\end{eqnarray}
Note that $a_\ell^{(I)}$ and $b_\ell^{(I)}$ are the usual scattering lengths and slope parameters.
Customarily these are given in
$M_\pi$ units.

\subsection{Parametrizations for S2, P, D, F and G waves}

The S2, P, D0, D2, F and G  waves are described by very simple expressions.
For the S2, P and D0 waves
we  use separate parametrizations for
the ``low energy region'', i.e. energies $s^{1/2}<s_M^{1/2}\sim 1\,\gev$,
and the  ``intermediate energy region'', which extends from 
the matching energy $s_M^{1/2}$ 
up to 1.42 GeV.
For each wave,  $s^{1/2}_M$ is typically the energy where inelastic processes 
cannot be neglected. Note that, above 1.42 GeV we will assume that
$\pi\pi$ amplitudes are given by Regge formulas, which
correspond to fits to experimental data (see \cite{Pelaez:2003ky} and KPY06 for details).

In the ``low energy region'', where the elastic approximation
is valid,  we use a {\it model independent} parametrization
for each partial wave $t_\ell^{(I)}$,
that ensures elastic unitarity:
\begin{eqnarray*}
t_\ell^{(I)}=\frac{\sqrt{s}}{2 k}\frac{1}{\cot\delta_\ell^{(I)}(s)-i}.
  \label{flI}
\end{eqnarray*}
To ensure maximal analyticity in the complex plane
$\cot\delta_\ell^{(I)}(s)$ is then expanded in powers of the conformal variable
\begin{eqnarray*}
  w(s)=\frac{\sqrt{s}-\sqrt{s_i-s}}{\sqrt{s}+
    \sqrt{s_i-s}}.
\label{eq:ws}
\end{eqnarray*}
where $s_i$ is a convenient scale for each wave, to be precised later,
always larger than the $s$
range where conformal mapping is used.
The use of a conformal variable allows for a very rapid convergence---at most 
two or three terms are needed in the expansion---so that each wave is 
represented by only 3 to  5 parameters,
corresponding to the coefficients of the expansion and the position of the zeros and poles
when we have found convenient to factorize them explicitly \cite{Yndurain:2007qm}. We remark again that the use of a conformal
expansion does not imply any model dependence. 

In the intermediate energy inelastic region, we have used purely 
polynomial expansions both for the phase shifts and inelasticities
in terms of the typical energy or momenta involved in the process.

All these simple parametrizations
have been fitted to a large number of experimental data 
on $\pi\pi$ phase shifts
or, in the case of the P wave, to the vector form factor data, which 
gives much more precise results.
In Appendix A, we provide the detailed parametrizations
for each partial wave, together with the resulting parameters 
and their uncertainties,
from now on denoted
by $p_i^{exp}$ and $\delta p_i$, respectively.

Let us remark that, on a first step, each partial wave has been fitted 
independently of each other, without imposing any constraint from dispersion relations,
and that is why we refer to such initial fits
as ``Unconstrained'' Fits to Data or UFD. In KPY08 we showed
that these UFD provided a good description of data,
and a fairly reasonable consistency in terms of dispersion relations.
Of course, the consistency is much better, remarkable indeed,
once we impose the dispersion relations as constraints to the fit,
but then all waves become correlated. 
The uncorrelated fits, apart from providing the starting point of our calculation,
and although they are less reliable than our final constrained results, could be of relevance
if new and more precise data becomes available for a given
partial wave, since then only that particular partial wave should
be modified, without affecting the others.

\section{S0 wave Parametrization} 
\label{sec:S0wave}

This is the only wave that changes in the new sets of unconstrained data fits.
This is due to three reasons that we will explain in separate subsections.

\subsection{On isospin violation in $K_{\ell 4}$ decays}
\label{sec:iviolation}

There has been a recent calculation 
\cite{Colangelo:2008sm} showing that, due to threshold
enhancements, isospin corrections
in $K_{\ell 4}$ decays \cite{Rosselet:1976pu,Batley:2007zz,ULTIMONA48} 
could be larger than naively expected. 
A leading order ChPT calculation 
has been provided to correct the phase shift determination in the isospin limit,
 that should be valid within the whole range of $K_{\ell 4}$ decays. 
Note that the  uncertainties in the previous UFD set in \cite{Yndurain:2007qm}  
were obtained taking into account
systematic errors on the data, including possible 
isospin corrections, but only of natural size.
Since the most recent data from $K_{\ell 4}$ decays play a 
relevant role in the S0 wave of our UFD set,
and the suggested isospin breaking effect is unnaturally large,
we will modify the S0 wave by correcting the $K_{\ell 4}$ data
 as suggested in \cite{Colangelo:2008sm},
so that it can be used in our isospin limit formalism. 
Note that this isospin correction was already made available in \cite{Batley:2007zz}
and again in the final NA48/2 results \cite{ULTIMONA48}.

\subsection{The $K\rightarrow2\pi$ data}
\label{sec:k2pidata}

Let us emphasize again that this is a data analysis, and, as such, 
it depends on whether we include or not certain
experimental results that are somewhat controversial. 
This is for instance the case of the phase shift difference obtained
from $K\rightarrow 2 \pi$ decay \cite{Aloisio:2002bs}
that we  used in KPY08:
\begin{eqnarray}
&&\delta_0^{(0)}(M_K^2)-\delta_0^{(2)}(M_K^2)=\nonumber\\
&&\qquad (57.27 \pm 0.82_{\rm exp.}\pm3_{\rm rad.}
\pm 1_{\rm ChPT\,appr.})^\circ \,.
\label{oldK2pi}
\end{eqnarray}
 The extraction of the $\pi\pi$ scattering phase from this decay 
is affected by large uncertainties that have to be estimated from ChPT.  
A similar value is obtained if using the Particle 
Data Group data and the prescription for radiative corrections in \cite{radiativeKl4}.  
In \cite{Yndurain:2007qm}  we took 
the simple linear sum of the errors quoted in \cite{Aloisio:2002bs}, 
which is larger than the usual quadrature addition. However,
the use of the datum above has been questioned in \cite{Colangelo:2007df}, 
also suggesting that it could be partly responsible
for the differences between our approaches in the intermediate energy region. 
It is true that this data point always lies somewhat above 
our parametrizations of KPY08, $51.7\pm1.2^\circ$ for the UFD and $50.4\pm1.1^\circ$ for the CFD,
and even more so from those in \cite{Bern}, $47.7\pm1.5^\circ$. 
While preparing this work, a re-analysis  has appeared \cite{Cirigliano:2009rr}
taking into account more precise experimental data and other improvements
including an update of the low-energy constants, yielding:
\begin{eqnarray}
\delta_0^{(0)}(M_K^2)-\delta_0^{(2)}(M_K^2)=
(52.5 \pm 0.8_{\rm exp.}\pm 2.8_{\rm theor.})^\circ.
\label{newK2pi}
\end{eqnarray}
This is still compatible with the value in Eq.~\eqref{oldK2pi}, but
seems in much better agreement with $\pi\pi$ scattering determinations.
However, this new extraction uses as an input the S0 phase shift value from 
a $\pi\pi$ scattering analysis using Roy equations and ChPT, obtained by
 the Bern group \cite{Bern}. Thus
it would be somewhat circular to use it as input in our approach.
Furthermore, we have studied
the alternative scenarios
with and without 
the $K\rightarrow2\pi$ value in our fits, finding that
the scenario without it is slightly preferred by dispersion relations.
For these reasons,  we will present results 
for fits removing the $K\rightarrow 2\pi$ controversial datum.
As a consequence, our new unconstrained fits have somewhat 
smaller errors than those in KPY08, which makes 
dispersion relations  harder to be satisfied.

\subsection{Improved parametrization and matching condition between low and
intermediate energies}
\label{sec:improvedparametrization}

In previous works,
only continuity, but not a continuous derivative, was imposed for the S0 phase shift at the matching point, then chosen at $s_M^{1/2}=932\,\mev$.
It has been suggested
\cite{Leutwyler:2006qp} that such a crude matching 
could explain the roughly $2\sigma$ 
level discrepancies in the S0 wave
 between the KPY08 analysis and that of the Bern group  \cite{Bern}
in the $450-800$ MeV region.  We have checked that the improved matching by itself
only affects the S0 wave sizably in the $f_0(980)$ region, although 
the effect is rather small below.
However, this improved matching adds together with the two effects in kaon
decays discussed above, to become a relatively larger
 effect 
that certainly improves the agreement with the predicted S0 wave in \cite{Bern}.

In this work we want to keep the same low energy conformal parametrization
of KPY08 or \cite{Yndurain:2007qm}.
However, to improve the flexibility of the parametrization we will keep
one more term in this expansion. Actually, it has been pointed out 
that the difference between the parametrization 
in KPY08
and that of \cite{Bern} could be due to the fact that 
our conformal parametrization
at low energies was not sufficiently flexible \footnote{We thank H. Leutwyer for this suggestion.}.
The additional parameter does not
improve significantly the fulfillment of dispersion relations nor the data fit, but the {\it output} of
the dispersion relations with one parameter less would violate very slightly the elastic unitarity condition
around 500 MeV. For that reason we keep this additional term, and use:
\begin{eqnarray}
\label{eq:S0lowparam}
&&  \cot\delta_0^{(0)}(s)=
\frac{s^{1/2}}{2k}\frac{M_\pi^2}{s-\frac{1}{2}z_0^2}\times\\
&&\quad\left\{\frac{z_0^2}{M_\pi\sqrt{s}}+B_0+B_1w(s)+B_2w(s)^2+B_3w(s)^3\right\},
\nonumber\\
&&w(s)=\frac{\sqrt{s}-\sqrt{s_0-s}}{\sqrt{s}+\sqrt{s_0-s}},
\qquad s_0=4M_K^2.
\end{eqnarray}
where the new values for the UFD parameters are:
\begin{eqnarray}
\label{eq:bestS0-UFDfit} 
B_0=7.26\pm0.23, B_1=-25.3\pm0.5,\qquad\nonumber\\
  B_2=-33.1\pm1.2, B_3=-26.6\pm2.3, z_0=M_\pi 
\end{eqnarray}
which are obtained with the same procedure as in \cite{Yndurain:2007qm} but now including
the additional $B_3$,
 the isospin corrections and getting rid of the $K\rightarrow2\pi$ data, as
already commented in subsections
\ref{sec:iviolation} and \ref{sec:k2pidata} above.
Namely, in this fit we have considered
the data on $K_{\ell 4}$ decays \cite{Rosselet:1976pu},
including the final 
$K_{\ell 4}$ decay data from NA48/2 \cite{ULTIMONA48} (which supersedes
\cite{Batley:2007zz}),
and a selection of all the existing and often 
conflicting $\pi\pi$ scattering data \cite{pipidata,pipidata2}. 
This selection corresponds to an average of the 
different
experimental solutions that passed a consistency test with 
Forward Dispersion Relations and other sum rules in the initial work PY05.
To this average  we assigned a large uncertainty to cover the difference 
between the initial data sets. 
For the sake of brevity we simply refer to that work, 
or the Appendix of Ref.~\cite{Yndurain:2007qm}, for a complete 
and detailed description of the data selection. 
Uncertainties in Eq.\eqref{eq:bestS0-UFDfit} come from data only.
In order to use the UFD by itself,  a systematic uncertainty due to
parametrization dependence \cite{Caprini:2008fc} should be taken into account. 
But as we have seen, possible parametrizations are strongly restricted 
by imposing dispersion relations and unitarity in their output, thus reducing
dramatically this source of systematic uncertanty. Hence, we will only quote
the data uncertainty for the CFD.
 Of course, since dispersion relations are
imposed within uncertainties, the residual parametrization dependence is reflected in the
error bars from the result of the dispersive representation, which we give in Table XII of Appendix D.

Despite this amplitude being used only in the 
physical region, we have explicitly factorized a zero at 
$s_{A}=z_0^2/2=M_\pi^2/2\simeq (98.7 \,\mev)^2$ for these
unconstrained fits.
This corresponds to the position of the so called Adler zero, 
required by chiral symmetry~\cite{Adler:1964um},
at leading order in ChPT. Note, however, that this zero lies
very close to the border of the convergence region 
of the conformal expansion (see Fig.~16 in KPY08), which is therefore 
not very
well described by the expression above. Hence, $z_0$ should not really
be interpreted as the exact position of the Adler zero, 
but just as another parameter of our parametrization.
Of course, 
the physical low energy region, which is the only one relevant for 
the dispersive representation, lies well inside the convergence
region of the conformal expansion, and is very well described 
by Eq.~\eqref{eq:S0lowparam}. 
Actually, we will show in Sect.~\ref{sec:thresholdand Adler} below 
that, when this parametrization is used inside the 
dispersive representation, one finds an Adler zero in the correct position.

Let us now turn to the intermediate energy region.
In previous works, a two-channel K-matrix formalism, following
the experimental reference in \cite{pipidata2}, was used to describe
the region around $K\bar K$ threshold. This is a rather popular formalism
to describe multichannel scattering of two-body states, but has several disadvantages
for our purposes. One, of course, is the use of only two channels $\pi\pi$ and $K \bar K$,
neglecting possible inelasticity contributions from
4$\pi$  or other channels. These are rather small, 
but since we aim at a precision determination, we should allow for 
more flexibility on the inelasticity, 
whereas the two channel K-matrix yields a strong relation between
phase and inelasticity. The second caveat is the huge correlations between
K-matrix parameters, which makes it very hard to improve by means of constrained fits,
as we will do later on. Finally, a very strong disadvantage is that 
 the phase dependence on the
K-matrix parameters is so complicated that it 
is not possible to 
make an analytic matching with the low energy parametrization, and a numerical
matching is much more ineffective and harder to implement. Let us note 
that some of
these caveats were  already removed when using some very 
naive polynomial parametrizations considered
in the Appendix B of KPY06. We will use those same parametrizations 
here but with additional terms 
in the expansion to compensate the loss of flexibility 
due to the improved matching conditions. In particular,
between the matching point and 1.42 GeV, we will use:
\begin{widetext}
\begin{equation}
  \label{eq:prenewparam}
  \delta_0^{(0)}(s)=\left\{ 
\begin{array}{ll}
\displaystyle{d_0+a\frac{\vert k_2\vert}{M_K}+b\frac{\vert k_2\vert^2}{M_K^2}+
c\frac{\vert k_2\vert^3}{M_K^3}},& (0.85 \,{\rm GeV})^2<s<4 M_K^2, \\
 \\
\displaystyle{d_0+B\frac{k_2^2}{M_K^2}+C\frac{k_2^4}{M_K^4}+D\,\theta(s-4M_\eta^2)\frac{k_3^2}{M_\eta^2}},&
4 M_K^2<s<(1.42 \gev)^2
\end{array}\right.
\end{equation}
\end{widetext}
where $k_2=\sqrt{s/4-M_K^2}$, $k_3=\sqrt{s/4-M_\eta^2}$ and $d_0$ is the phase shift at
the two kaon threshold.
Note, however, that we have lowered the matching point to
$s_M^{1/2}=850\,\mev$, since we have found empirically 
that this helps improving dispersion relation fulfillment,
as the slope is somewhat smaller there. 
As a final remark,
we have added a term proportional to the $\eta$ momentum, to 
reflect the opening of the $\eta\eta$ channel, which has shown to have some
relevance in the description of the data \cite{Achasov}.
In this respect we want to clarify a common source of confusion about
Roy (or GKPY) equations: These relations include {\it all possible coupled
  channels contributions}, or at least are consistent with them, as long as
they are in agreement with the experimental inelasticicity. 
 This simple term is purely phenomenological, and given the size of the experimental errors this additional term is more
than enough to just describe the cusp due to the presence of this channel. However, it yields very slightly, but favorable, difference in the fulfillment
of dispersion relations.

\begin{figure*}
  \centering
\hspace*{-.5cm}
\includegraphics[scale=0.41]{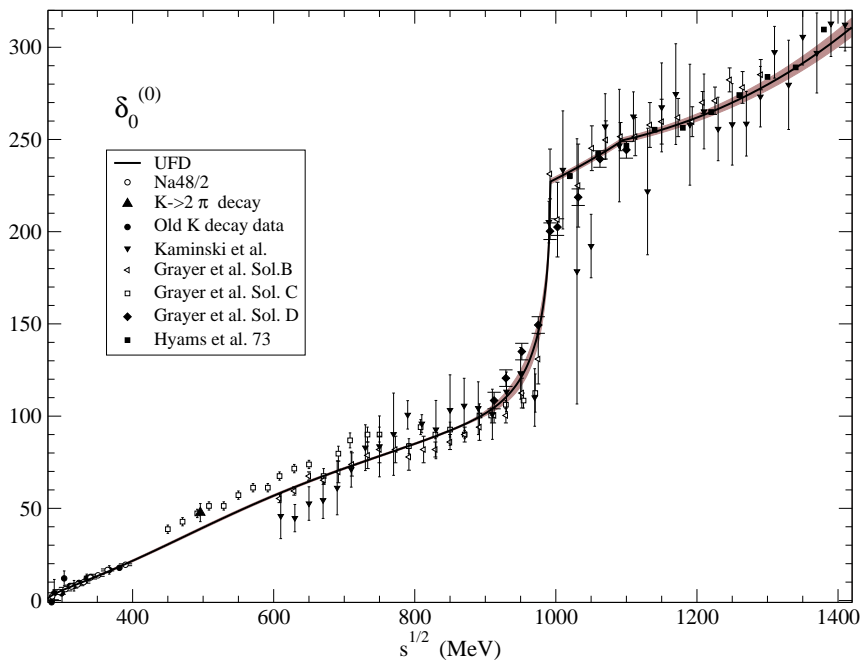}
  \caption{ The new S0 wave unconstrained
fit (UFD), where the dark band covers the uncertainties,
  versus the existing phase shift data from \cite{pipidata,pipidata2}.
Note that the $K\rightarrow2\pi$ point has been excluded from 
the fit as explained in the text.
}
\label{fig:S0-UFD-data}
\end{figure*}

By defining $\delta_M=\delta(s_M)$ and $\delta_M'=d\delta(s_M)/ds$, which are obtained from 
Eq.~\eqref{eq:S0lowparam}, and $k_M=\vert k_2(s_M) \vert $, it is
 rather straightforward to impose continuity and a continuous derivative 
for the phase shift 
at $s_M$, to find:
\begin{widetext}
\begin{equation}
  \label{eq:newparam}
  \delta_0^{(0)}(s)=\left\{ 
\begin{array}{ll}
\displaystyle{d_0\left(1-\frac{\vert k_2 \vert}{k_M}\right)^2+\delta_M \, \frac{\vert k_2 \vert}{k_M}\left(2-\frac{\vert k_2 \vert}{k_M}\right)+
 \vert k_2 \vert (k_M-\vert k_2 \vert)
\left(8\delta_M'+c \frac{(k_M-\vert k_2 \vert)}{M_K^3}\right)
},& (0.85 \,{\rm GeV})^2<s<4 M_K^2, \\
 \\
\displaystyle{d_0+B\frac{k_2^2}{M_K^2}+C\frac{k_2^4}{M_K^4}+D\,\theta(s-4M_\eta^2)\frac{k_3^2}{M_\eta^2}},&
4 M_K^2<s<(1.42 \gev)^2.
\end{array}\right.
\end{equation}
\end{widetext}

\begin{figure}
  \centering
\hspace*{-.4cm}
\includegraphics[scale=0.3]{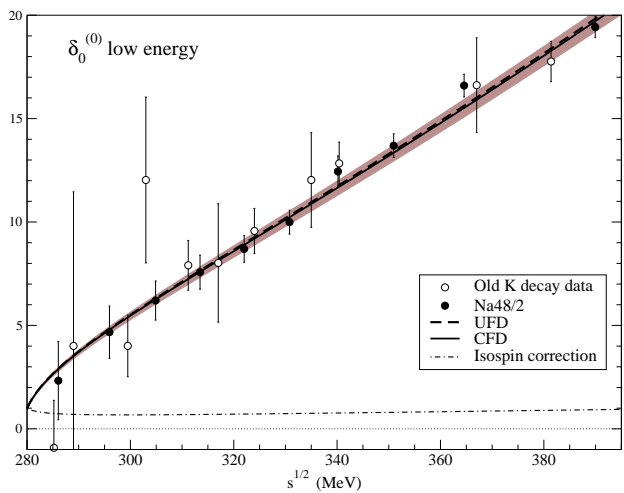}
\vspace*{.2cm}
  \caption{The new S0 wave unconstrained
fit (UFD), where the dark band covers the uncertainties,
  versus the ``old'' phase shift data from $K_{\ell4}$ decays
 \cite{Rosselet:1976pu} together with the final NA48/2 results,
which supersedes the data from the same experiment~\cite{Batley:2007zz} that
 we used in KPY08. We are also showing the isospin violation correction \cite{Colangelo:2008sm},
which has been included in the data shown here.  Finally, we show the
results of the CFD parametrization to be explained in Sect.~\ref{sec:CFD},
which is almost indistinguishable from the UFD curve.
\label{fig:phaselowenergies}
}
\end{figure}

As previously commented,
with the exception of the $K\rightarrow 2\pi$ datum,
the inclusion of isospin corrections to  $K_{\ell 4}$ data explained above,
and our use of the final  NA48/2 results \cite{ULTIMONA48},
our treatment
and selection of data for the phase is exactly the same one followed in the previous 
works \cite{Kaminski:2006yv} and \cite{Yndurain:2007qm}, 
so we will not repeat them here.
In Table \ref{tab:S0parameters} of the Appendix,
we provide the values for the $d_0, c, B, C$ and $D$ parameters resulting
from the unconstrained fit to those data.
In Fig.~\ref{fig:S0-UFD-data}  
we show the resulting phase from the unconstrained data
fit to the S0 wave phase shift up to 1420 MeV, and in
Fig.~\ref{fig:phaselowenergies} we show the low energy region in detail, including the
isospin violation correction  \cite{Colangelo:2008sm} that we have subtracted from all
the $K_{\ell 4}$ data. Note that this correction amounts to slightly less than 1 degree  in the region from threshold to 400 MeV, which is not much
at high energies, but very relevant close to threshold.

\begin{figure*}
  \centering
\hspace*{-.4cm}
\includegraphics[scale=0.41]{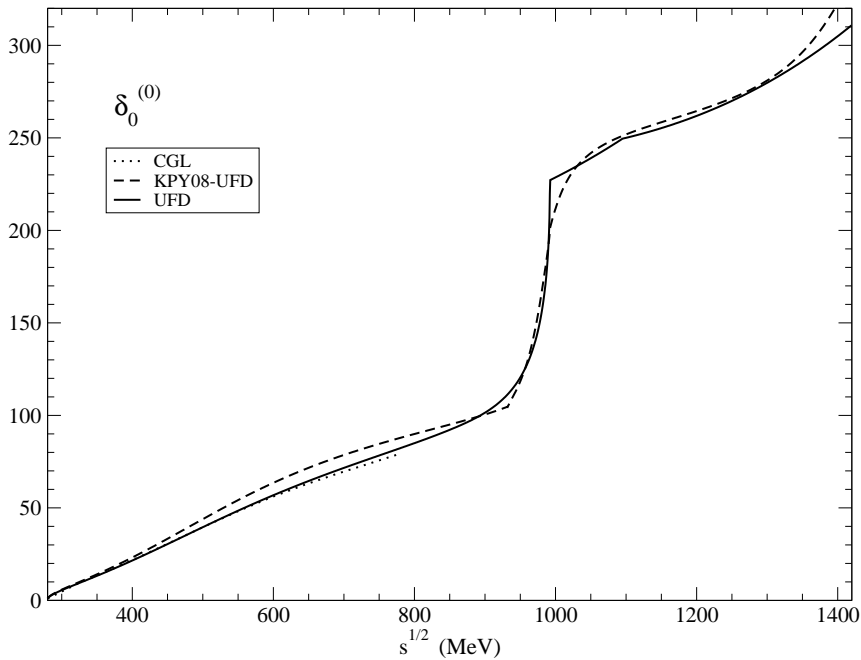}
\vspace*{.2cm}
  \caption{Fit to the S0 wave phase shift, 
with the improved continuous derivative matching (UFD, continuous line) 
versus the simpler one used in KPY08. We also show the phase predicted in 
\cite{Bern} (CGL).
\label{fig:S0-UFD-comparison}
}
\end{figure*}

In Fig.~\ref{fig:S0-UFD-comparison} we show a comparison of the phase shift 
resulting from the new UFD
with the improved matching versus the one obtained
in KPY08. The changes at low energy are due to the update on the $K_{\ell 4}$ data and 
their isospin corrections, 
together with the fact that we now discard the $K\rightarrow 2\pi$ datum. 
The bump in the 500 to 800~MeV region observed in KPY08
has almost disappeared. 
Thus, the improvement on the 
 data and its corrections reduces almost completely the disagreement of our 
UFD description with the phases  in \cite{Bern}---the line labeled 
CGL in the plot---although our central values are still 
larger in the 550-800~MeV region.  Furthermore, as we will see later, for the constrained fits 
we are in an even better agreement with \cite{Bern}.
The changes above the matching point are sizable for the phase, 
mostly around the sharp phase increase usually 
associated with the $f_0(980)$ resonances,
as can be seen in Fig.~\ref{fig:S0-UFD-comparison} 
where the central value for the
 new phase is compared with that in KPY08.
Note the much smoother behavior in the matching region for 
the new UFD parametrization and the more dramatic $\bar K K$ threshold effect.

Concerning the S0 wave inelasticity, 
we approximate it to 1 up to the two-kaon threshold, 
and use the following parametrization above that energy:
\begin{eqnarray}
\eta_0^{(0)}(s)=\exp\bigg[\frac{-k_2(s)}{s^{1/2}}\left(\tilde\epsilon_1+
\tilde\epsilon_2\,
\frac{k_2}{s^{1/2}}+\tilde\epsilon_3\,\frac{k_2^2}{s}\right)^2\\\nonumber
-\tilde\epsilon_4 \theta(s-4M_\eta^2)\frac{k_3(s)}{s^{1/2}}\bigg],
\label{eq:inelasticityS0}
\end{eqnarray}
for $4 M_K^2<s< (1.42 \,{\rm GeV})^2$.
By neglecting the term proportional to the $\eta$ momentum, which is numerically very small as seen in
the Appendix A.1,
and by re-expanding the above equation in powers of $k_2/s^{1/2}$ up to third
order, we recover the polynomial expression in KPY06, but the definition
above ensures the $0\leq\eta_0^{(0)}\leq1$ physical condition, whereas 
the simple polynomial in KPY06 did not.

\begin{figure}
  \centering
\hspace*{-.5cm}
\includegraphics[scale=0.3]{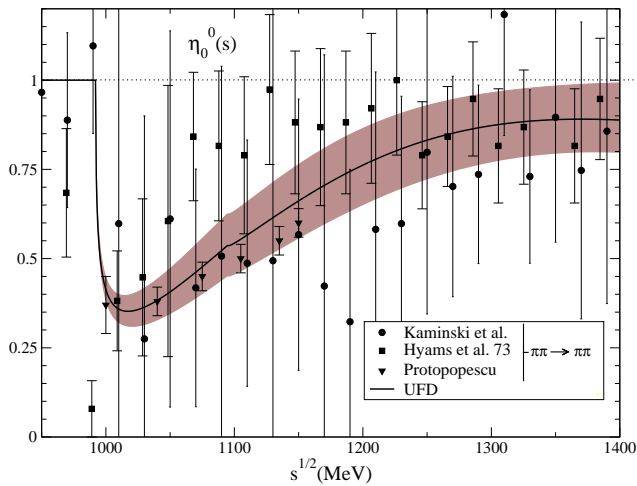}
  \caption{The new S0 inelasticity fit (UFD set)
to the $\pi\pi\rightarrow\pi\pi$ scattering data of 
Hyams et al. (1973) and Protopopescu et al. As explained in the text, we do not fit the Kaminski et al. data
\protect{\cite{pipidata}} although our fit is compatible with them.
The dark band covers our uncertainties.
For all data sets see Fig.~\protect{\ref{fig:UFD-CFD-elaasticity-data}}.
}
\label{fig:S0-UFD-inel-data}
\end{figure}

For the inelasticity data, we follow again the same selection as in previous
 works of this series, 
but we do not include now 
the data from Kaminski et al. \cite{pipidata} in the $\chi^2$ calculation; 
we only consider 
the 1973 data of Hyams et al. \cite{pipidata} 
and Protopopescu et al. \cite{pipidata}. The reason is that the main source of uncertainty is systematic, and if we include the large number of points of Kaminski et al. with their huge statistical errors, the outcome of the fit has much smaller errors than the original systematic uncertainties.
By keeping only the other two sets, which are incompatible, 
we obtain a fit with a large $\chi^2/d.o.f.$, and by rescaling the uncertainties 
in the inelasticity parameters we mimic the 
dominant systematic uncertainties much better. Of course, our results 
are still in very good agreement with Kaminski et al. 
Was the systematic uncertainty not dominant, this would not be necessary.
In Table \ref{tab:S0parameters} of the Appendix,
we provide the values for the $\tilde\epsilon_i$ parameters, and
in Fig.~\ref{fig:S0-UFD-inel-data} we show the results of the 
unconstrained 
fit to the S0 wave inelasticity data up to 1420 MeV.

\begin{figure}
  \centering
\hspace*{-.5cm}
\includegraphics[scale=0.3]{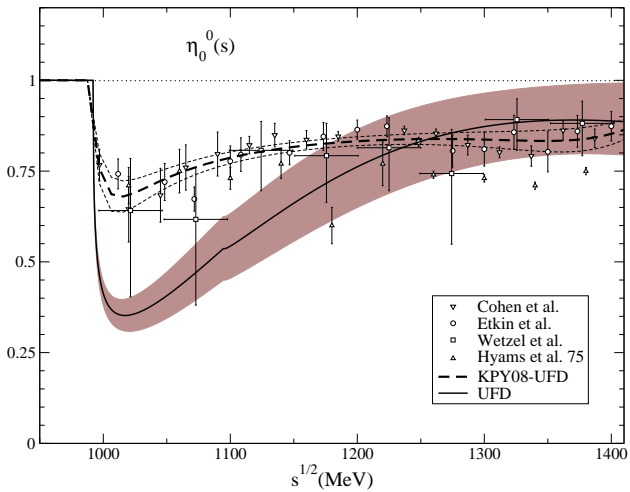}
  \caption{Fit to the S0 wave inelasticity (UFD)
with the improved continuous derivative matching (continuous line) 
versus the simpler one used in KPY08 (dashed line). 
The dark band covers the uncertainties of the former, whereas the 
dotted curves enclose the uncertainties of the latter.
Note that the 
drop in the inelasticity right above 1 GeV has become much deeper.
In contrast to Fig. \ref{fig:S0-UFD-inel-data},
we only show the data coming from $\pi\pi\rightarrow K\bar K$ and the
$\pi\pi\rightarrow\pi\pi$ on which is based the KPY08 fit.
For all data sets see Fig.~\ref{fig:UFD-CFD-elaasticity-data}.
\label{fig:S0-UFD-inel-comparison}
}
\end{figure}

Finally, let us remark that the inelasticity is the scattering parameter 
that suffers the biggest change with respect to the KPY08-KPY06 
parametrization, as can be seen in Fig.~\ref{fig:S0-UFD-inel-comparison}. 
The new parametrization shows a big dip in the 
inelasticity between 1 and 1.1 GeV, whereas the KPY08 one does not.
As already commented in PY05, this is a longstanding controversy 
(see, for instance \cite{Au:1986vs} and references therein)
between different sets of data coming from pure $\pi\pi\rightarrow\pi\pi$
scattering
versus those coming from $\pi\pi\rightarrow\bar K K$ analysis.
Actually, in PY05 (see Fig.~6 there) we considered both possibilities:
we found that Forward Dispersion Relations favored
the ''non-dip solution'' very slightly, but we kept
the ''dip-solution'' in order to use the phase and inelasticity
coming from the same experiment. 
In KPY06 we found a similar situation
but since the K-matrix slightly preferred again the ``non-dip solution'',
this time we decided to use it.
However, in terms of fulfillment, 
the difference is minute for FDRs, and even more so for standard Roy equations,
since, as we have already commented and we will see in detail below,
the uncertainties in the
subtraction constants become so large above 500 MeV that we cannot
use them to discard any of the two scenarios. The existing set of dispersion 
relations did not allow us to make a really conclusive statement about the
inelasticity in the 1 GeV region.

One of the main results of this work is the derivation and  use of once subtracted 
Roy-like dispersion relations, the GKPY equations presented in Sect.~\ref{sec:GKPY} below, which are more precise in the 1 GeV region and clearly favor the
 solution with a dip, thus helping to settle this dip versus non-dip controversy.

\section{Dispersion Relations and sum rules}
\label{sec:Dispersive}

From the theoretical side, $\pi\pi$ scattering is very special due to 
the strong constraints from  isospin, crossing and chiral symmetries, 
but mostly
from analyticity. The latter allows for a very rigorous
dispersive integral formalism that
relates the $\pi\pi$ amplitude 
at any energy with an integral over the whole energy range, 
increasing precision 
and providing information
on the amplitude even at energies where data are poor,
or in the complex plane. 

Let us emphasize once more that the dispersive approach
 is model independent, since
it makes the data parametrization irrelevant once 
it is included in the integral.
The previous works \cite{Kaminski:2006qe,Yndurain:2007qm} of this series
made use of two complementary 
dispersive approaches, Forward Dispersion Relations and Roy equations,
that we briefly review next, before introducing the new set of once subtracted Roy-like equations.

\subsection{Forward Dispersion Relations (FDR)}

They are calculated at $t=0$, so that the 
unknown large-$t$ behavior of the amplitude is not needed.  
There are two symmetric and one antisymmetric isospin combinations
to cover the isospin basis. For further convenience we will write them
as a difference $\Delta_i(s)$ that should vanish if the dispersion relation
is satisfied exactly. In particular,
  the two symmetric ones, for
 $\pi^0\pi^+$ and $\pi^0\pi^0$, have {\it one subtraction} and imply the vanishing of
\begin{eqnarray}
\Delta_i(s)\equiv\real F_i(s,0)-F_i(4M_{\pi}^2,0)-\frac{s(s-4M^2_\pi)}{\pi}\times
\nonumber \\
  \pepe\int_{4M_{\pi}^2}^\infty
  \frac{(2s'-4M^2_\pi)\,\imag F_i(s',0)\,\dd s'}{s'(s'-s)(s'-4M_{\pi}^2)(s'+s-4M_{\pi}^2)}, \;
\label{FDR1}
 \end{eqnarray}
 where $F_i$ stands for the
 $F_{0+}(s,t)$ or $F_{00}(s,t)$ amplitudes,
and ``P.P.'' stands for the principal part of the integral. They are very precise, since
all the integrand contributions 
 are positive.
 The antisymmetric isospin combination $I_t=1$ does not require subtractions
and implies the vanishing of the following difference:
 \begin{eqnarray} 
\Delta_{(I_t=1)}(s)\equiv && F^{(I_t=1)}(s,0)-
 \frac{2s-4M^2_\pi}{\pi} \times  \label{FDR2}\\
&&\pepe\int_{4M^2_\pi}^\infty\dd s'\,
 \frac{\imag F^{(I_t=1)}(s',0)}{(s'-s)(s'+s-4M^2_\pi)}. 
\nonumber
 \end{eqnarray}
All FDRs are calculated up to $\sqrt{s}=1420$~MeV.

\subsection{Roy Equations}

These are an infinite set of coupled equations 
\cite{Roy:1971tc}, equivalent to non-forward 
dispersion relations plus $t-s$ crossing symmetry.
They are well suited to study poles of resonances and
scattering data, since they are 
 written directly in terms of partial waves $t^{(I)}_\ell$ 
of definite isospin $I$ and angular momentum $\ell$.
Remarkably, S. M. Roy managed to rewrite the
complicated left cut contribution 
as a series of integrals over the physical region.
 In the original work
of Roy and all applications until now, the convergence of the
integrals was ensured by making two subtractions.

As we did with FDR, we will recast each one of 
the Roy Equations as the difference 
\begin{align}
  \nonumber
  & \Delta_\ell^{(I)}(s) \equiv
  \mbox{Re } t_{\ell}^{(I)}(s) - ST_\ell^I(s) - DT_{\ell}^{I}(s) \\
  & - \sum_{I'=0}^{2} \sum_{\ell'=0}^{1}
  \mbox{P.P.} \int_{4\mps}^{s_{max}} ds'
  K_{\ell \ell^\prime}^{I I^\prime}(s,s') \mbox{Im } t_{\ell'}^{I^\prime}\!(s'),
  \label{RoyEquations}
\end{align}
that should vanish when the equation is exactly satisfied.
Roy equations provide as output the real part of partial waves below
1115~MeV. Although, in principle, one could consider output
for waves up to higher $\ell$, 
in this work we are interest in 
results for $\ell=0,1$ only. Hence, we have separated those waves explicitly
below $s_{max}$.

As it was done in KPY08, below $s^{1/2}_{max}=1420$~MeV, we consider 
the imaginary parts from all our $\ell \leq4$
partial wave parametrizations as input. Above that energy,
we take into account all waves together parametrized with Regge 
theory---see Appendix~\ref{sec:Regge}. 
The $K_{\ell \ell^\prime}^{I I^\prime}(s,s')$ are known kernels, and thus 
we will refer to the integral
terms as ``kernel terms'' or $KT(s)$.
The  ``driving terms'', $DT_{\ell}^{I}(s)$, have the same structure as the kernel terms,
but their input  
contains both the contribution from $\ell=2,3$ partial waves up to $s_{max}^{1/2}=1420$~MeV,
and the Regge parametrizations above.
We have explicitly checked that the $\ell=4$ contribution below $s_{max}$ is irrelevant,
so that we will refer just to waves up to $\ell=3$.
Finally, the so-called subtraction terms are given by:
\begin{align}
  ST_\ell^I(s) = & 
  a_{0}^{0} \delta_{I0} \delta_{\ell 0} + a_{0}^{2} \delta_{I2}
  \delta_{\ell 0} + \frac{s-4M_\pi^2}{12M_\pi^2} \times
  \\ &
  (2a_{0}^{0}-5a_{0}^{2})
  (\delta_{I0}\delta_{\ell 0} + \frac{1}{6}\delta_{I1}\delta_{\ell 1} 
  - \frac{1}{2} \delta_{I2}\delta_{\ell 0}).
  \nonumber
\end{align}

It is very relevant to 
remark once more that these equations have two subtractions,
as can be seen by the presence of the term proportional to 
$(s-4M_\pi^2)(2a_{0}^{0}-5a_{0}^{2})$. This strong energy dependence 
of $ST(s)$ makes these twice subtracted Roy Equations very suitable
for low energy studies, and even more so when complemented
with theoretical predictions of the scattering lengths coming from ChPT~\cite{Bern}.

Roy Equations are valid up to $\sqrt{s}\leq8M_\pi\simeq1120$~MeV. 
However, we will see that the uncertainties in the scattering lengths,
 when propagated
to high energies, become too large above roughly 450 MeV,
due to the term proportional to $s$. For this reason, in KPY08 it did not
make sense to deal with the complications of a
 precise description around $\bar KK$ threshold and thus we 
implemented them up to $2 M_K$. One of the main novelties
of the present work is that, since the once-subtracted Roy-like
equations explained below will have much smaller uncertainties
in the $\bar KK$ threshold region, we have now implemented these new equations,
 together with
the standard Roy equations, up to 1115 MeV.

\subsection{Two Sum Rules}

Apart from FDRs and Roy equations, 
two sum rules that 
relate high energy (Regge) parameters 
for $t\neq0$ to low energy P and D waves, have been considered
throughout previous works.

The first sum rule (PY05) is   nothing but the vanishing of the following
difference
\begin{eqnarray}
I\equiv \int_{4M^2_\pi}^\infty\!\!\dd\! s\,
 \frac{\imag F^{(I_t=1)}(s,4M^2_\pi)-\imag F^{(I_t=1)}(s,0)}{s^2}
\nonumber \\
\!\!\!\!\!-
  \int_{4M^2_\pi}^\infty\dd \!s\,\frac{8M^2_\pi[s-2M^2_\pi] \,\imag F^{(I_s=1)}(s,0)}{s^2(s-4M^2_\pi)^2},\quad
\label{eq:Isumrule}
\end{eqnarray}
where the contributions of the S waves cancel and only the P and D  waves 
contribute (we also include F and G waves, but they are negligible). 
At high energy, the integrals are dominated by the rho reggeon exchange.

The second sum rule we consider is given in Eqs.~{(B.6) and (B.7)} of 
the second reference in \cite{Bern}, which requires the vanishing of

\begin{eqnarray}
 J\equiv\int_{4M^2_\pi}^\infty\dd s\,\Bigg\{
 \frac{4\imag F'^{(0)}(s,0)-10\imag F'^{(2)}(s,0)}{s^2(s-4M^2_\pi)^2}
\nonumber\\
 -6(3s-4m^2_\pi)\,\frac{\imag F'^{(1)}(s,0)-\imag F^{(1)}(s,0)}{s^2(s-4M^2_\pi)^3}
 \Bigg\}.\quad\label{eq:Jsumrule} 
\end{eqnarray}

Here, $F'^{(I)}(s,t)\equiv\partial F^{(I)}(s,t)/\partial\cos\theta$.
At high energy, the integral is dominated by isospin zero Regge trajectories.

\subsection{GKPY Equations}
\label{sec:GKPY}

The main novelty of this work is that we present and use a new set
of Roy-like dispersion relations for $\pi\pi$ scattering amplitudes. 
For brevity, we will call them GKPY equations,  as we have already
done when presenting some partial and preliminary results in several conferences
\cite{MonteCarlo,Kaminski:2008rh}.
In brief, their derivation follows the same steps as for Roy equations,
starting from fixed $t$ dispersion relations for a complete isospin basis, 
that S.~M.~Roy subtracted twice to ensure
that the integrals converged when extended to infinity.
However, by using the complete set of isospin amplitudes 
$F_{00}$, $F_{0+}$ and
$F^{(I_t=1)}$, it is easy to see that one subtraction is enough.
Actually, the two first amplitudes are $s-u$ symmetric and the
contributions from the $s$ and $u$ channels, that would be divergent by themselves alone,
cancel when considered simultaneously. The $F^{(I_t=1)}$ amplitude is dominated 
by the rho Regge exchange and neither the left nor the right cut are divergent
with one subtraction. We provide the detailed derivation in Appendix~\ref{app:gkpy},
which leads to the vanishing of the following difference:
\begin{align}
  \nonumber
 & \Delta_\ell^{GKPY\,(I)} \equiv
  \re t_{\ell}^{(I)}(s) - \overline{ST}_\ell^I - \overline{DT}^{I}_\ell(s) \\ & -
\sum_{I'=0}^{2} \sum_{\ell'=0}^{1}
  \mbox{P.P.} \int_{4\mps}^{s_{max}} ds'
  \overline K_{\ell\ell'}^{I I'}(s',s) \im t_{\ell'}^{(I')}(s').
\label{1SEquations}
\end{align}
The subtraction terms $\overline{ST}_\ell^I$ are linear combinations of scattering
lengths $a_0^I$, and can be found in Appendix~\ref{app:gkpy}.
A very relevant observation for this work is that,
in contrast to the standard Roy Equations,
\emph{the subtraction terms in GKPY do  not depend on} $s$.

The integral and driving terms $\overline{DT}_\ell^I(s)$ in Eq.~(\ref{1SEquations})
are analogous to the kernel and driving terms in Roy equations,
but the integrals contain the $\overline K_{\ell\ell'}^{II'}$ kernels, instead of the
$K_{\ell\ell'}^{II'}$. 
The explicit expressions for $\overline K_{\ell\ell'}^{II'}$ are lengthy and
we provide them in Appendix~\ref{app:kernels}.
Note that, as the once subtracted GKPY equations have kernel terms
that behave as $\sim 1/s^2$ at higher energies, instead of the
$\sim 1/s^3$ behavior in Roy equations, the weight of the high energy
region is larger. Nevertheless, the 
contribution to the driving terms coming from energies above 1.42 GeV
is generically smaller than the 
contribution coming from the D and F waves below 1.42 GeV, which means that 
their influence is still under control.

\subsection{Roy versus GKPY Equations}
\label{sec:decomposition}

\begin{figure}[t]
\begin{center}
\hspace*{15pt}
\includegraphics[width=5.8cm,angle=-90]{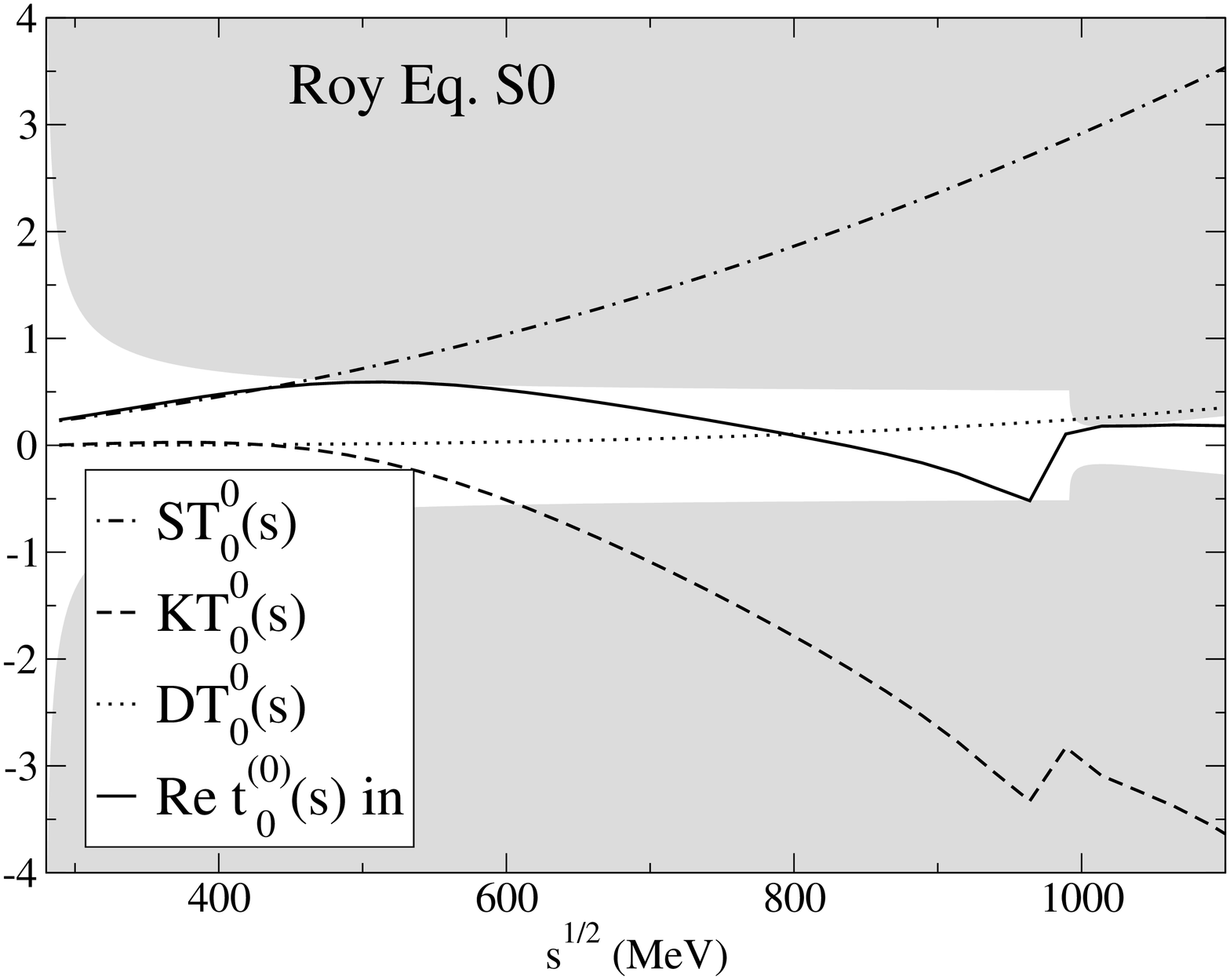}
\hspace*{15pt}
\vspace*{10pt}
\includegraphics[width=5.8cm,angle=-90]{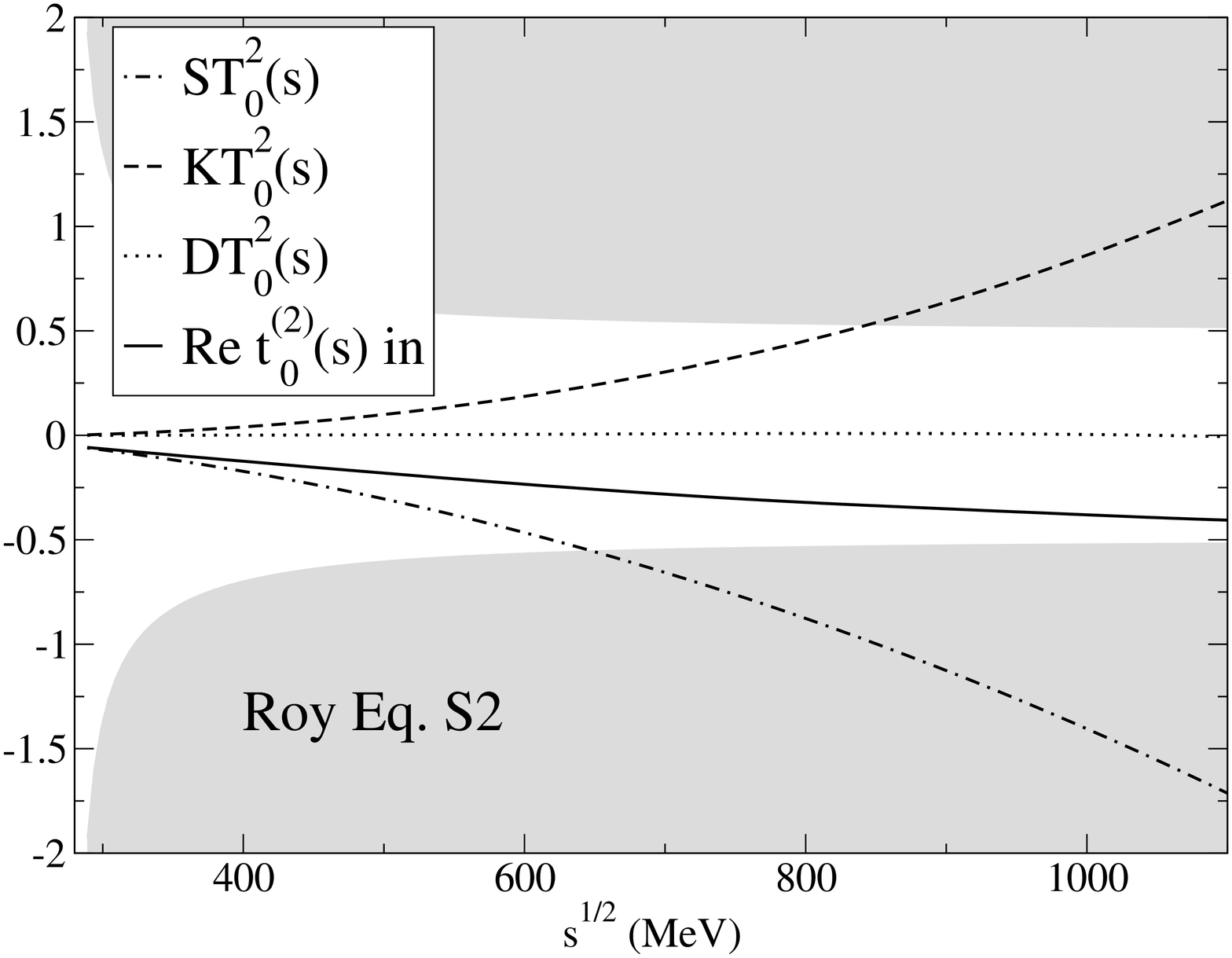}
\hspace*{15pt}
\vspace*{10pt}
\includegraphics[width=5.8cm,angle=-90]{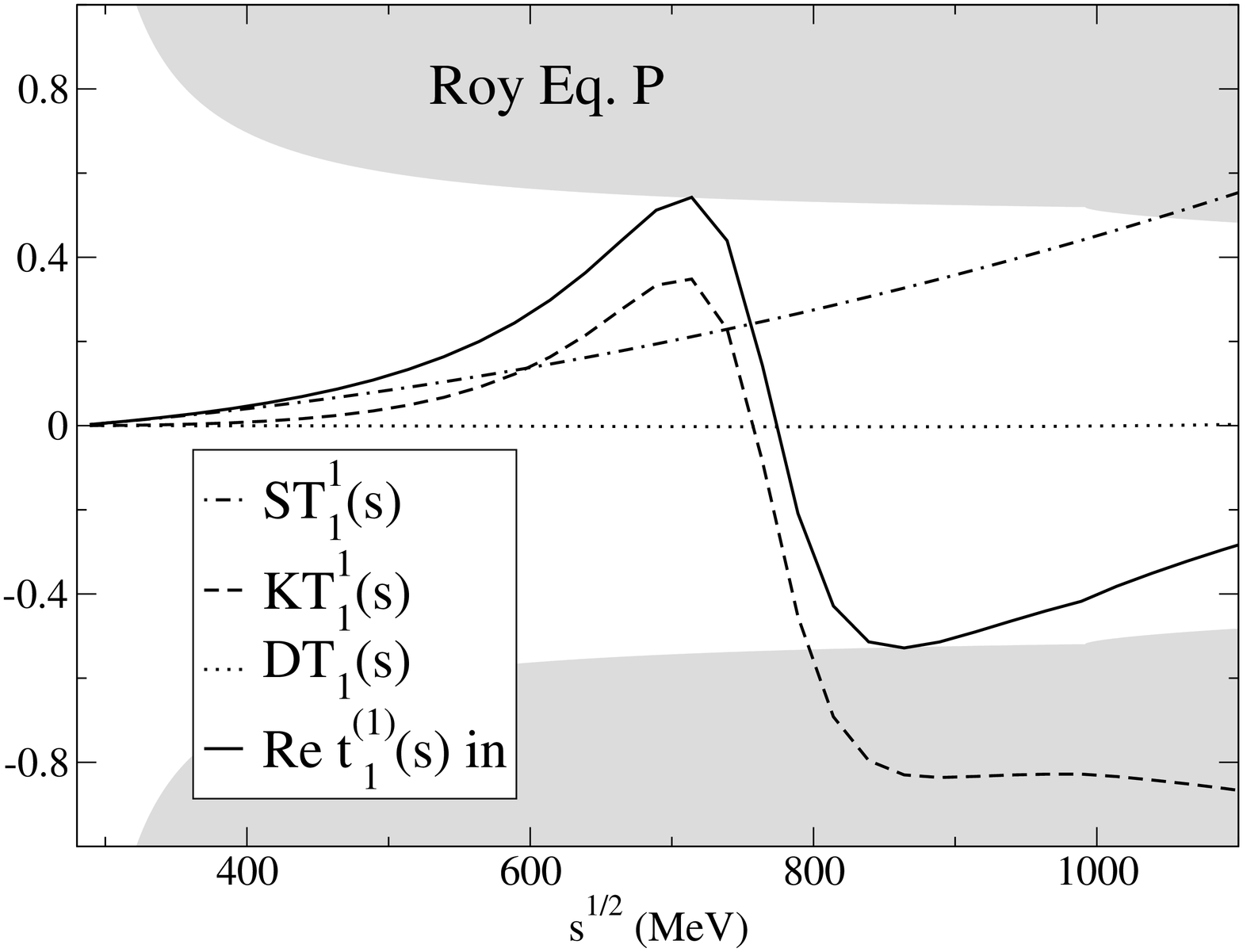}
\end{center}
\caption{ Using the UFD set as input, we show the decomposition of Roy equations
into the subtracting term $ST$, kernel term $KT$, and driving term $DT$ for the  
S0, P and S2 waves. 
Note the different scales used on each plot.
}\label{RoyDecompositions}
%\vspace*{-20pt}
\end{figure}

\begin{figure}[t]
\begin{center}
\hspace*{15pt}
\includegraphics[width=5.8cm,angle=-90]{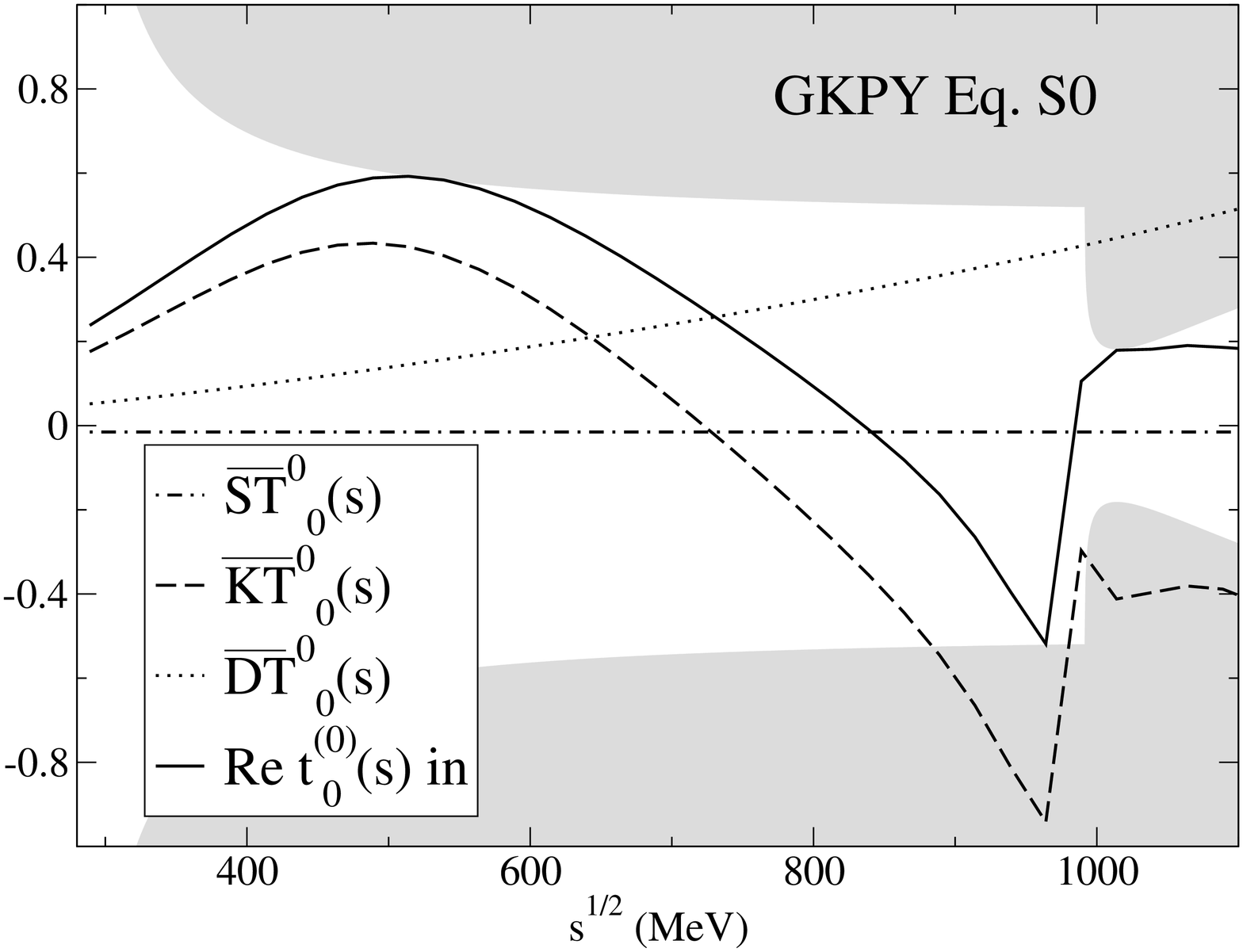}
\hspace*{15pt}
\vspace*{10pt}
\includegraphics[width=5.8cm,angle=-90]{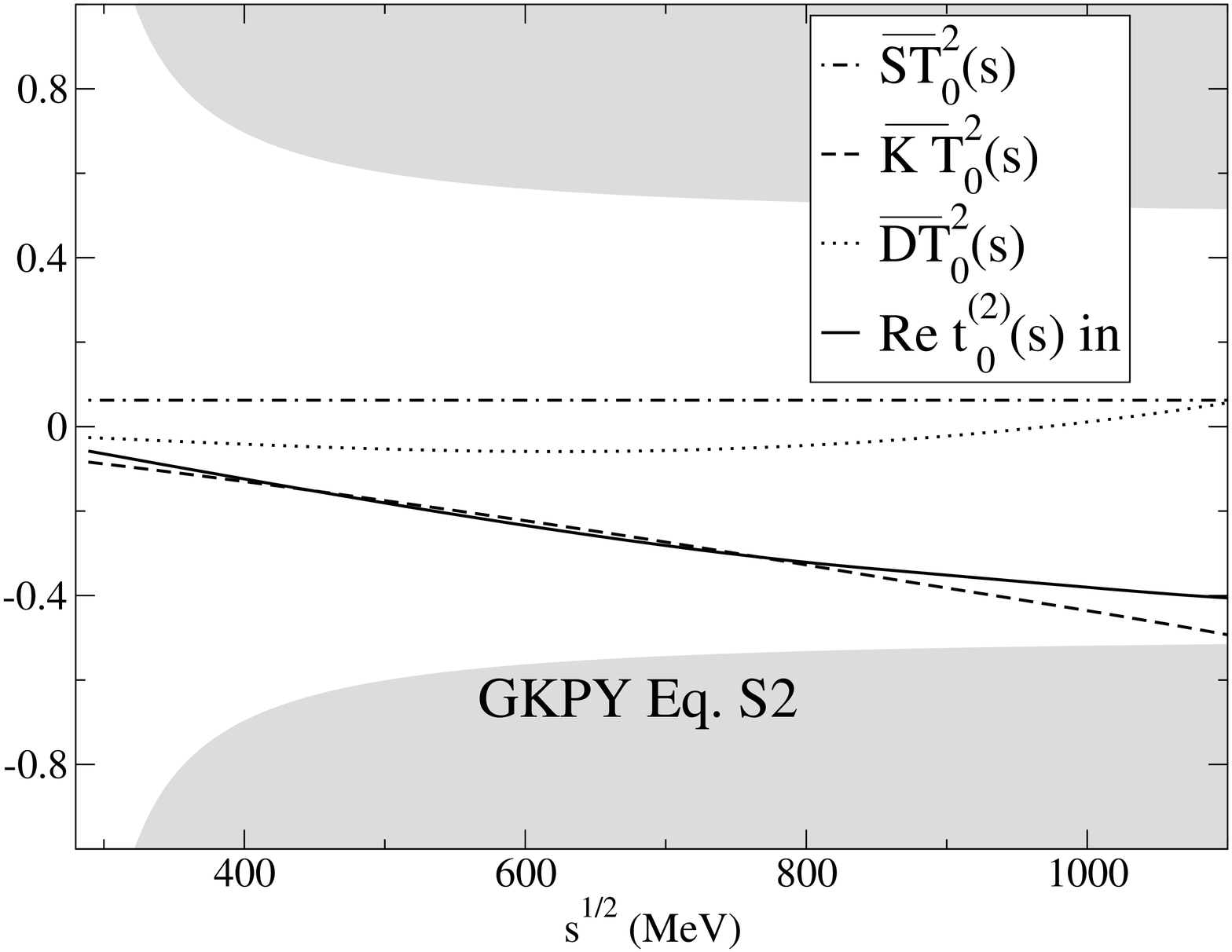}
\hspace*{15pt}
\vspace*{10pt}
\includegraphics[width=5.8cm,angle=-90]{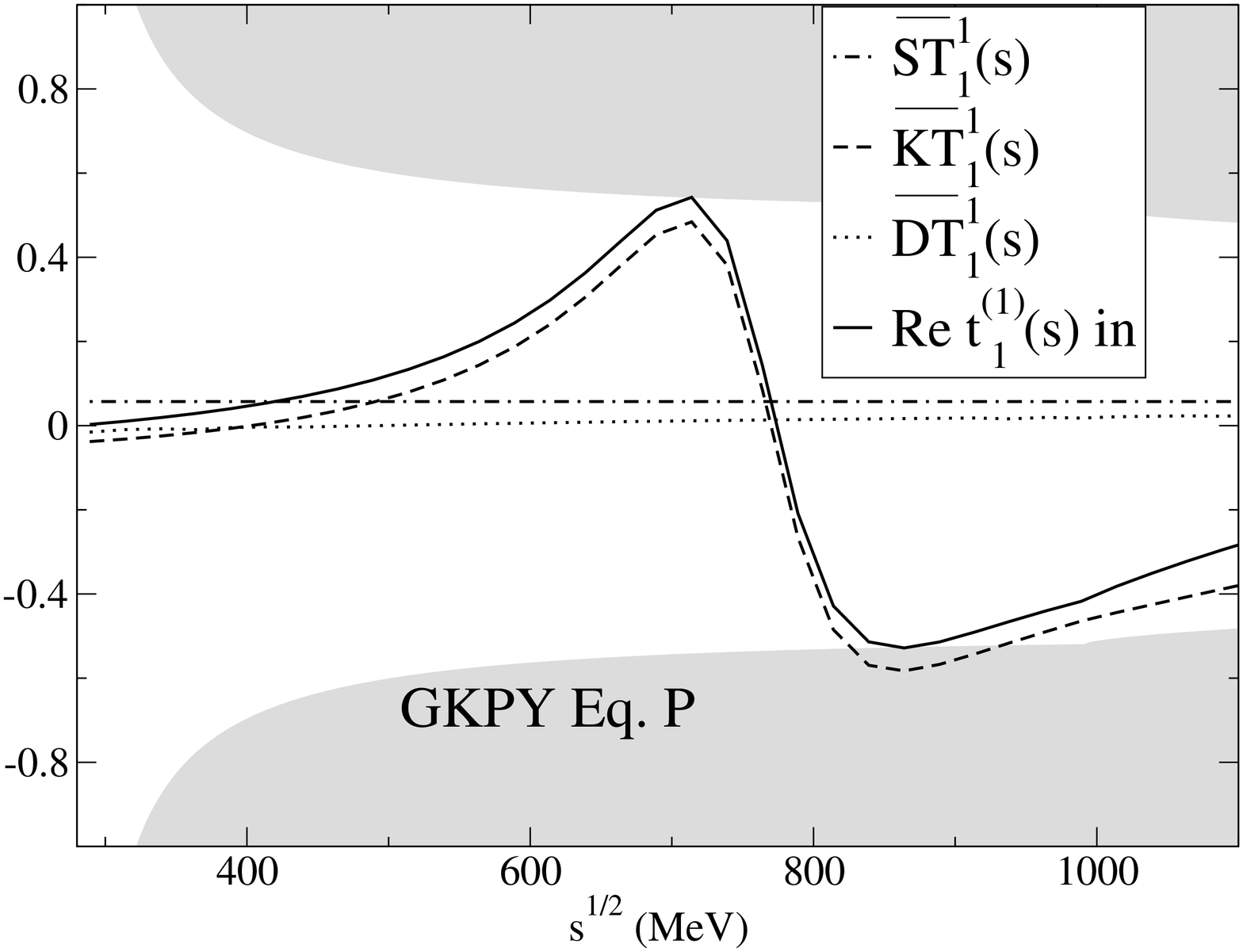}
\end{center}
\caption{
Using the UFD set as input, we show the decomposition of GKPY equations
into the subtracting term $\overline{ST}$, kernel term $\overline{KT}$, and driving term $\overline{DT}$ for the  
S0, P and S2 waves. Note the different scales used here and in
Fig.~\protect{\ref{RoyDecompositions}}.
}
\label{GKPYDecompositions}
\end{figure}

Fig.~\ref{RoyDecompositions} presents a decomposition of  Roy equations 
 for the S0, P and S2 waves
 into four parts: 
the ``in'' part, that represents 
what our parametrizations give for $\real t_\ell^{(I)}$,
the subtracting terms $ST(s)$,
the kernel terms $KT(s)$ and the driving terms $DT(s)$.
Note that, for these equations to be satisfied exactly,
the first contribution should equal the sum of the other three.
The numerical calculations have been performed by taking the
UFD amplitudes described in
the previous Sections as input. 
For illustration,
we have drawn as a gray area the region that violates the unitarity bound
 $\vert {\rm Re}\, t\vert \le \eta s^{1/2}/4k$, (note that $\eta=1$ in the elastic region).
For comparison, we present in Fig.~\ref{GKPYDecompositions} the
same decompositions for the GKPY equations. Note the very different
scales on both sets of Figures.

As can be seen in Fig.~\ref{RoyDecompositions}, the $ST(s)$ and
$KT(s)$ terms in Roy equations become huge at higher energies  and
suffer a large cancellation against each other. 
This cancellation is particularly strong
for the S0 wave, where, for a sufficiently large
energy, both terms are much larger than the unitarity
bound. For instance,   they are larger by roughly a factor of four at 750 MeV,
and of eight at 1100 MeV.

In contrast, as seen in Fig.~\ref{GKPYDecompositions} 
for the GKPY equations, Eq.~\eqref{1SEquations},
the $\overline{ST}$ terms are constant, and in fact
much smaller than the $\overline{KT}(s)$ terms, which are clearly the dominant
ones. Therefore, no big cancellations between any two terms are
needed in order to reconstruct the total real part of the amplitude.
Moreover, we have checked that the high energy part, 
which has been parametrized by means of Regge theory, 
corresponds to somewhat less than half of the total  $\overline{DT}(s)$ contribution. 
Therefore, although the $\overline{DT}(s)$ terms in the GKPY equations are
larger than in Roy equations due to the fact that there is one subtraction
less, the contribution coming from the amplitudes above 1420~MeV 
is still small compared with the dominant term $KT(s)$.
Thus, the high energy behavior is still well under control.

Note that, to keep the plots clear, we have only provided central values
for the moment. In the next section we will provide the total uncertainties 
(the uncertainties of each separated contribution were presented in a conference \cite{Kaminski:2008rh} using a very preliminary UFD set).
For our purposes it is enough to remark that 
 uncertainties follow a similar pattern to these central values.
In particular, the $ST(s)$ term in Roy equations for scalar waves
has a large uncertainty due to the poor
experimental knowledge of the $a^2_{0}$ scattering length, that becomes
 larger and larger, proportionally to $s$, 
as the energy grows, becoming dominant above roughly  450 MeV. 
 In contrast, since the GKPY $\overline{ST}$
term is constant and there are no large cancellations, the resulting
GKPY equations have a much smaller uncertainty in that region.
Actually, the errors for the GKPY equations in the three waves
come almost completely from the $\overline{KT}(s)$ terms.
At low energies, the effect is reversed and Roy equations provide a much more
stringent constraint than GKPY.
Therefore, and as we will show next, they become 
complementary ways of checking our data parametrizations at different energies.

\subsection{Consistency check of Unconstrained Fits}
\label{sec:consistency}

In order to provide a consistency measure for our parametrizations
with respect to the dispersive relations and sum rules
presented in the previous sections, we will make use 
(as we did in previous works) of a
quantity similar to an averaged $\chi^2/( d.o.f.)$ distribution.
In particular, we can consider that a
 dispersion relation $i$ is well satisfied at a point $s_n$
if the difference $\Delta_i$, defined in Eqs.~\eqref{FDR1}, \eqref{FDR2}, \eqref{RoyEquations} and \eqref{1SEquations},  
is smaller than its uncertainty $\delta\Delta_i$. 
Thus, when 
the average discrepancy verifies
\begin{equation}
\bar{d}_i^2\equiv\frac{1}{\hbox{number of points}}
\sum_n\left(\frac{\Delta_i(s_n)}{\delta\Delta_i(s_n)}\right)^2\leq1,
\label{avdiscrep}
\end{equation}
we consider that the corresponding dispersion relation 
is well satisfied within uncertainties in the energy region spanned by the 
points $s_n$.
 In practice, the values of $s_n^{1/2}$ are taken at intervals of 25 MeV
between threshold and the maximum energy where we study each dispersion relation
(1420 MeV for FDR and 1115~MeV for Roy and GKPY equations).
In addition, we have added a point below
threshold at $s=2 M_\pi^2$ for the $F_{00}$ and $F_{0+}$ FDRs.

Similarly, we define discrepancies for the sum rules in Eqs.
\eqref{eq:Isumrule} and
\eqref{eq:Jsumrule}, as follows: 
\begin{equation}
  \bar{d}^2_I=\left(\frac{I}{\delta I}\right)^2,
\quad
\bar{d}^2_J=\left(\frac{J}{\delta J}\right)^2.
\label{eq:IJdiscrep}
\end{equation}

In order to calculate the uncertainties 
$\delta\Delta_i(s_n)$, $\delta I$, $\delta J$, 
we have followed two approaches: On the one hand we have simply added in quadrature
the effect of varying  
each parameter independently in our parametrizations from $p_i$ 
to $p_i\pm\delta p_i$.
 The errors are symmetric since, in order to be conservative,
 we have always taken the largest
variation as the final error when changing the sign of $\delta p_i$.
 This is rather simple but does not
take their correlations into account.
On the other hand, we have also estimated the uncertainties using a
Monte Carlo Gaussian sampling \cite{MonteCarlo} of all CFD parameters 
(within 6 standard deviations). The uncertainties are then
slightly  asymmetric, corresponding 
to the independent left and right widths of the generated distribution
for $10^5$ events. This is, of course, much more time consuming, although
in this way we can keep part of the correlations 
in the results.
However, we have checked that both methods yield very similar results,
because the errors coming from each individual parameter are
small and the number of parameters is large. The difference between 
using one method or another is almost negligible \cite{MonteCarlo} and thus, for simplicity,
 we are providing
numbers and figures with the first one, that would be much easier to reproduce
should someone use our parametrizations.

In Table~\ref{tab:UFDdiscrepancies} we show the 
averaged squared discrepancies $\bar d_i^2$  that result
when we use the UFD set described in Sects.~\ref{sec:UFD} 
and \ref{sec:S0wave}. We are showing these discrepancies up to 
two different energy regions,
932 MeV and 1420 MeV for FDRs, and up to 992 MeV and 11115 MeV 
for both Roy and GKPY equations (note that we have kept the same definition of energy regions
as in KPY08, so that we can compare easily with the results obtained there).
Let us remark  that these discrepancies are ``squared distances'',
similar to a $\chi^2$, and so we will abuse the language and talk about average ``standard deviations'', which correspond to the {\it square root} of $\bar d_i^2$.
Still,  one has to keep in mind that these dispersion 
relations have not been fitted yet. 

Let us first concentrate in the low energy part below 932 MeV or 992 MeV.
We can observe that FDRs are reasonably 
 well satisfied: discrepancies are never beyond 1.3 standard deviations. 
Roy equations are also well satisfied, with a discrepancy below 1.2 standard deviations.
However, the GKPY equations are much more demanding:
The UFD set satisfies 
the S2 wave equation fairly well, but it does not satisfy the S0 and P wave
relations so well. Still, no
dispersion relation lies beyond 1.6 standard deviations.
This is not too bad, given the fact that we have not fitted the 
dispersion relations, but there is clear room for improvement. 
Let us recall that this is just how experimental data satisfy these
constraints, there is no theory on the UFD set. 

\begin{table}
% \hspace*{.2cm}
\begin{tabular}{|l|c|c|c|c|}
\hline
$\qquad\bar d_i^2$ & {  new UFD} & old UFD & {  new UFD} & old UFD\\
\hline
\hline
FDRs&\multicolumn{2}{c|}{\small $s^{1/2}\leq 932\,$MeV}&\multicolumn{2}{c|}{\small $s^{1/2}\leq 1420\,$MeV}\\
\hline
$\pi^0\pi^0$& {0.31} & 0.12 &  {2.13}& 0.29 \\
$\pi^+\pi^0$& {1.03} & 0.84 &  {1.11}& 0.86\\
$I_{t=1}$& {1.62}  & 0.66 &  {2.69}& 1.87\\
\hline
\hline
Roy Eqs.&\multicolumn{2}{c|}{\small $s^{1/2}\leq 992\,$MeV}&\multicolumn{2}{c|}{\small $s^{1/2}\leq 1100\,$MeV}\\\hline
S0&  {0.64}& 0.54  &{0.56}& 0.47 \\
S2&  {1.35}& 1.63  &{1.37}& 1.68 \\
P&   {0.79}& 0.74  &{0.69}& 0.65 \\
\hline
\hline
GKPY Eqs.&\multicolumn{2}{c|}{\small $s^{1/2}\leq 992\,$MeV}&\multicolumn{2}{c|}{\small $s^{1/2}\leq 1100\,$MeV}\\
\hline
S0& {1.78} & 5.0 &{2.42} & 8.6 \\
S2&  {1.19} & 0.49&{1.14} & 0.58\\
P & {2.44} & 3.1 &{2.13} & 2.7 \\
\hline
\hline
Average& 1.24 & 1.46 & 1.58 &1.97\\
\hline
%\hline 
\end{tabular}
%\vspace{5pt}
\caption{ Average discrepancies $\bar d_i^2$ of the 
unconstrained data fits (UFD set)  for each 
dispersion relation. We compare the results
of the parametrization obtained in this work (new UFD) with
those in KPY08 (old UFD set). The huge discrepancies seen in 
KPY08 for GKPY equations all come from energies above $\sim500\, \mev$. 
This is the main reason to improve our unconstrained S0 fit
as explained in Sect.~\ref{sec:improvedparametrization}.
\label{tab:UFDdiscrepancies}}
%\vspace{-20pt}
\end{table}

If we now also include the region above 932 MeV for FDRs or 
above 992 MeV for Roy and GKPY equations,
we find that the agreement deteriorates considerably:
four relations lie between 1.4 and 1.65 average standard deviations, but not beyond that. 
Fortunately we will get 
much better fulfillment of dispersion relations in all regions by
allowing for a small variation of 
the parameters in the constrained fits to be discussed below.

\begin{figure}
\begin{center}
%\vspace*{-20pt}
\includegraphics[width=7.3cm]{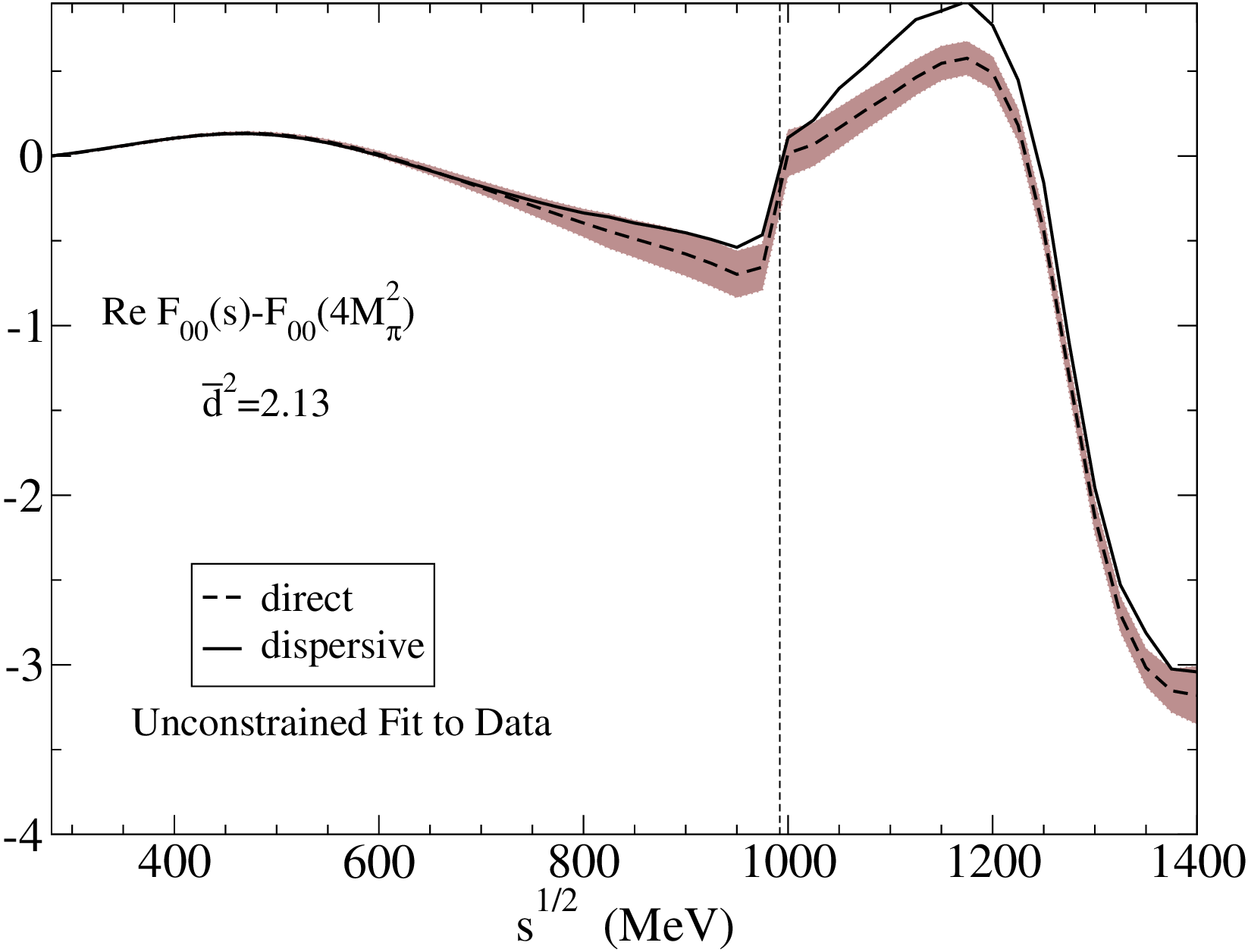}

\vspace{.2cm}
\includegraphics[width=7.3cm]{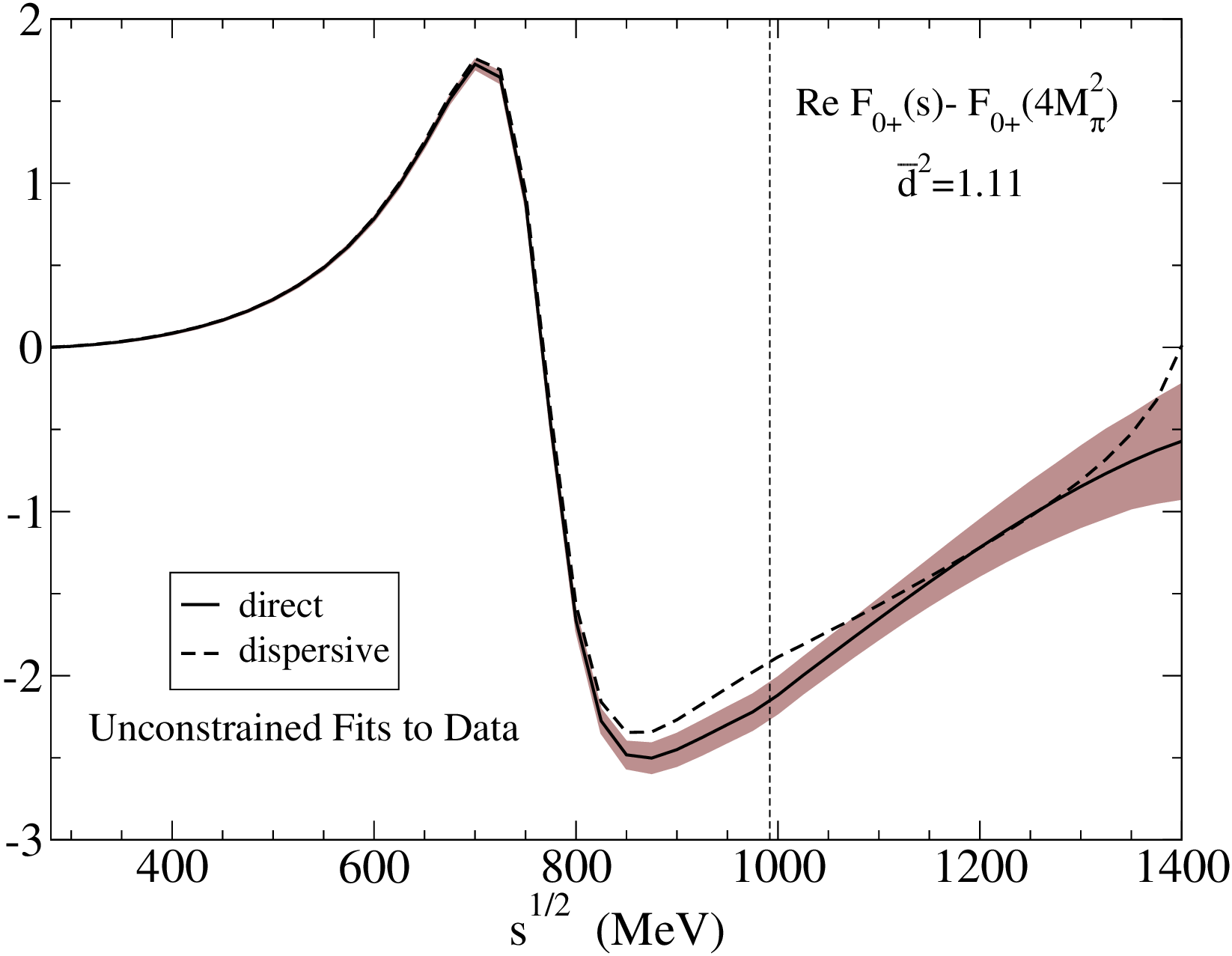}

\vspace{.2cm}
\includegraphics[width=7.3cm]{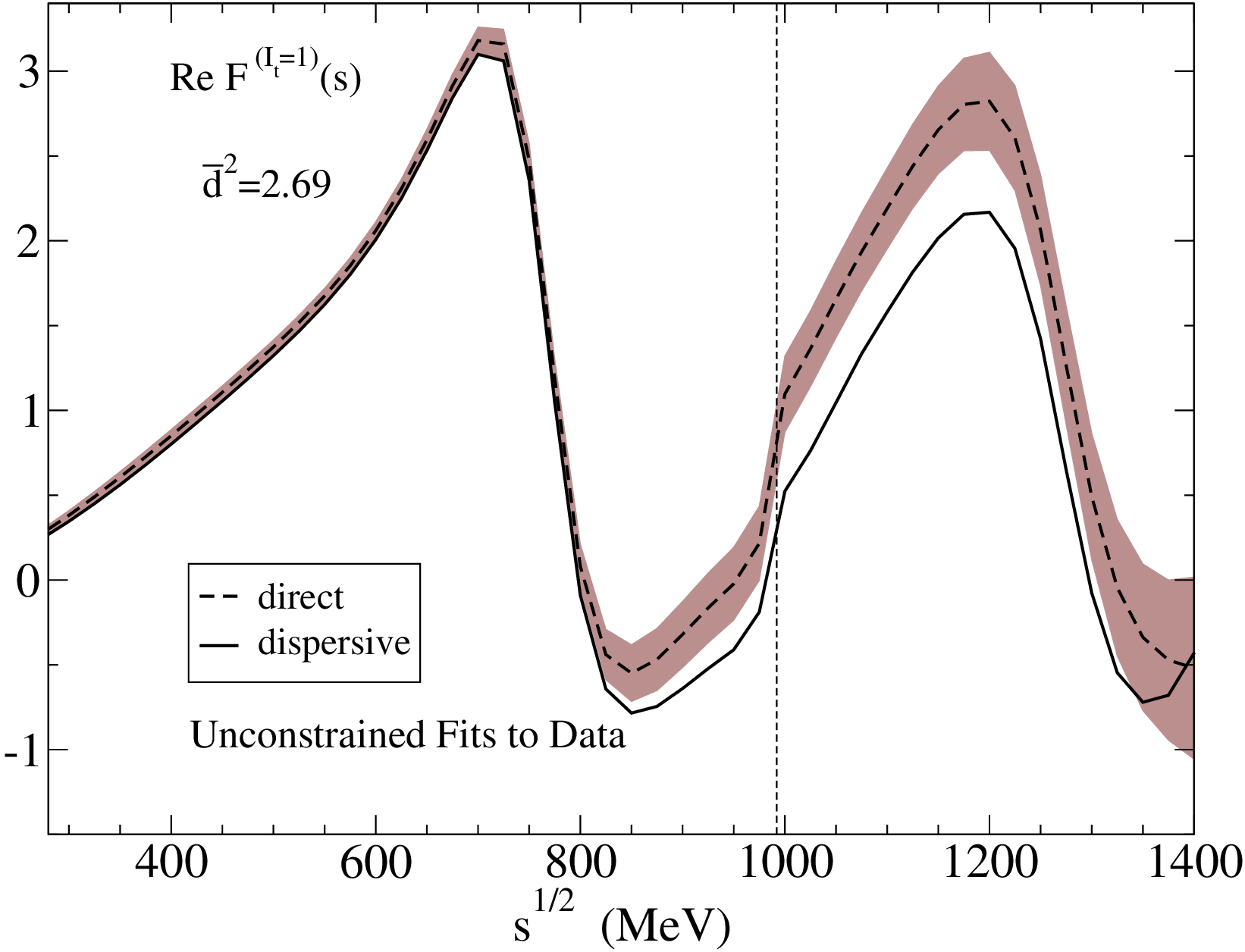}
\end{center}
\vspace*{-18pt}
\caption{Results for Forward Dispersion Relations. Dashed lines: real part,
  evaluated directly with the UFD parametrizations. Continuous lines: the
  result of the dispersive integrals. The dark bands cover the uncertainties
  in the difference between both. From top to bottom: (a) the $\pi^0\pi^0$
  FDR, (b) the $\pi^0\pi^+$ FDR, (c) the FDR for $I_t=1$ scattering. The
  dotted vertial line stands at the $\bar{K}K$ threshold.
\label{fig:UFD-FDR-plots}}
\vspace*{-10pt}
\end{figure}  

Let us also remark that the two sum rules,
Eqs.~\eqref{eq:Isumrule} and \eqref{eq:Jsumrule},
are  satisfied within 1.9 and 0.3 standard deviations.
Even for the first one, this is still a fair  agreement, because,
in practice, both of them correspond to a one-order of 
magnitude cancellation between the low and high energy contributions
to the sum rules, which, in these 
UFD set are determined from uncorrelated data fits.

Also in Table~\ref{tab:UFDdiscrepancies} we show
the average discrepancies for the old UFD set in KPY08.
With regard to FDRs and Roy equations, it is evident that the new UFD fit
is doing worse than the one in KPY08.
Nevertheless, one should keep
in mind that the new S0 wave 
has reduced its uncertainty at low energies by somewhat more than
10\%, because the published NA48/2 data are more precise and also because
we are discarding the controversial $K\rightarrow2\pi$ datum.
For that reason, one would have expected the averaged squared discrepancies
to look now bigger by 
as much as 20 or 30\% whenever the S0 wave contributes significantly
to the dispersion relation. With this correction in mind, 
the deterioration is not so significant. Nevertheless, we want to insist that this is 
basically 
due to the new results of NA48/2 and our 
getting rid of the $K\rightarrow 2 \pi$ datum. The data have changed.

Why do we then claim to have improved the S0 wave in this work?
The answer comes from GKPY equations, which,
as we already explained,
are much more precise than Roy equations above roughly 450~MeV for the S0 wave, 
given the present experimental input. 
It is clear that the KPY08 UFD parametrization
satisfies the S0 GKPY equation very poorly at any energy and 
is not satisfying the low energy P GKPY equation very well.
For that reason, we have improved the matching and the data selection,
so that our new UFD parametrization, that will be our starting point 
for the constrained fits, satisfies GKPY equations much better without
spoiling FDR and Roy equations.  The improvement due to
the  new unconstrained S0 wave fit
is obvious  from Table~\ref{tab:UFDdiscrepancies},
particularly in the S0 GKPY equation. 
Up to 1100~MeV, the old UFD set from KPY08 had an
averaged squared discrepancy of 8.6, whereas the new UFD set
has 2.42. This huge improvement on the S0 wave has been
compensated by some deterioration 
in other relations at high energy, 
so that the averaged discrepancy up to high energies 
is  reduced only from 1.97 to 1.58.
Note that the change in the inelasticity parameter, 
that now shows a much 
bigger dip in the 1000 to 1100 MeV region, as shown in 
Fig.~\ref{fig:S0-UFD-inel-comparison}, plays a relevant 
role in this dramatic improvement. This dip structure is thus
favored by the GKPY equations, something 
that could not be seen with standard Roy equations since their
uncertainties in that region are huge. We will discuss this in detail in
Sect.~\ref{sec:nodip}.
At low energies, the average 
squared discrepancy has been reduced very little,
from 1.46 down to 1.24.
Of course, let us remark once again that our uncertainties are now 10-15\%
smaller in the S0 wave at low energies, so that the improvement
is actually bigger than it seems just from the numbers in the table. 

Let us mention here that the inclusion of the new terms parametrizing 
a crude dependence on the $\eta$ momentum above $\eta\eta$ threshold, 
help reducing the average squared distances by 6\%, namely, from 1.68 to 1.58. In particular, the averaged
squared discrepancies $\bar d_i^2$ for the S0 GKPY equation decrease from 3.02
to 2.42 and for the $F_{00}$ FDR equation from 2.35 to 2.13.

\begin{figure}
\begin{center}
\includegraphics[width=7.4cm]{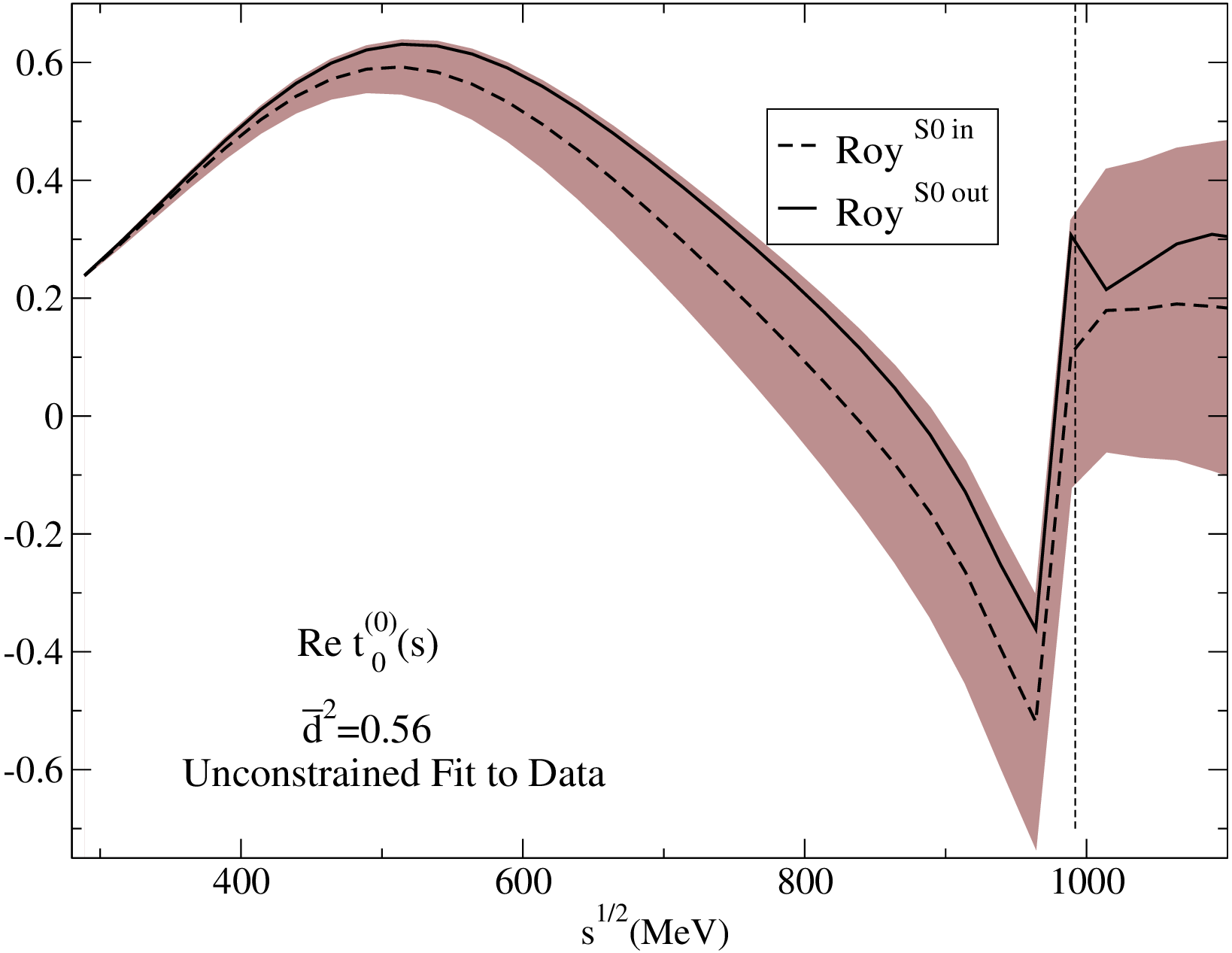}
\includegraphics[width=7.4cm]{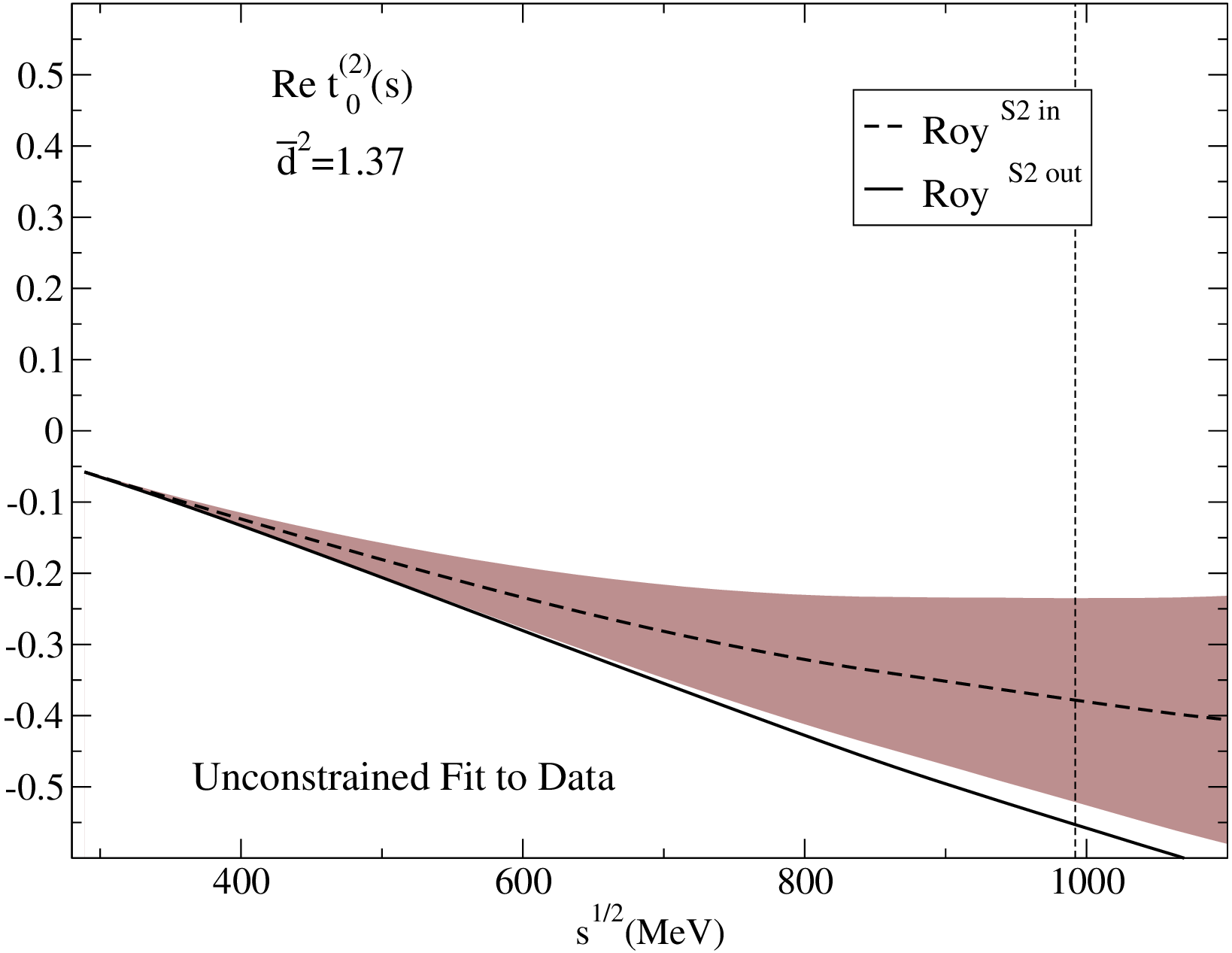}
\includegraphics[width=7.4cm]{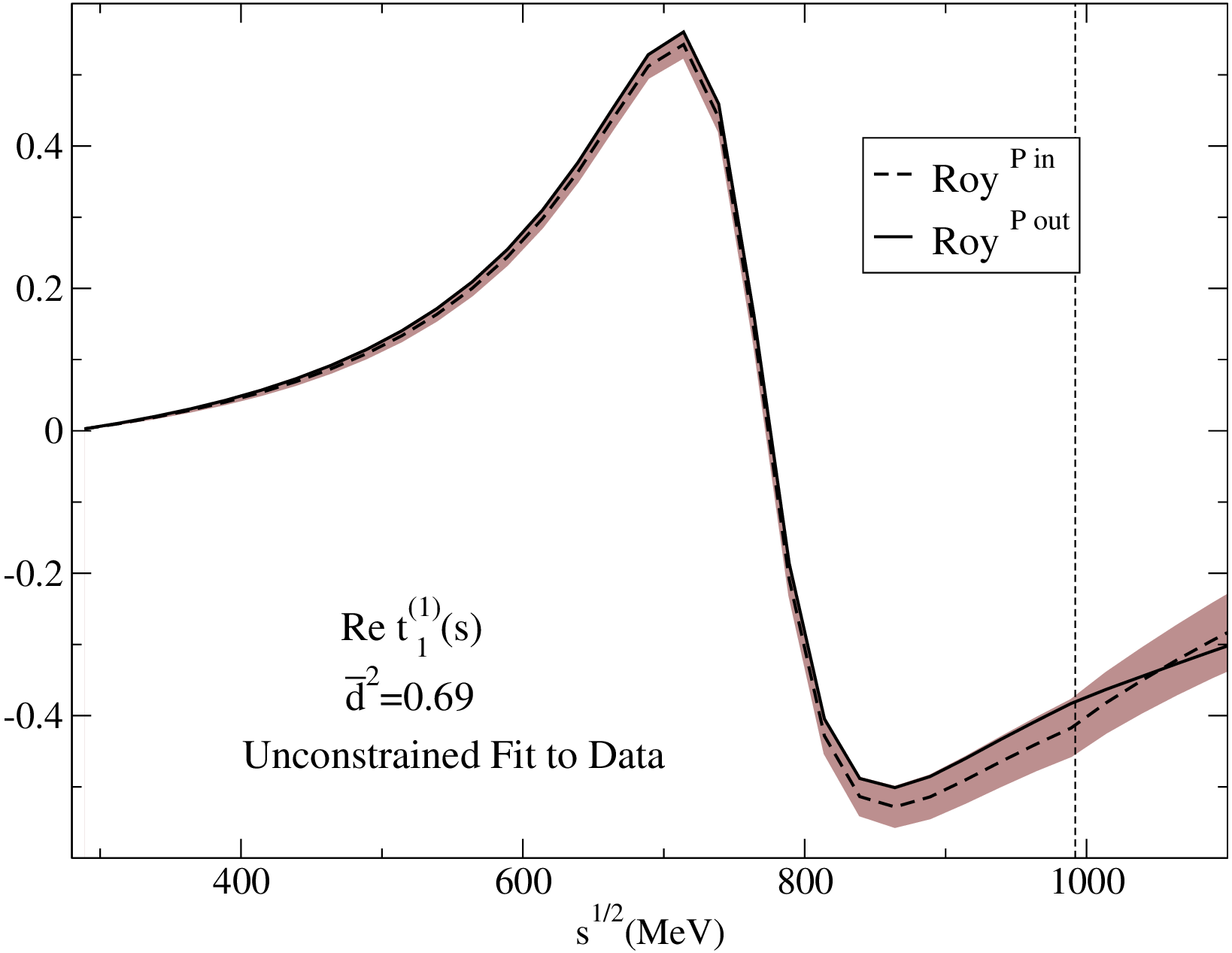}
\end{center}
\vspace{-16pt}
\caption{Results for Roy Equations. Dashed lines (``in''): real part, 
evaluated directly with the UFD parametrizations. 
Continuous lines (``out''): the result of the dispersive representation. 
The gray bands cover the uncertainties in the difference between both. 
From top to bottom: (a) S0 wave, (b) S2 wave, (c) P wave. The
  dotted vertial line stands at the $\bar{K}K$ threshold.
\label{fig:UFD-Roy-plots}}
\vspace*{-15pt}
\end{figure}

\begin{figure}
\begin{center}
\includegraphics[width=7.4cm]{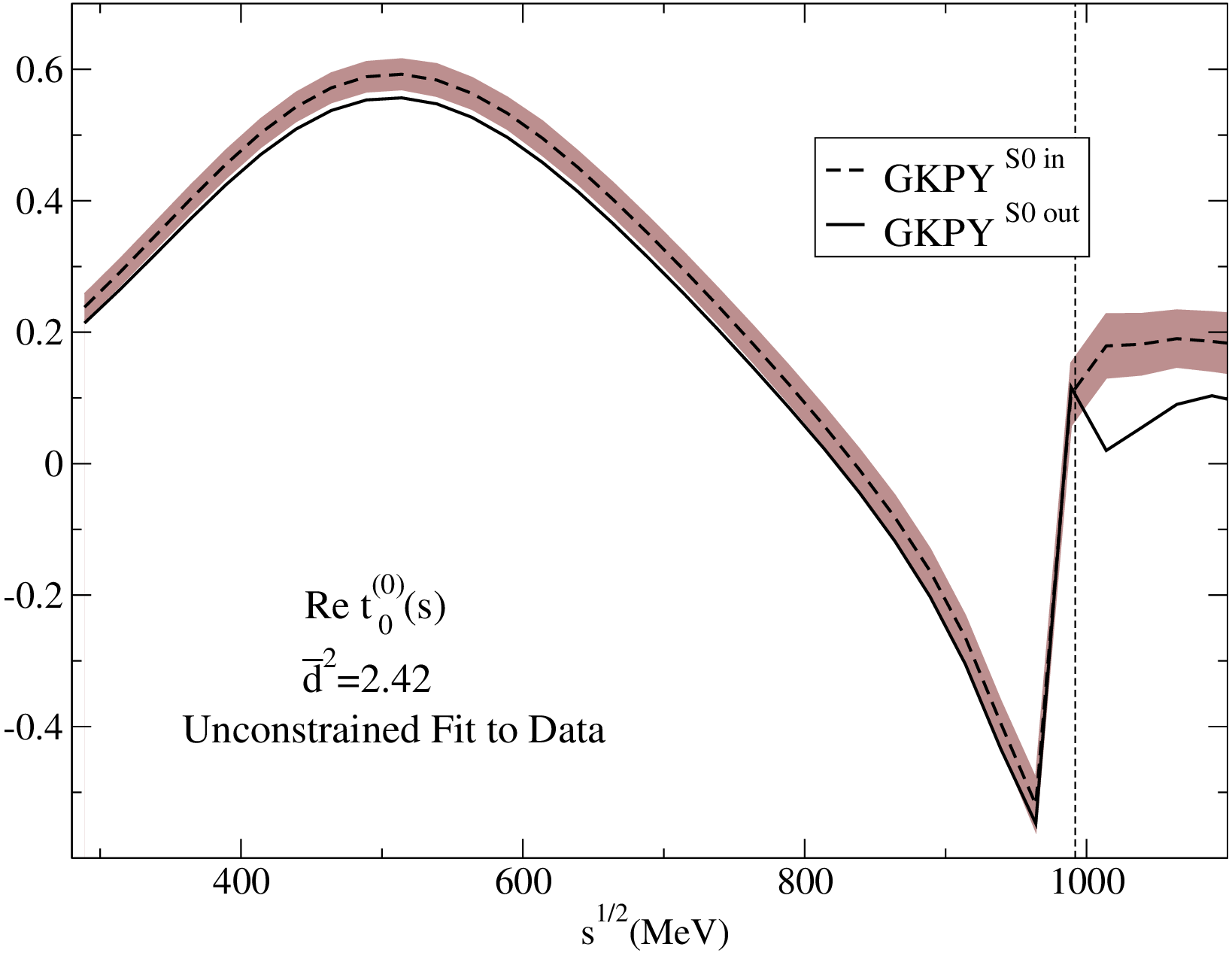}
\includegraphics[width=7.4cm]{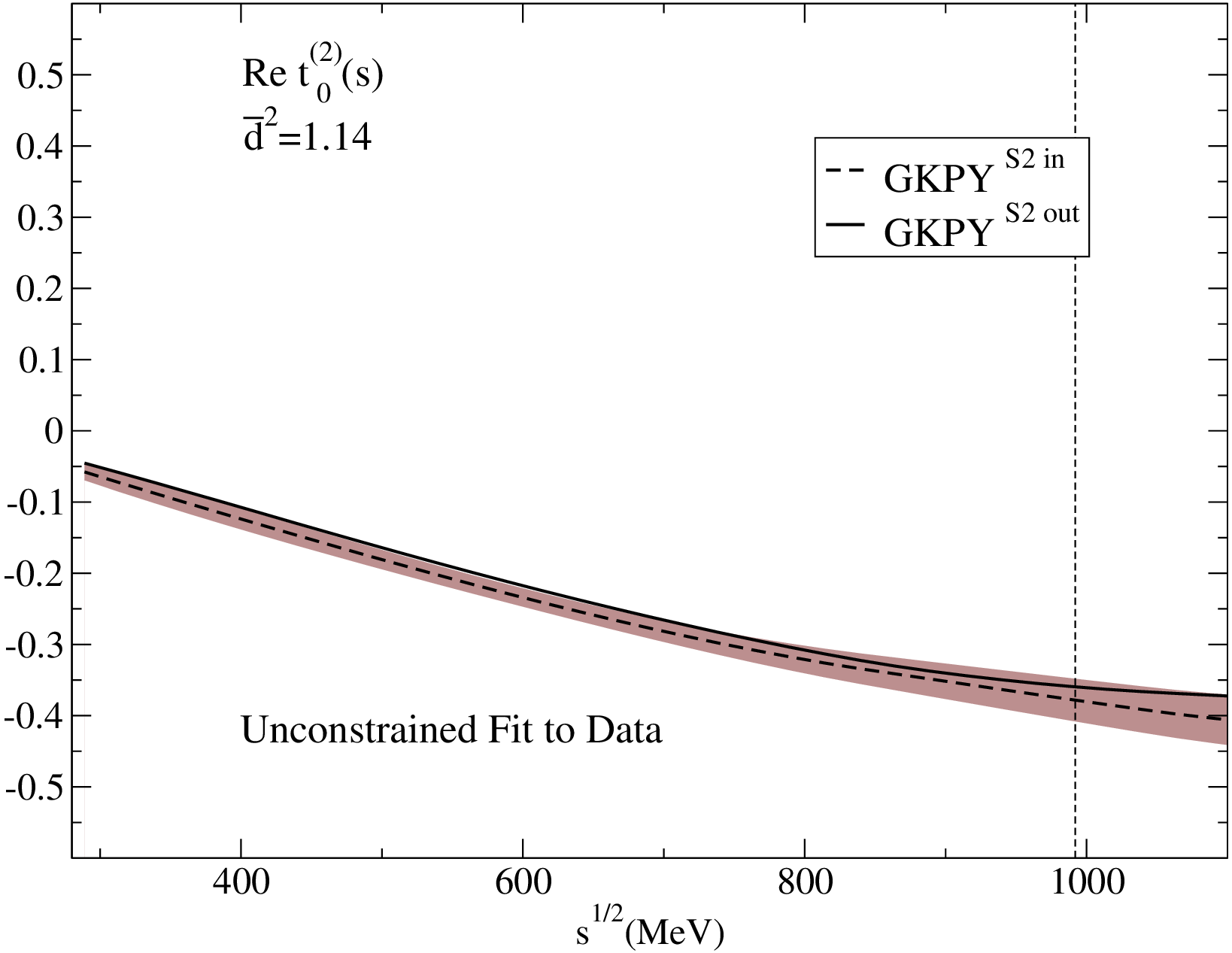}
\includegraphics[width=7.4cm]{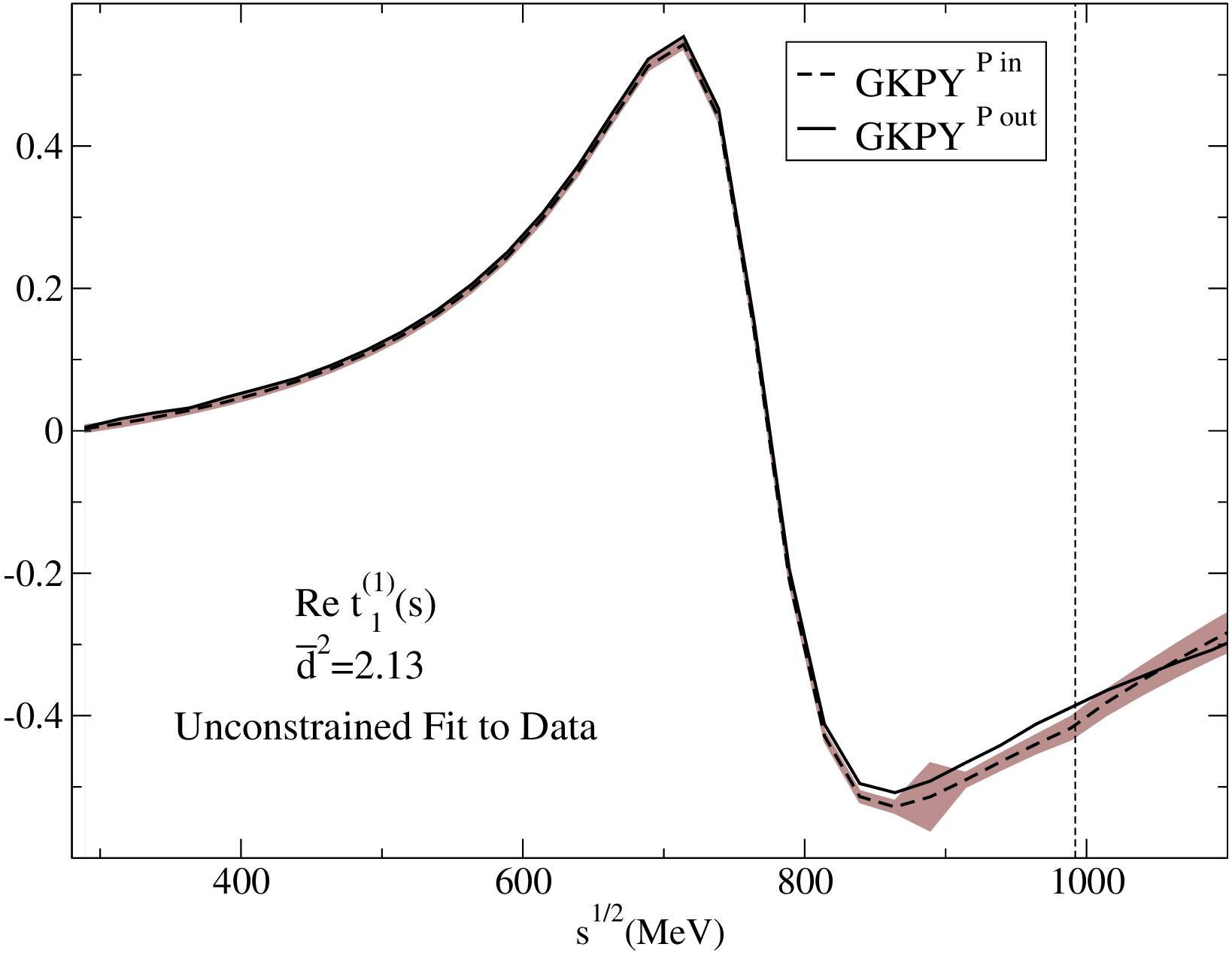}
\end{center}
\vspace{-16pt}
\caption{Results for GKPY equations. Dashed lines (``in''): real part, 
evaluated directly with the UFD parametrizations. 
Continuous lines (``out''): the result of the dispersive representation. 
The gray bands cover the uncertainties in the difference between both. 
From top to bottom: (a) S0 wave, (b) S2 wave, (c) P wave.
Note how these uncertainties are much smaller above 450 MeV
than those from standard Roy equations shown in Fig.~\ref{fig:UFD-Roy-plots}. The
  dotted vertial line stands at the $\bar{K}K$ threshold.
\label{fig:UFD-GKPY-plots}}
\vspace*{-15pt}
\end{figure}

Up to now, we have studied the overall uncertainties, but 
in Fig.~\ref{fig:UFD-FDR-plots} we show to what extent FDRs are satisfied 
by the UFD set, as a function of energy. Of course, the best 
fulfillment is found at lower energies.
In Fig.~\ref{fig:UFD-Roy-plots} we show how the usual, twice subtracted
Roy equations are satisfied by the UFD set. Here, as we did in Sect.~\ref{sec:decomposition},
we denote by ``in''  what our parametrizations give for $\real t_\ell^{(I)}$, 
whereas we denote by ``out'' the result of the dispersive representation
from Roy equations, namely, the subtraction constant terms, plus the kernel terms,
plus the driving terms  in Eq.~\eqref{RoyEquations}.
Finally, in Fig.~\ref{fig:UFD-GKPY-plots} we show how the new, once subtracted,
GKPY equations are satisfied by the UFD set. We follow the same ``in'' and ``out'' notation as for Roy equations.
 
Comparing Fig.~\ref{fig:UFD-Roy-plots}
with Fig.~\ref{fig:UFD-GKPY-plots}, it is clear that, given the present experimental input,
 the uncertainty band for GKPY
equations is much smaller than that for Roy equations above 450 MeV, whereas the opposite
occurs at lower energies. Therefore, as we have emphasized repeatedly,
 the new GKPY equations represent a much 
stronger constraint in the intermediate energy region than standard Roy equations.

In summary, with the new S0 unconstrained fit, all dispersion relations are satisfied in the different
energy regions within less than 1.6 standard deviations 
in the low energy regime, 
and 1.7 including the intermediate energies. 
This is a fairly reasonable fulfillment, 
given the fact that the information
 about analyticity has not been included as a constraint in the UFD description.
Nevertheless, it is obvious that there is room for improvement, 
which is what we will do by obtaining constrained data fits in the next section.

\section{Fits to Data Constrained by Dispersion Relations}
\label{sec:CFD}

 In previous works (PY05, KPY08) we had improved the consistency  of our
description of $\pi\pi$ scattering amplitudes
 by imposing FDR and Roy equations fulfillment  within
uncertainties.
As we have just seen in the previous section, 
the GKPY equations provide 
a much more stringent constraint in the intermediate energy region
than standard Roy equations  and thus it now makes sense to impose
 the new GKPY equations {\it as an additional constraint} in a new set of Constrained Fits to Data (CFD set).

\subsection{Minimization procedure}

Our goal is then to obtain a fit to data, by changing the UFD parametrizations slightly,
that fulfills each dispersion relation within errors. 
As we did in \cite{Kaminski:2006qe}, 
we will now use the average discrepancies $\bar d_i^2$, defined in Eqs. \eqref{avdiscrep}
and \eqref{eq:IJdiscrep},
to obtain these constrained fits, 
by minimizing:
\begin{equation}
 \sum_i W^2_i \bar{d}_i^2
 +\bar{d}^2_I+\bar{d}^2_J+\sum_k\left(\frac{p_k-p_k^{\rm exp}}{\delta p_k}\right)^2,
\label{tominimize}
 \end{equation}
where $i$ runs over the three FDRs, the three Roy  and the three GKPY equations.
Here, we denote by $p_k^{exp}$ 
all the parameters of the UFD parametrizations
for each wave or Regge trajectory. In this way
we force the previous data parametrizations 
to satisfy dispersion relations and 
sum rules within uncertainties. 
In KPY06 and KPY08 a common weight of $W_i^2\sim9$ 
was estimated from
the typical number of degrees of freedom needed to describe 
the shapes of the output. This value ensured that every single
dispersion relation was fairly well described by 
the KPY08 constrained data fits
up to the matching energy used in that work, namely, 932~MeV.  

However we are now
considering partial waves up to 
1115 MeV. For most waves, this extension does not
alter significantly their shape and $W_i=3$ 
is still a good weight. Nevertheless, we have less points in 
the region above 932 MeV
and if we want the fit to give not just a good average $\bar d_i^2$, but
also a good description for each wave, some of these waves need 
further weight on the high energy region, in particular if 
 their UFD $\bar d_i^2$ 
 was larger than 2. 
For this purpose, we have increased
$W_i$ up to 3.5 for the high energy parts of the
$F_{00}$, $F^{(I_t=1)}$, as well as 4.2 for the GKPY P wave in the whole energy region.
Finally, we have increased $W_i$ up to 7 for the high energy 
part of the S0 GKPY equation. 
The latter was to be expected,
since in this region there is a lot more of structure,
both in the phase and inelasticity, due to the presence of the $f_0(980)$.
These  values are not arbitrary, 
since they have been obtained by
increasing each $W_i$ gradually, starting
from 3, until the $\bar d_i^2$ are below or very close to one
{\it uniformly} throughout the whole energy range, for all 
dispersion relations obtained from the constrained fit.
This uniformity is very relevant to avoid dispersive constraints
being  badly satisfied in some small energy 
region despite the averaged $\bar d^2_i$
still remaining below~1.

Before proceeding further, let us recall that, strictly speaking,
 the quantity that we minimize
in Eq.~\eqref{tominimize} is {\it not} a $\chi^2$, but that each individual
$\bar d_i^2$ is a measure of how well each dispersion relation is satisfied.

\subsection{Variation of the S2 Adler zero}
\label{sec:ADlerS2}

As we have seen in Section~\ref{sec:improvedparametrization}, 
in the parametrization of each scalar wave
we explicitly factorized a zero in the subthreshold region.
These are  
the Adler zeros required by chiral 
symmetry constraints \cite{Adler:1964um}. Actually, we fixed
them to $\sqrt{s_A^{S0}}\equiv\sqrt{M_\pi^2/2}\simeq 99\,$MeV
and $\sqrt{s_A^{S2}}\equiv\sqrt{2M_\pi^2}\simeq 197\,$MeV, which are
their current algebra values (leading order ChPT). 
Of course, once these UFD parametrizations are used inside the S0 and S2 
Roy or GKPY  equations we can also obtain 
the {\it dispersive result} for the S0 and S2 Adler zeros, 
that we provide in Table~\ref{tab:Adlerzeros}. 

In order to determine the
positions of Adler zeros better when making constrained fits in KPY08,
we allowed them to change within 
the dispersive uncertainties obtained from the UFD set.
However, in this work we will not insist on $z_0/\sqrt{2}$  reproducing
the S0 wave Adler zero very precisely. 
The reason is that, as we see in Table~\ref{tab:Adlerzeros},
the uncertainties in $\sqrt{s_A^{S0}}$ obtained either from Roy or GKPY equations 
are huge, and setting $z_0$ free introduces a spurious and extremely correlated
source of error. 
In addition, in KPY08,
the $z_0$ central value moved in the wrong direction~\footnote{We thank H. Leutwyler for his comments and suggestions
on this issue.}. 
In addition, as already explained in Sect.~\ref{sec:improvedparametrization},
the S0 wave Adler zero lies close to the border of the conformal circle,
i.e., $w(s_A^{S0})\simeq-1$, where the conformal expansion coverges very slowly.
We simply have to accept that our S0 wave conformal expansion is not very accurate around the Adler zero.
Of course, this  is irrelevant for the integrals in the physical region and has a negligible influence
in the set of constrained fits we will obtain next.

% \afterpage{
% \begin{widetext}
\begin{table}
\centering
  \begin{tabular}{lcccc}
 \hline
 \hline
   & Roy Eqs. &  GKPY Eqs.&  Roy Eqs. & GKPY Eqs.~\\
   & with UFD &  with UFD &  with CFD & with CFD \\
$\sqrt{s_{A}^{S0}}$ & $112\pm 24$   & $120\pm30$   & $83\pm32$   & $85\pm34$   \\
$\sqrt{s_{A}^{S2}}$ & $189\pm 11$   & $200\pm6$   & $200\pm10$   & $201\pm5$   \\
 \hline
 \hline
\end{tabular}
\vspace{.2cm}
   \caption{Adler zero positions $\sqrt{s_A}$, in MeV, for the S0 and S2 waves,
obtained from Roy or GKPY equations using either the parametrizations form the UFD or CFD sets.
}\label{tab:Adlerzeros}
\end{table}
% \end{widetext}
% }

In contrast, the S2 Adler zero obtained from the dispersive representation 
 moves very little 
from its current algebra value and its uncertainty is rather small. 
The reason for this difference in uncertainties is, for a good part,
that the S0 wave Adler zero
lies very close to the left cut, whereas
the S2 Adler zero is not so far from threshold and is
quite well determined when data is used  as input of 
either Roy or, even better, 
GKPY equations. For that reason, we still allow the S2 Adler
zero to vary when making the constrained fits, using as a starting point
 the weighted average of the values 
obtained from the UFD set inside Roy and GKPY equations,
namely, $\sqrt{s_{A}^{S2}}=197.7\pm5.1\,\mev$.

\subsection{The Constrained Fits to Data (CFD)}

The resulting parameters for the CFD are gathered in the tables of the
Appendix A.
It is reassuring to observe that, except for the S0 wave at intermediate energies, the values of the parameters do not change much
from the UFD to the CFD sets, as could be expected, since, as 
we saw in Table~\ref{tab:UFDdiscrepancies}, the UFD fulfillment
of dispersive constraints only needed some improvement, but not a radical change.
In particular, the GKPY equation for the S0 wave
is very well satisfied 
 in the CFD at the expense of an average change of 0.82 standard deviations
in the high energy parameters and almost no change in the low energy ones. 
Certainly, most of this change is concentrated
in the parameters $c$ and $\tilde \epsilon_1$ in Eqs.~\eqref{eq:newparam} and \eqref{eq:inelasticityS0}. 
We will discuss below that the resulting phase after this 
change still describes the  phase shift and inelasticity data
fairly well,  
but tends to make the $f_0(980)$ somewhat wider.
The D2 wave is the one that deviates most  from its unconstrained 
parametrization, 
but its parameters are, on average, within 
1.4 standard deviations of their UFD value.
This could be expected, as was already commented in our previous
works \cite{Pelaez:2004vs,Kaminski:2006qe}, since, together with the S0 at high energy,
it is probably the one where data has the worst quality.
The parameters of the other waves, or those of the Regge parametrizations, do not 
deviate---on the average---beyond 0.6  standard deviations from their UFD values.
In Table~\ref{tab:ponderated phases} in Appendix~\ref{sec:ponderatedphases} we
provide the S0, P and S2 phase-shifts that result from using the CFD set
inside the dispersive representation.

\begin{table}
% \hspace*{.2cm}
\begin{tabular}{|l|c|c|}
\hline
FDRs&{\small $s^{1/2}\leq 932\,$MeV}&{\small $s^{1/2}\leq 1420\,$MeV}\\
\hline
$\pi^0\pi^0$& 0.32 & 0.51\\
$\pi^+\pi^0$& 0.33 & 0.43\\
$I_{t=1}$& 0.06 & 0.25\\
\hline
\hline
Roy Eqs.& $s^{1/2}\leq 992\,$MeV & $s^{1/2}\leq 1100\,$MeV \\
\hline
S0& 0.02&0.04\\
S2& 0.21&0.26\\
P& 0.04& 0.12\\
\hline
\hline
GKPY Eqs.& $s^{1/2}\leq 992\,$MeV & $s^{1/2}\leq 1100\,$MeV \\
\hline
S0& 0.23&0.24\\
S2& 0.12&0.11\\
P& 0.68& 0.60\\
\hline
\hline
Average&0.22&0.28\\
\hline 
\end{tabular}
%\vspace{5pt}
\caption{ Average discrepancies $\bar d_i^2$ of the 
Constrained Fits to Data (CFD)  for each dispersion relation. \label{tab:CFDdiscrepancies}}
%\vspace{-20pt}
\end{table}

\begin{figure}
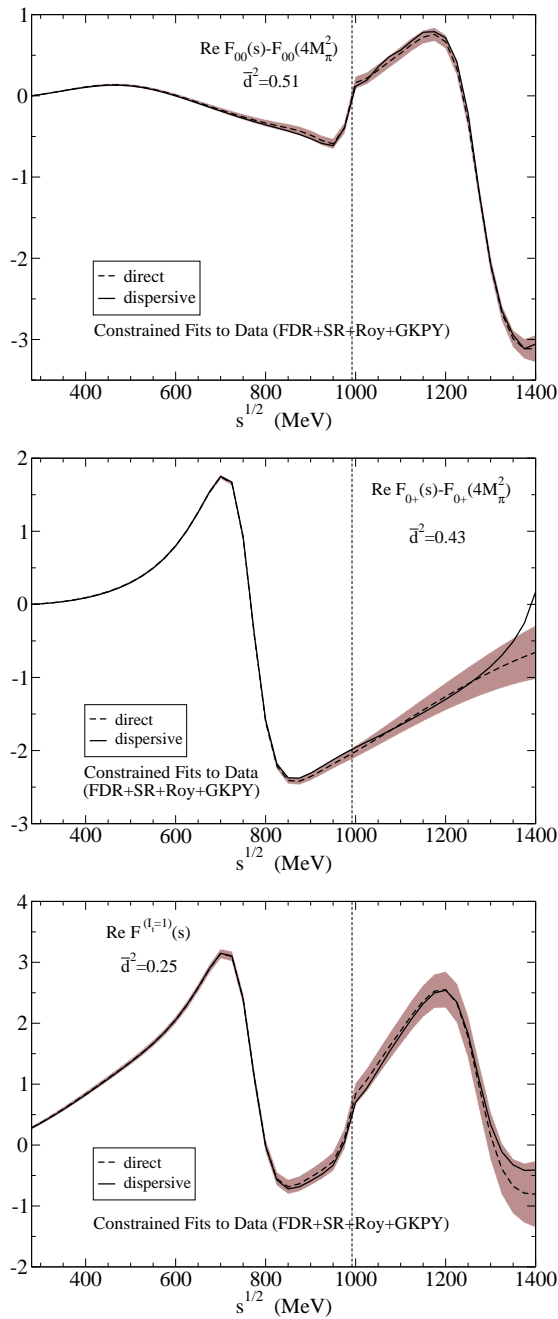

\begin{center}
%\vspace*{-20pt}
\includegraphics[width=7.3cm]{CFD-disp00.eps}

\vspace{.2cm}
\includegraphics[width=7.3cm]{CFD-disp0p.eps}

\vspace{.2cm}
\includegraphics[width=7.3cm]{CFD-dispolsson.eps}
\end{center}
\vspace*{-18pt}
\caption{Results for forward dispersion relations. Dashed lines: real part, evaluated directly with the CFD parametrizations. Continuous lines: the result of the dispersive integrals. The dark bands cover the uncertainties in the difference between both. From top to bottom: (a) the $\pi^0\pi^0$ FDR, (b) the $\pi^0\pi^+$ FDR, (c) the FDR for $I_t=1$ scattering.
\label{fig:CFD-FDR-plots}}
\vspace*{-10pt}
\end{figure}  
 
In Table~\ref{tab:CFDdiscrepancies} we list the averaged discrepancies that result
when we use the constrained fits (CFD) inside the dispersion relations. Let us remark that
all discrepancies are now below one, 
and very similar both for the low energy region
and also when including the high energy region.
This shows a 
remarkable average consistency and homogeneity for this new
set of data parametrizations. Let us recall that we only constrain our fits to satisfy 
dispersion relations up to
1420 MeV for FDR and 1115 MeV for Roy and GKPY equations.
Consequently, we expect the dispersive representation to be  somewhat worse
satisfied in the region near the maximum energy under consideration. This is indeed observed 
since the average squared discrepancies 
are somewhat smaller below 1 GeV than up to the maximum 
energy, where we usually find the point satisfying the dispersion relations worse. 

Furthermore, as already commented, the updated selection and treatment
of the S0 wave data has decreased
the S0 wave uncertainties by roughly 10 to 15\%.
This means that the consistency shown by the average 
discrepancies in Table~\ref{tab:CFDdiscrepancies} 
is even  better than it looks when comparing with 
similar results given in KPY08 for FDR and Roy equations, since we are getting a very good
consistency with slightly smaller uncertainties.

\begin{figure}
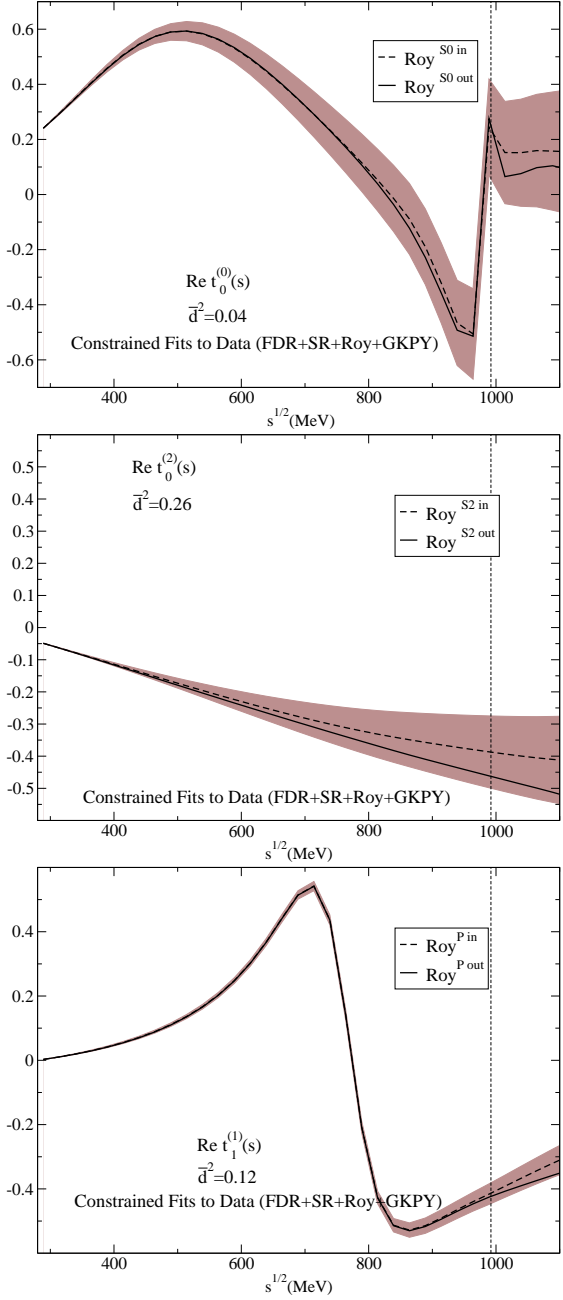

\begin{center}
%\vspace*{-15pt}
\includegraphics[width=7.4cm]{CFD-roys0.eps}
\includegraphics[width=7.4cm]{CFD-roys2.eps}
\includegraphics[width=7.4cm]{CFD-royp.eps}
\end{center}
\vspace{-16pt}
\caption{Results for Roy equations. Dashed lines (``in''): real part, 
evaluated directly with the CFD parametrizations. 
Continuous lines (``out''): the result of the dispersive representation. 
The gray bands cover the uncertainties in the difference between both. 
From top to bottom: (a) S0 wave, (b) S2 wave, (c) P wave.
\label{fig:CFD-Roy-plots}}
\vspace*{-15pt}
\end{figure}

\begin{figure}
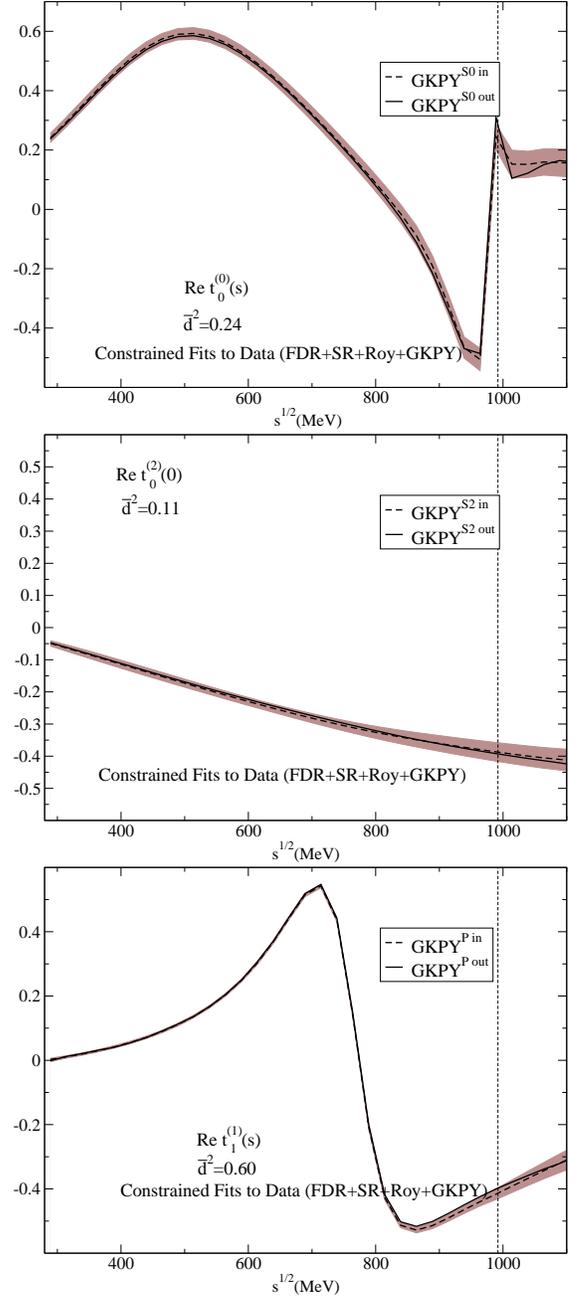

\begin{center}
%\vspace*{-15pt}
\includegraphics[width=7.4cm]{CFD-gkpys0.eps}
\includegraphics[width=7.4cm]{CFD-gkpys2.eps}
\includegraphics[width=7.4cm]{CFD-gkpyp.eps}
\end{center}
\vspace{-16pt}
\caption{Results for GKPY equations. Dashed lines (``in''): real part, 
evaluated directly with the CFD parametrizations. 
Continuous lines (``out''): the result of the dispersive representation. 
The gray bands cover the uncertainties in the difference between both. 
From top to bottom: (a) S0 wave, (b) S2 wave, (c) P wave.
Note how these uncertainties are much smaller above 450 MeV
than those from standard Roy Eqs. shown in Fig.~\ref{fig:CFD-Roy-plots}
\label{fig:CFD-GKPY-plots}}
\vspace*{-15pt}
\end{figure}

As we did for the UFD set, we now show in Figs.~\ref{fig:CFD-FDR-plots}, 
\ref{fig:CFD-Roy-plots} and \ref{fig:CFD-GKPY-plots}, how well the CFD set
satisfies FDR, Roy and GKPY equations respectively. The improvement 
in the consistency of the CFD set over the UFD is evident by comparing these plots with their UFD counterparts in Figs.~\ref{fig:UFD-FDR-plots}, 
\ref{fig:UFD-Roy-plots} and \ref{fig:UFD-GKPY-plots}.

Finally, the two sum rules in Eqs.~\eqref{eq:Isumrule} and \eqref{eq:Jsumrule}
are also remarkably well satisfied,
within 0.93 and 0.1 standard deviations, respectively.
In particular, the 1.9 standard deviations 
for the sum rule in Eq.~\eqref{eq:Jsumrule} using the UFD set are reduced dramatically,
 and this implies now a two orders of magnitude
cancellation between the low and high energy contributions.

\section{Threshold  parameters and Adler zeros}
\label{sec:thresholdand Adler}

Apart from the additional GKPY equations, 
the main novelty of this work
is the S0 wave improvement, both in its parametrization and data
analysis. Thus, naively, one may not expect a big variation 
in the low energy part of the other waves
with respect to previous works.

However, let us recall that, as we did in KPY08,
 we calculate most threshold parameters
from sum rules.  Thus, the changes in the
S0 wave can also affect the calculation of 
these low energy parameters for other waves. In  
particular, when using sum rules with one subtraction, the intermediate
energy part of our parametrizations, now constrained by GKPY equations,
also plays a relevant role in our final results. 
In this section 
we will thus recalculate all these threshold parameters with the new CFD set.
Actually, we will find that not only the S0 wave, but the D wave threshold parameters
suffer sizable modifications.

Finally, in previous works we did not use the dispersive or
sum rule techniques to determine with precision 
the position of Adler zeros, 
which are required by chiral symmetry 
in the subthreshold region of the S0 and S2
waves, and are therefore of interest for Chiral Perturbation Theory.
Also in this section we will determine them 
using the Roy and GKPY equations
with the CFD set as input for the integrals.

\subsection{Sum rules for threshold parameters}

We list in Table \ref{tab:thresholdparameters} 
the values of threshold parameters  for
all the partial waves we considered in this analysis, namely: S0, S2, P, D0, D2
and F. In addition, we provide values for 
$a_0^{(0)}-a_0^{(2)}$, $2a_0^{(0)}-5a_0^{(2)}$  and
$\delta_0^{(0)}(M_K^2)-\delta_0^{(2)}(M_K^2)$, 
since these parameters are of relevance for 
pion atoms, scalar threshold parameters, and kaonic decays. 
In the second and third columns, we provide the results from the UFD and CFD sets.
We already commented that the CFD 
parametrizations change only very slightly compared to the UFD,
and this is well corroborated by the fact that all the UFD and CFD results
on Table \ref{tab:thresholdparameters}  are compatible with one another
within roughly one standard deviation.

In the fourth column, we use the very reliable CFD set inside several sum rules, 
that we detail next only very briefly, since they had already been given
in detail in KPY08. First, we use the well known Olsson sum rule:
\begin{equation}
  2a_0^{(0)}-5a_0^{(2)}= 3M_\pi\int_{4M_\pi^2}^\infty \dd\! s\,
\dfrac{\imag F^{(I_t=1)}(s,0)}{s(s-4M_\pi^2)},
\label{eq:OlssonSR}
\end{equation}
which 
is dominated at high energies by the $\rho$-Regge exchange, and can 
thus have only one subtraction.
Apart from the normalization, 
this is just the FDR in Eq.~\eqref{FDR2}, but evaluated at threshold.

Next, for $\ell\geq1$, we use the Froissart-Gribov representation:
\begin{align}
a_\ell=&\,\dfrac{\sqrt{\pi}\,\gammav(\ell+1)}{4M_{\pi}\gammav(\ell+3/2)}
\!\int_{4M_{\pi}^2}^\infty\dd\!s\,\dfrac{\imag F(s,4M_{\pi}^2)}{s^{l+1}},\cr
b_\ell=&\,\dfrac{\sqrt{\pi}\,\gammav(\ell+1)}{2M_{\pi}\gammav(\ell+3/2)}
\!\int_{4M_{\pi}^2}^\infty\dd\!s\,
\Big\{\dfrac{4\imag F'_{\cos\theta}(s,4M_{\pi}^2)}{(s-4M_{\pi}^2)s^{\ell+1}}\nonumber\\
&-
\dfrac{(\ell+1)\imag F(s,4M_{\pi}^2)}{s^{\ell+2}}\Big\},\hspace{-.5cm}
\label{eq:FG}
\end{align}
with $\imag F'_{\cos\theta}\equiv (\partial/\partial\cos\theta_s)\imag F$, where $\cos\theta_s$
is the angle between the initial and final pions.
For amplitudes with fixed isospin in the $t$ channel, an extra factor of 2 
(due to identity of particles) has to be added to the left hand side of the equation above. 

In addition, we use the following sum rule that we 
derived in \cite{Pelaez:2004vs}:
\begin{eqnarray}
  b_1\!\!&=&\!\!\!
\dfrac{2}{3M_\pi }\int_{4M^2_\pi}^\infty\!\!\!\!\dd\!s\Bigg\{
\tfrac{1}{3}\left[\dfrac{1}{(s-4M^2_\pi)^3}-\dfrac{1}{s^3}\right]
\imag F^{(I_t=0)}(s,0)\nonumber\\
&&+\tfrac{1}{2}\left[\dfrac{1}{(s-4M^2_\pi)^3}+\dfrac{1}{s^3}\right]
\imag F^{(I_t=1)}(s,0)\nonumber\\
&&-\,\tfrac{5}{6}\left[\dfrac{1}{(s-4M^2_\pi)^3}-
\dfrac{1}{s^3}\right]\imag F^{(I_t=2)}(s,0) \Bigg\},\label{eq:b1SR}
\end{eqnarray}
together with another two sum rules, derived in \cite{Kaminski:2006qe},
involving either the S0 and S2 slopes:
\begin{align}
\label{eq:twoSR1}
&    b_0^{(0)}+2b_0^{(2)}=\\
&\lim_{s\to{4M^2_\pi}^+}\!\!\pepe\!\!
\int_{4M^2_\pi}^\infty\!\!\!\!\!\dd s'\,\dfrac{6M_\pi\,(2s'-4M^2_{\pi})\imag F_{00}(s')}
{s'(s'+s-4M^2_\pi)(s'-4M^2_\pi)(s'-s)},\nonumber
\end{align}
or the S2 slope parameter and the P wave scattering length:
\begin{align}
\label{eq:twoSR2}
&3a_1^{(1)}+b_0^{(2)}=\\
&\lim_{s\to{4M^2_\pi}^+}\!\!\pepe\!\!
\int_{4M^2_\pi}^\infty\!\!\!\!\!\dd s'\,\dfrac{4M_\pi\,(2s'-4M^2_{\pi})\imag F_{0+}(s')}
{s'(s'+s-4M^2_\pi)(s'-4M^2_\pi)(s'-s)}.\nonumber
\end{align}

\afterpage{
\begin{widetext}
\begin{longtable}{lccccc}
\hline
\hline
     &UFD& CFD& Sum rules with CFD & {\bf Best values} & KPY08 values\\
\hline
$a_{0}^{(0)}$             & 0.218  $\pm$ 0.009 & 0.221  $\pm$ 0.009 & &  {\bf 0.220$\pm$ 0.008}$^e$  & 0.223 $\pm$ 0.009\\ 
$a_{0}^{(2)}$             & -0.052 $\pm$ 0.010 & -0.043 $\pm$ 0.008 & & {\bf -0.042$\pm$ 0.004}$^e$ & -0.044$\pm$ 0.004\\ 
$a_{0}^{(0)}-a_{0}^{(2)}$  &  0.270 $\pm$ 0.009 &  0.264 $\pm$ 0.009& & {\bf 0.262 $\pm$ 0.006}$^e$ & 0.267 $\pm$ 0.009 \\ 
$2a_{0}^{(0)}-5a_{0}^{(2)}$ & 0.696 $\pm$ 0.054 &  0.657 $\pm$ 0.043 & 0.648$\pm$ 0.016$^a$& {\bf 0.650 $\pm$ 0.015} & 0.668 $\pm$ 0.017\\ 
$\delta_{0}^{(0)}(M_K^2)
-\delta_{0}^{(2)}(M_K^2)$ &  47.4 $\pm$ 0.9$^\circ$ & 47.3 $\pm$ 0.9$^\circ$ & & {\bf 47.3 $\pm$ 0.9}$^\circ$ & 50.9 $\pm$  1.2$^\circ$\\ 
$b_{0}^{(0)}$             & 0.276  $\pm$ 0.007 & 0.278 $\pm$ 0.007  & 0.278$\pm$ 0.008$^d$ & {\bf 0.278 $\pm$ 0.005} & 0.290 $\pm$ 0.006\\ 
$b_{0}^{(2)}$             & -0.085 $\pm$ 0.010 & -0.080 $\pm$ 0.009 & -0.082$\pm$ 0.004$^d$ & {\bf -0.082 $\pm$ 0.004} & -0.081 $\pm$ 0.003\\
$a_{1}$(x$10^3$)         &   37.3 $\pm$ 1.2   & 38.5  $\pm$ 1.2 & 37.7 $\pm$ 1.3$^b$ & {\bf 38.1 $\pm$ 0.9} & 38.1  $\pm$ 0.9\\  
$b_{1}$(x$10^3$)         &   5.18 $\pm$ 0.23  & 5.07 $\pm$ 0.26 & 6.0 $\pm$ 0.9$^b$, 5.48 $\pm$ 0.17$^c$  & {\bf 5.37 $\pm$ 0.14}  & 5.12  $\pm$ 0.15\\ 
$a_{2}^{(0)}$(x$10^4$)    &  18.7 $\pm$ 0.4  &  18.8 $\pm$ 0.4 & 17.8  $\pm$ 0.3$^b$ & {\bf 17.8  $\pm$ 0.3}  & 18.33 $\pm$ 0.36\\ 
$a_{2}^{(2)}$(x$10^4$)    &  2.5 $\pm$ 1.1   & 2.8   $\pm$ 1.0 &  1.85 $\pm$ 0.18$^b$ & {\bf 1.85 $\pm$ 0.18}  &  2.46 $\pm$ 0.25\\ 
$b_{2}^{(0)}$(x$10^4$)    & -4.2 $\pm$ 0.3   & -4.2 $\pm$ 0.3  & -3.5 $\pm$ 0.2$^b$ & {\bf -3.5 $\pm$ 0.2}  & -3.82 $\pm$ 0.25\\ 
$b_{2}^{(2)}$(x$10^4$)    & -2.7 $\pm$ 1.0   & -2.8 $\pm$ 0.8  & -3.3 $\pm$ 0.1$^b$ & {\bf -3.3 $\pm$ 0.1}  & -3.59 $\pm$ 0.18\\ 
$a_{3}$(x$10^5$)         & 5.2  $\pm$ 1.3   & 5.1  $\pm$ 1.3  & 5.65 $\pm$ 0.23$^b$ & {\bf 5.65 $\pm$ 0.21}  & 6.05  $\pm$ 0.29\\  
$b_{3}$(x$10^5$)         & -4.7 $\pm$ 2.6   & -4.6 $\pm$ 2.5  & -4.06 $\pm$ 0.27$^b$ & {\bf -4.06 $\pm$ 0.27}  & -4.41 $\pm$ 0.36
\\\hline\hline  
 \caption{
 Threshold parameters in the customary $M_\pi=1$ units and  the $\delta_{0}^{(0)}(M_K^2)
 -\delta_{0}^{(2)}(M_K^2)$ phase difference. 
 The values in the second and third columns are obtained directly from the
 UFD and CFD parametrizations, respectively. The fourth column
 is obtained using the CFD set inside sum rules: $^a$from
 Eq.~\eqref{eq:OlssonSR}, $^b$from Eq.~\eqref{eq:FG}, $^c$from Eq.~\eqref{eq:b1SR},
 $^d$from Eqs.~\eqref{eq:twoSR1} and \eqref{eq:twoSR2}. In addition, for the scalar scattering lengths$^e$ best values,
 we have re-fitted their CFD values
 constrained to satisfy the Olsson sum rule, Eq.~\eqref{eq:OlssonSR},
which is also used to obtain the best value for their difference and 
its uncertainty, Eqs.~\eqref{eq:xy} and \eqref{eq:besta0a2}.
 } 
\label{tab:thresholdparameters}
\end{longtable}
\end{widetext}
}

Note that, as explained in \cite{Kaminski:2006qe}, the limits above are
to be taken for $s>4M_\pi^2$. In practice, for the value
of $a_1$ we simply use its Froissart-Gribov representation
and we are left with a sum rule representation for both 
$ b_0^{(0)}$ and $b_0^{(2)}$.

The results for all these sum rules are listed 
in the fourth column of Table~\ref{tab:thresholdparameters}.  

The fifth column, that contains what we consider our best values, 
is obtained as follows: For $2a_0^{(0)}-5a_0^{(2)}$, $b_0^{(0)}$, $b_0^{(2)}$,
$a_1$ and $b_1$, we take the average between the 
sum rules above and the direct value of the CFD set, since they are basically independent. 
However, for the D0, D2 and F waves, in 
order to stabilize the fits,
 we had already constrained the value of the threshold
parameters by means of the Froissart-Gribov representation
in the UFD set (see \cite{Pelaez:2004vs}). Hence, in those cases, 
it makes no sense to average either the UFD or CFD 
direct result with the Froissart-Gribov for $a_2^{(0)}$, $a_2^{(2)}$,
$b_2^{(0)}$ and $a_3$, which is therefore considered our best result.
The only exception are $b_2^{(2)}$ and $b_3$, since those values were not
constrained in the initial UFD, but their uncertainty from the CFD is 
an order of magnitude larger 
than from the sum rule, which value is the one we quote as the best one.

\begin{figure}[h]
\begin{center}
%\vspace*{-20pt}
\includegraphics[width=8.4cm]{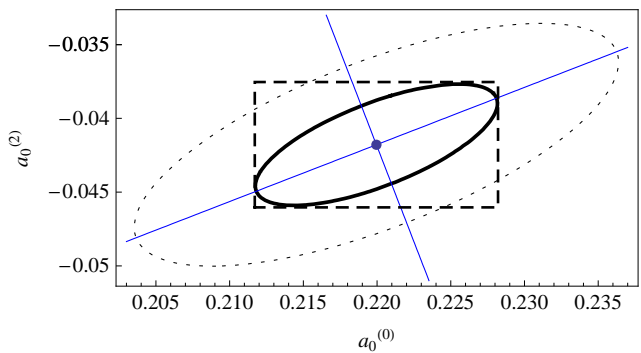}
\end{center}
\vspace*{-20pt}
\caption{The one and two standard deviation 
ellipses (thick and dashed lines, respectively)
in the $(a_0^{(0)},a_0^{(2)})$ plane. The square covers the uncertainties
of our best results in Eq.\eqref{eq:besta0a2}, obtained from the
uncorrelated expressions in Eq.\eqref{eq:xy}.
\label{fig:Ellipse}}
\end{figure}

Let us remark that the S0 and S2 scattering lengths, which are of
special interest for ChPT, are refined by
re-fitting them again to the CFD direct results and the Olsson sum rule simultaneously.
Obviously, the resulting errors are strongly correlated
and the corresponding correlation ellipse is shown in Fig.~\ref{fig:Ellipse}.
The uncertainties can be uncorrelated by using two new variables, $x$, $y$
defined as:
\begin{eqnarray}
&&a_0^{(0)}=0.220+0.130\,x+0.337\,y,\cr
&&a_0^{(2)}=-0.042-0.337\,x+0.130\,y,\cr
&&a_0^{(0)}-a_0^{(2)}=0.262+0.467\,x+0.206\,y,\cr
&&x=\,0\pm0.076,\quad y=0\pm0.023,
\label{eq:xy}
\end{eqnarray}
which give the numbers listed in the tables as our ``Best values'':
\begin{eqnarray}
&&a_0^{(0)}=0.220\pm0.008,
\label{eq:besta0a2}\\
&&a_0^{(2)}=-0.042\pm0.004, \nonumber\\
&&a_0^{(0)}-a_0^{(2)}=0.262\pm0.006,
\nonumber
\end{eqnarray}
in units of $M_\pi$.

For the sake of comparison, we list in the sixth column 
our results from KPY08,
where we did not impose the GKPY equations nor did we add the several improvements
to the amplitudes and the data implemented in this work. 
Note that the low energy parameters are quite consistent with our previous results, i.e.
many central values lie within one standard deviation of
our KPY08 results, and most of them overlap within one standard deviation. 
There are, of course,
the expected exceptions: first, the 
$\delta_0^{(0)}(M_K^2)-\delta_0^{(2)}(M_K^2)$ central value changes
by 3 standard deviations, mostly due to the fact that we have discarded here the 
controversial $K\rightarrow \pi\pi$ datum. Next, the S0 slope $b_0^{(0)}$
changes by two standard deviations, and this is mostly due to the inclusion of
the isospin violation correction in the low energy $K_{\ell 4}$ data. One could have expected
that the scattering length $a_0^{(0)}$ may have suffered a large shift for the same
reason, but it has only decreased about a third of a standard deviation. Hence, most of  the
change due to the $K_{\ell 4}$ isospin correction is concentrated on the slope parameter.
In addition, as we already anticipated, both D wave scattering lengths have decreased by roughly 2 standard deviations.

Although it will be commented in detail in the discussion section, 
let us note that
these new results
are in much better agreement with the results in  \cite{Bern}, 
 than were those in KPY08.

As commented in Sec.~\ref{sec:CFD}, we can also check here that
the new uncertainties are slightly smaller, but only by 10-15\%, 
than in KPY08, due to: discarding the $K\rightarrow\pi\pi$ conflicting input,
keeping the S0 Adler zero fixed, and the more precise NA48/2 published data.
The $a_0^{(0)}-a_0^{(2)}$ uncertainty in \eqref{eq:besta0a2}
has decreased by almost 50\%,
although this is not only due to our improvement of the S0 wave, but mainly
to the fact that we are now calculating it differently, using
Eqs.~\eqref{eq:xy}.

\subsection{Determination of Adler zeros}
\label{sec:Adlerzeros}

As already explained, chiral symmetry requires the  existence of zeros in the amplitude
close to $s=0$ for the scalar waves S0 and S2 \cite{Adler:1964um}.
We have explicitly factorized them in our amplitudes at
$s_A^{S0}=z_0^2/2$
and $s_A^{S2}=2 z_2^2$, see Eqs.~\eqref{eq:S0lowparam} or \eqref{eq:AppendixS0lowparam} and \eqref{eq:S2lowparam}.
As a starting point,  we have first fixed them to 
the ChPT leading order estimate  by 
setting $z_0=z_2=M_\pi$ for the UFD parametrizations.
We  then used these parametrizations inside Roy or GKPY equations 
to recalculate the position of these Adler zeros, which were listed in the first two columns
of Table \ref{tab:Adlerzeros}. 

In previous works, we allowed the $z_0$ and $z_2$ parameters to change in the CFD set, expecting them to be
accurately fixed by imposing the dispersion relations.
Unfortunately, as discussed in Sect.~\ref{sec:ADlerS2}, this does not work for the S0 wave. The reason is that its Adler zero
is very close to the left cut, in a region where, on the one hand, neither Roy nor GKPY
equations provide a precise determination of the zero position
(see Table~\ref{tab:Adlerzeros}) and, on the other hand, 
the conformal
expansion converges badly.
For that reason, we have simply 
kept the S0 parameter $z_0$ fixed to $M_\pi$ both on the UFD and CFD sets.
Being so far from the threshold region, this effect is irrelevant inside the 
dispersive integrals. Thus, only the S2 Adler zero is allowed to change when obtaining the CFD set, but only within the UFD uncertainties
obtained from Roy and GKPY equations.

In this section we go one step further and 
we finally provide, in the last two columns of Table~\ref{tab:Adlerzeros},
the value of the S0 and S2 wave Adler zeros obtained when the
CFD set is used inside Roy and GKPY equations. The CFD S0 zero is closer
to its expected position (around 80 MeV) than the UFD result, 
but note that the uncertainty gets worse because
of this displacement towards the left cut. In summary, we do not have enough precision  to pin
down the location of this S0 Adler zero accurately.

In contrast, the S2 Adler zero is determined quite precisely by GKPY
equations (and  to a lesser extent by Roy equations), and the resulting 
$z_2$ parameter, if allowed to vary, is almost identical to its UFD
determination. Thus, as explained in Section~\ref{sec:ADlerS2}, we have allowed
$z_2\sqrt{2}$ to vary within the weighted average between the GKPY and Roy equation results of the UFD set.
The resulting Adler zero, when read directly from the CFD parametrization is
$\sqrt{s_{A}^{S2}}=z_2\sqrt{2}=201\pm5\,\mev$,
which is almost identical to the values obtained by using the CFD set inside Roy or GKPY equations---listed in Table~\ref{tab:Adlerzeros}.
This confirms that it is correct to identify the Adler zero with the $z_2\sqrt{2}$ term in our S2 wave conformal parametrization. 

 \begin{figure*}
   \centering
 \hspace*{-.2cm}
 \includegraphics[scale=0.41]{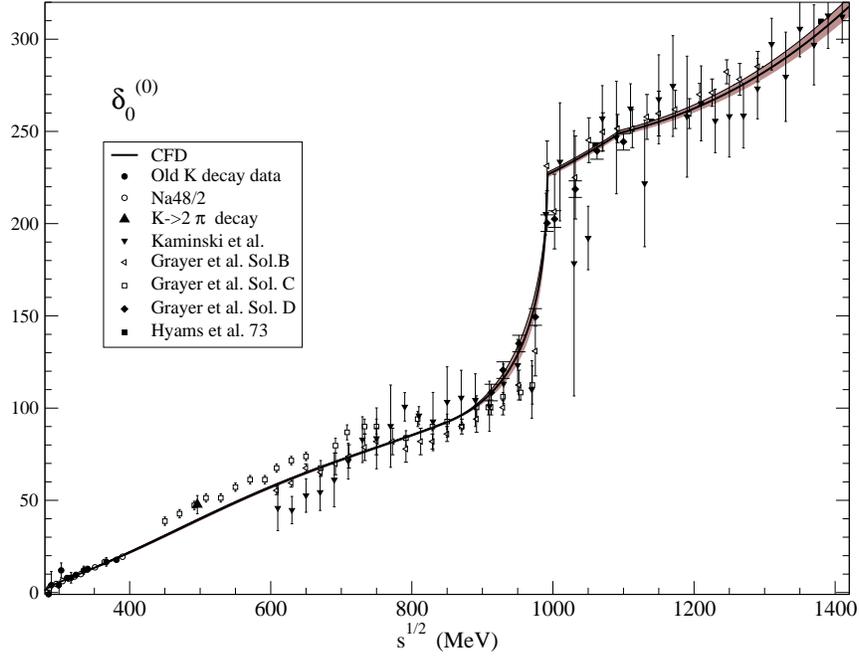}
   \caption{  The new constrained
fit (CFD) for the S0 wave 
  versus the existing phase shift data from 
\protect{\cite{pipidata,pipidata2}}. The dark band covers the uncertainties,
 }
 \label{fig:CFD-datos}
 \end{figure*}

\section{Discussion}

First of all, we want to remark that ours is just a data analysis,
and we are not predicting the value of any observable, just determining
them from experiment. 
In contrast to other approaches \cite{Bern}, we are  {\it not solving} 
FDR, Roy or GKPY equations, but just imposing 
them as constraints on the data analysis.
Actually, all these equations have been obtained
 with several approximations, for instance, they are 
obtained in the isospin limit, and we only expect them to describe 
the real world up to some uncertainty of the order of 3\%. 
In addition, all Roy equations studies we are aware of---including this one---neglect 
any inelasticity to four or more pion states below 
the two kaon threshold. This is certainly a very small effect, but  is nevertheless
an approximation.

Being a data analysis, our parametrizations change when the data changes.
In particular, in this work we have updated the NA48/2
data \cite{Batley:2007zz} with their final results \cite{ULTIMONA48}, which
have smaller uncertainties. In addition, we have incorporated
the threshold-enhanced isospin correction in \cite{Colangelo:2008sm}
 to all $K_{\ell 4}$ data. Moreover, we have discarded the controversial 
$K\rightarrow\pi\pi$ datum \cite{Aloisio:2002bs}. 
Furthermore, the increased precision 
provided by the once-subtracted dispersion relations that we have introduced in this
work, requires an improved parametrization with a continuous derivative matching.
This additional constraint and the requirement that the {\it output}
 of the dispersion relations
should satisfy the elastic unitarity bound---which is automatic in the input parametrizations---, has made us also add an additional parameter
to the S0 wave parametrization at low energies.
As we will see below, the S0 wave parametrization at intermediate energies favors the ``dip-scenario'' for the S0 
inelasticity between 1000 and 1100 MeV.
In this discussion section we will show in detail 
the new CFD set, particularly the S0 wave, 
comparing it to other works,  and we will discuss
the consequences of these modifications.

 \begin{figure*}
   \centering
 \hspace*{-.2cm}
 \includegraphics[scale=0.41]{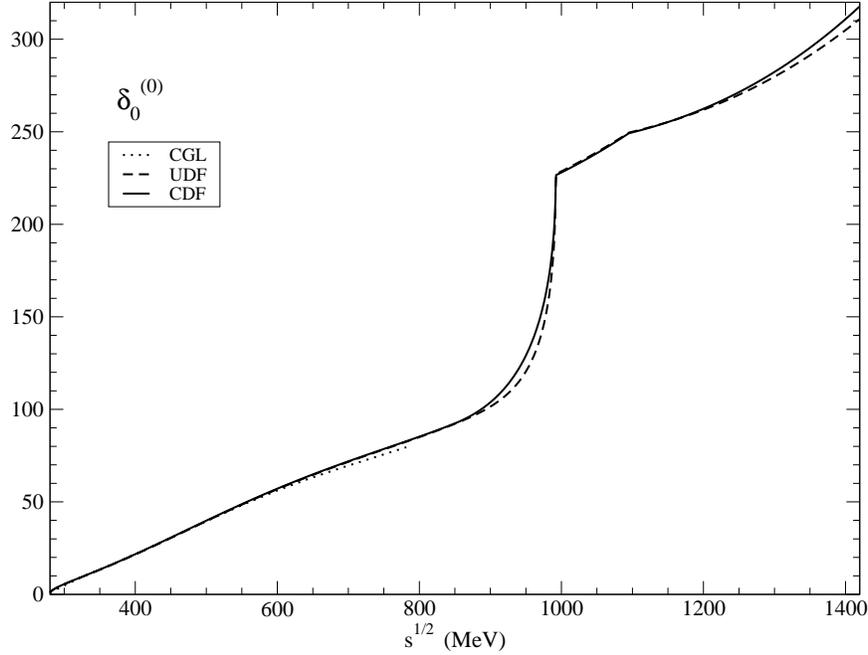}
 \vspace*{.5cm}
   \caption{ Comparison between the phase of the constrained (CFD) and unconstrained (UFD) Fit to Data for the S0 wave. We also plot the phase from 
the Roy equations analysis in~\cite{Bern}.
 }
\label{fig:UFD-CFD-CGL}
 \end{figure*}

\subsection{The new CFD S0 wave}

In Fig.~\ref{fig:CFD-datos}  we show the resulting CFD S0 wave from threshold up to 1420~MeV,
versus the data from different sets in the literature 
\cite{pipidata,pipidata2}.
Note the smooth matching at 850 MeV 
and the kink at $K\bar{K}$ threshold. 
This is in contrast with our old KPY08 results, 
already shown in Fig.~\ref{fig:S0-UFD-comparison}, which have a
spurious kink
at the matching point (932 MeV in that work), 
and a much less pronounced kink at $K\bar K$ threshold.
The difference between the UFD and CFD S0 wave phase shift at low energies, 
which we showed in Fig.~\ref{fig:phaselowenergies}, is almost
imperceptible.

To ease the comparison of this CFD results with the UFD set for all energies, 
we have plotted
their central values
together in Fig.~\ref{fig:UFD-CFD-CGL}.   
It can be noted that the change above 
$K\bar K$ threshold
is again almost imperceptible up to 1200~MeV.
The 
only sizable differences between the phase of the UFD and CFD parametrizations
are above 1200~MeV, where our parametrizations are less reliable since Roy and GKPY Eqs.
only extend up to 1115~ MeV,
and on the sharp phase rise in the 900~MeV to $2 m_K=992\,$MeV region due to 
the $f_0(980)$ resonance, which
   is clearly less steep in the CFD case than in the UFD. The latter
is one of the reasons 
   why the CFD solution satisfies GKPY equations well within uncertainties, but the UFD 
   lies somewhere around 2 standard deviations away (see Tables \ref{tab:UFDdiscrepancies}
   and \ref{tab:CFDdiscrepancies},
   respectively).

In addition,  we are also showing in Fig.~\ref{fig:UFD-CFD-CGL}
the results from \cite{Bern}, which are in 
good agreement with ours, but
 lie slightly lower only above 550~MeV (see discussion below). Actually, our
 CFD solution does not show the ``hunchback`` between 500 and 900 MeV seen in KPY08, as already shown in Fig.~\ref{fig:S0-UFD-comparison}.

\begin{figure}
   \centering
 %\hspace*{-.1cm}
 \includegraphics[scale=0.3]{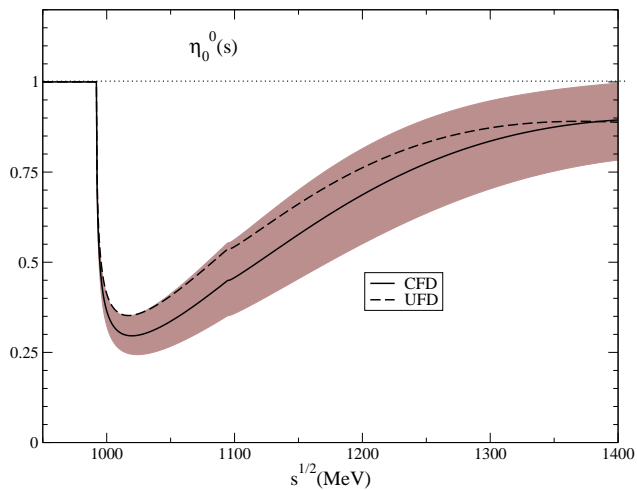}
   \caption{ Comparison between the UFD and CFD S0 wave inelasticity.
The gray area corresponds to the CFD uncertainty. A similar size area should
be associated to the UFD result, but for clarity we only show its central value.
Note the dip structure between 1 and 1.1 GeV.
 \label{fig:UFD-CFD-elasticity}
 }
 \end{figure}

\begin{figure}
   \centering
 %\hspace*{-.1cm}
 \includegraphics[scale=0.3]{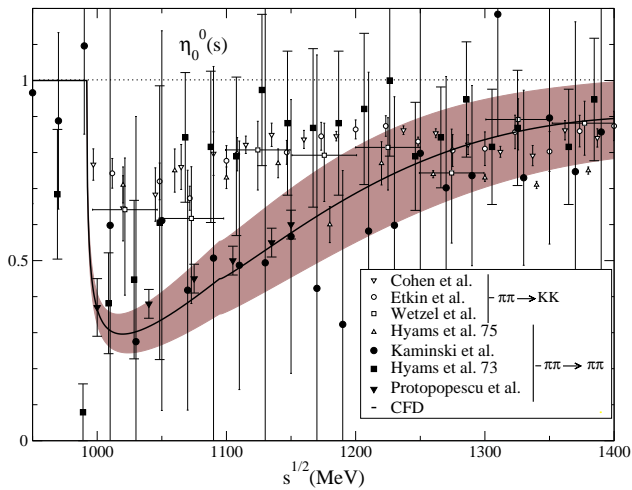}
   \caption{ CFD S0 wave inelasticity versus experimental data.
 \label{fig:UFD-CFD-elaasticity-data}
 }
 \end{figure}

Concerning the S0 inelasticity, we show in Fig.~\ref{fig:UFD-CFD-elasticity}
the difference between the UFD and CFD sets. 
It can be noticed that the difference 
lies essentially within the uncertainties (gray area),
although the ``dip'' structure above 1000~MeV becomes even deeper in the CFD set.
Finally,
in Fig.
\ref{fig:UFD-CFD-elaasticity-data} we show the CFD inelasticity versus all
the existing experimental data.

Since the UFD set was already providing a good description of the inelasticity
data obtained from $\pi\pi\rightarrow\pi\pi$ experiments, as shown
in Fig.~\ref{fig:S0-UFD-inel-data},  so it does the CFD. For the same reason,
it also fails to reproduce the inelasticity data from $\pi\pi\rightarrow K \bar K$, as we had already shown 
for the UFD case in Fig.~\ref{fig:S0-UFD-inel-comparison}.
Note that this is due to the fact that both our UFD and CFD solutions show a ``dip'' structure between 1 and 1.1 GeV, 
which is seen in the data coming from 
$\pi\pi\rightarrow\pi\pi$, but not in those coming from 
$\pi\pi\rightarrow K \bar K$. This is a 
longstanding problem (see \cite{Au:1986vs} and references therein) 
that we will address in the next subsection, 
showing that the ``non-dip'' 
scenario is not able to satisfy the dispersive representation well
even when allowing for a large deviation from the phase shift data.

\subsection{S0 inelasticity: the ``non-dip'' scenario is disfavored}
\label{sec:nodip}

\begin{figure}
   \centering
 \hspace*{-.1cm}
 \includegraphics[scale=0.3]{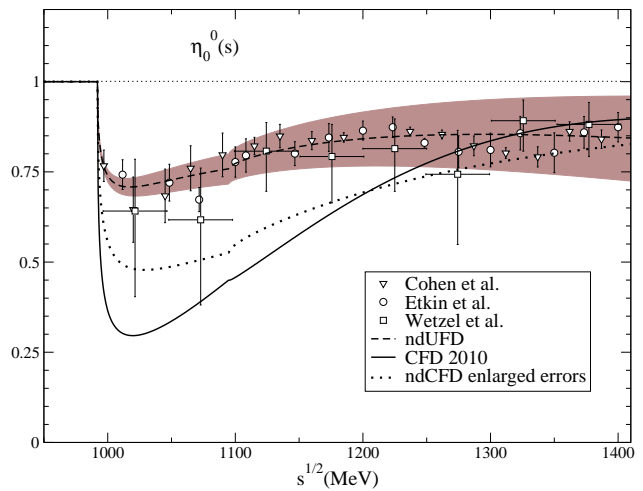}
   \caption{ S0 wave inelasticity versus the ``non-dip'' 
$\pi\pi\rightarrow K\bar{K}$ data. 
We first show the ``ndUFD'' set obtained from a fit to these ``non-dip''
data. Next, we show the  ``ndCFD'' set obtained with enlarged errors to try to 
fulfill dispersion relations. 
This constrained fit satisfies the dispersive constraints  better but
does not describe these ``non-dip'' data,
coming closer to the best CFD set, which actually 
describes the alternative ``dip'' 
data from $\pi\pi\rightarrow\pi\pi$.
 \label{fig:wrong-inel}
 }
 \end{figure}

In order to show how much the ``non-dip'' scenario is disfavored, we
will first repeat the same procedure of this whole paper, 
but starting from the S0 inelasticity fitted
to the ``non-dip'' data,
as shown in Fig.~\ref{fig:wrong-inel},
while keeping the same UFD
parametrization for all other waves and for the S0 phase.
We will refer to this set as ``ndUFD''.
The resulting averaged discrepancies 
$\bar d^2_i$ are relatively similar to those in
Table~\ref{tab:UFDdiscrepancies} for our UFD, except for the S0 wave 
GKPY equations up to $\sqrt{s}\leq1100\,$MeV,  whose averaged $\bar d^2_i$
rises from 2.42 to 4.77.
This already disfavors the ``non-dip'' scenario.

Of course, the ``dip scenario'' UFD set was not doing very well either,
but we were able to improve it by constraining the fit to data
with dispersion relations, i.e., the CFD set. One could wonder if a similar quality
fit can also be obtained by imposing the dispersive constraints, but starting
from the ``ndUFD''. 
Thus, we followed again the procedure  described in previous sections,
but to arrive now  to a ``ndCFD'' set. 
Surprisingly, the S0 inelasticity barely changes, but the improvement
comes from a bigger variation of the phase in the two-kaon subthreshold region.
The resulting
 average discrepancies $\bar d^2_i$ come in general
larger than for our CFD set, sometimes by a factor of two, but still below 1. 
This may look like an agreement, but one should not be 
misguided now by these relatively low averaged $\bar d^2_i$ because, contrary to the CFD set
where discrepancies are below 1
 uniformly over the whole energy region, 
for the ndCFD set they are larger in the $f_0(980)$ resonance region.

In particular, in the interval between 950 and 1050 MeV, for the CFD set, 
the GKPY S0 equation have $\bar d^2=1.02$,
whereas the ndCFD set has $\bar d^2=3.49$.
This averaged discrepancy is unacceptable now, since
this time we are using the dispersion relations as constraints of our fits.
In addition, the crossing sum rule in Eq.~\eqref{eq:Isumrule} grows to $\bar d_I^2=2.0$.

\begin{figure}
   \centering
 %\hspace*{-.1cm}
 \includegraphics[scale=0.3]{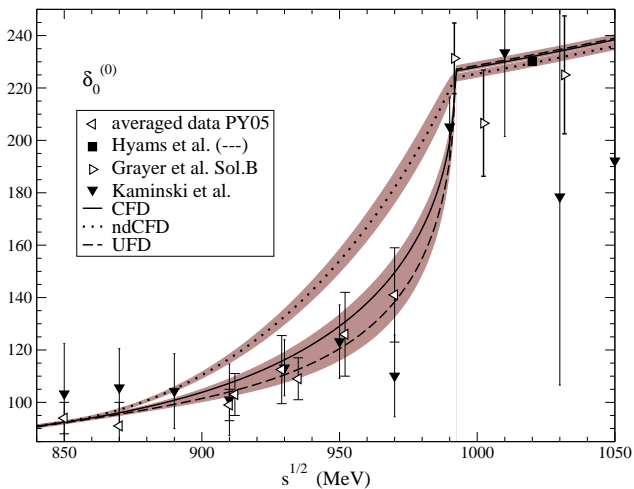}
   \caption{ Comparison of the UFD, CFD and ndCFD solutions for the S0 phase in the
850 to 1050~MeV region. Note that the ndCFD parametrization is largely inconsistent with data,
despite the fact we are plotting the PY05 averaged data, 
that includes our estimations of the large dominant systematic uncertainties.
 \label{fig:phasenodip}
 }
 \end{figure}

Furthermore, 
as we show in Fig.~\ref{fig:phasenodip}, in the region from 900 MeV up to 
$K \bar K$ threshold,
the resulting phase of this ndCFD scenario lies above all
data points with a $\chi^2/{\rm \# points}=3.4$, which is a very bad fit. 
In contrast, the CFD set 
has $\chi^2/{\rm \# points}=0.98$ in this region and is just a small modification
from the UFD phase, which has $\chi^2/{\rm \# points}=0.63$.
Moreover, the ndCFD parameters lie far from the original ndUFD ones,
with the $c$ parameter more than 6 standard deviations away from its ndUFD value. 
These numbers  clearly show
the incompatibility
of the ndCFD set with the S0 wave $\pi\pi\rightarrow\pi\pi$ phase-shift scattering data.
This disagreement cannot be mended by adding systematic uncertainties, 
since  in this region we had already included large systematic
uncertainties (see KPY08 and PY05 for details) and all points 
have total uncertainties of more than 10 degrees.

One could wonder if our minimization procedure, that was good enough to reach
$\bar d_i^2<1$ for the dip scenario, is badly tuned
for the ``non-dip'' one. This, of course is the role of the $W_i$ weights in 
Eq.~\eqref{tominimize}. For this reason we have repeated the above procedure
adding additional weight to the GKPY S0 wave equation above 900 MeV.
 The resulting ``ndCFD2'' yields
 $\bar d^2=2.06$ for the GKPY S0 equation. Besides, the crossing sum rule in
Eq. ~\eqref{eq:Isumrule} is also $\bar d_i^2=1.43$. 
Although they still disfavor this solution, these numbers by themselves are not too bad.
However, the phase shift data between 950 and 1050 MeV
has  $\chi^2/{\rm \# points}=5.9$,
so that it is described even worse than with the 
previous ndCFD.

Since we cannot fix the dispersive constraints without spoiling the data phase description,
 we have allowed, as a final check, for larger errors in the 
inelasticity parameters of the ``non-dip'' 
scenario, and applied the dispersive constraints.
In so doing, we can obtain  $\bar d_i^2<1$ 
uniformly over all energy regions 
for all GKPY equations except for the S0 wave
between 950 to 1050~MeV, for which we obtain $1.42$. However, 
the central value of the inelasticity for the resulting 
constrained ``non-dip''  fit 
starts developing a dip as seen in Fig.~\ref{fig:wrong-inel}.
Therefore, we describe neither the ``non-dip'' nor the ``dip'' scenario.

In conclusion, the non-dip scenario, even when constrained with dispersion relations,
is not able to describe the data 
and  simultaneously satisfy Forward Dispersion Relations, Roy and  
GKPY equations, plus certain crossing sum rules.

\subsection{Comparison with other works.}

The results listed in Table \ref{tab:thresholdparameters} 
for threshold parameters
 are remarkably compatible with the predictions of \cite{Bern}
using Chiral Perturbation Theory and Roy equations:
$$
a_0^{(0)}=0.220\pm0.005,\qquad a_0^{(2)}=-0.0444\pm0.0010.
$$
The agreement with that reference has also improved a great deal 
since the  $\delta_0^{(0)}(M_K^2)-\delta_0^{(2)}(M_K^2)=(47.3\pm0.9)^\circ$ value, obtained directly from our CFD set, 
is now completely consistent with their value of $(47.7\pm1.5)^\circ$. Essential for that agreement is, of course,
not to consider the $K\rightarrow \pi\pi$ datum.
Also, all the D wave threshold parameters are now in good agreement
with those used in \cite{Bern}.
The remaining differences with respect to that work are rather small: the largest one is a 2.1 standard deviation disagreement,
with respect to their predicted value $b_1=(5.67\pm0.13)\times 10^{-3}$.
In general, and up to 500 MeV, the results of \cite{Bern} fall within roughly one standard deviation
of our analysis. For instance, at the kaon mass, our CFD S0 wave phase shift is
$\delta_0^{(0)}(M_K)=39.1\pm0.6^\circ$, identical to theirs to the last digit, but our S2 wave is $\delta_0^{(2)}(M_K)=-8.2\pm0.6^\circ$,
$0.3$ degrees more than theirs, which is half an standard deviation.
This good agreement does not deteriorate much above that energy. For instance, 
at 800 MeV, which is their matching point
between the calculated phase shifts and their input, they use an
input value of $\delta_0^{(0)}=82.3\pm3.4^\circ$.
In contrast, we obtain $\delta_0^{(0)}=85.2\pm0.5^\circ$ directly from the CFD set,
whereas we find $\delta_0^{(0)}=85.7\pm1.6^\circ$ when using the same CFD set inside GKPY equations,
that is, one  of their standard deviations.
Above 800 MeV their amplitudes are part of the 
input and not solutions of Roy equations.

Finally, we would like to remark that our best values
for the scalar scattering lengths in  Eq.~\eqref{eq:besta0a2},
are in very good agreement with the experimental results from
pionic atoms \cite{Adeva:2007zz,Adeva:2005pg}, that yield:
\begin{eqnarray*}
&&  a_0^{(0)}-a_0^{(2)}=0.280\pm0.013{\rm (St.)}\pm0.008{\rm (Syst.)}\,M_\pi^{-1},\\
&&  a_0^{(0)}-a_0^{(2)}=0.264^{+0.033}_{-0.020}\,M_\pi^{-1},
\end{eqnarray*}
or  $K_{3\pi}$ decays \cite{Cabibbo:2005ez}:
\begin{eqnarray*}
 a_0^{(0)}-a_0^{(2)}&=&0.2571\pm0.0048{\rm (St.)}\qquad\\
&&\qquad \pm0.0025{\rm (Syst.)}\pm 0.0014{\rm (Ext.)}\,M_\pi^{-1}.
\end{eqnarray*}
Had we used them as additional constraints with the statistical and systematic
errors added linearly as we did with other decays, the difference with our
best results would have been barely modified.

As we commented in Sect.~\ref{sec:k2pidata}, the  phase difference
$\delta_0^{(0)}(M_K^2)-\delta_0^{(2)}(M_K^2)=(52.5\pm0.8_{\rm exp}\pm2.8_{\rm exp})^\circ$ 
has been recently re-analyzed \cite{Cirigliano:2009rr}. 
This is a considerable shift from the previous
value of   $(57.27 \pm 0.82_{\rm exp.}\pm3_{\rm rad.}\pm 1_{\rm ChPT\,appr.})^\circ $,
in much better agreement with ours and other previous dispersive analyses. 
Note that the new number is also in good
agreement with our results in Table~\ref{tab:thresholdparameters}.

\section{Summary}

In this work, we have presented the derivation of a once-subtracted set of Roy-like dispersion relations---the GKPY equations.
We have shown and explained that above 450 MeV, and up to 1115 MeV, they
provide stronger constraints 
on $\pi\pi$ scattering amplitudes than other existing
sets of dispersion relations.

We have then applied these new equations as constraints in our fits to 
data---together with the standard Roy equations and 
Forward Dispersion relations---in order to obtain a precise description of $\pi\pi$ scattering
amplitudes. In contrast to previous works, we have extended the Roy and GKPY 
equations analysis from 932 MeV up
to their applicability limit of 1100 MeV. Forward Dispersion Relations are considered up to 1420 MeV.

We have also made use of the final and very precise 
data on $K_{\ell 4}$ decays from NA48/2, including the isospin violation corrections proposed in \cite{Colangelo:2008sm},
and  we have removed a conflicting data point
from $K\rightarrow2\pi$ decay.
With these changes in the data selection, most of the disagreement 
with previous Roy equation calculations \cite{Bern} has disappeared below 800 MeV.
The largest discrepancy that remains is on the P wave
slope parameter, but just at the two standard deviation level.

In addition, we have improved our S0 wave parametrization to
ensure a continuous matching between the 
low and intermediate energy parametrizations. Both parametrizations have been made more flexible, which
allows the phase and inelasticity to include contributions from states different
from $\pi\pi$ and $K\bar K$, above the $K\bar K$ threshold. 

There are two sets of fits to data: 
unconstrained (UFD), or constrained  with dispersion relations (CFD).
In the UFD set each wave is independent of all others, 
but dispersion relations are satisfied
only up to the two sigma level (in the sense 
explained in the text). In contrast, the CFD waves 
are all correlated, but they fulfill all 
dispersion relations under consideration 
within less than one standard deviation in the whole energy region. The CFD set can be considered as a very 
precise parametrization of experimental data consistent with the requirements 
of analyticity, unitarity and crossing symmetry. 
Using this CFD set as an input in different sum rules 
and the dispersion relations themselves, we have also provided
a precise determination of phases in the elastic regime, 
threshold parameters and Adler zeros. 

In addition, and concerning the 
conflicting data for the S0 wave inelasticity between the two-kaon threshold and 1100 MeV, 
the use of the new GKPY equations has allowed us to show that 
the sudden drop around 1050 MeV in the S0 wave inelasticity, or ``dip solution'',
is clearly favored with respect to the ``non-dip'' solution. Actually, for the ``non-dip'' 
inelasticity scenario to fulfill dispersion relations, it would require a very poor
description of the phase shift data, even when allowing for large systematic uncertainties.

In conclusion, we  provide fits to data in terms 
of simple and ready to use parametrizations for the S0, S2, P, D0, D2 
and F partial waves, between threshold and 1420~MeV. 
Additional simple Regge parametrizations are given above that energy.
In particular, the CFD set satisfies remarkably well all the analyticity
 and crossing symmetry constraints in the form of once and twice subtracted Roy equations and Forward Dispersion Relations.

\section*{Acknowledgments:} 
At the early stages of this collaboration, we suffered the 
devastating loss of one of the authors, F. J. Yndur\'ain, whose contributions
were essential for this work. He was an example of humanity and scientific dedication.
We would like to dedicate this work to his memory.

We also thank I. Caprini, G. Colangelo, J. Gasser and H. Leutwyler for many discussions and suggestions on possible improvements
of our parametrizations, as well as D. V. Bugg for comments on the S0 wave
inelasticity, and B. Kubis for his comments on $K_{3 \pi}$ decays.
This work is partly
supported by DGICYT contracts FIS2006-03438 and FPA2005-02327, 
Santander/Complutense contract PR27/05-13955-BSCH and the EU Integrated
Infrastructure Initiative Hadron Physics Project under contract
RII3-CT-2004-506078.

\appendix
\section{Partial wave parametrizations}

In the following, we provide the parametrizations we use for each partial wave,
and then, the parameters for the UFD and CFD sets.
For brevity, we do not explain again why a 
specific parametrization for each wave
has been chosen, since such details  
can be found in KPY08 \cite{Kaminski:2006qe}.
In what follows we use $M_\pi=139.57$ MeV, $M_K=496$ MeV and
$M_\eta=547.51$ MeV.

\subsection{S0 wave}
This wave has been thoroughly discussed in the main text.
However, for the sake of completeness, we  
repeat here the form of the parametrizations
and provide the values of the parameters for the UFD and CFD sets in
Table~\ref{tab:S0parameters}.

For this wave we have set the
 matching point between the intermediate and low energy parametrizations at
$s_M^{1/2}= 0.85\,\gev$.
Thus, at low energies $s\leq s_M$, we use:
\begin{eqnarray}
\label{eq:AppendixS0lowparam}
&&  \cot\delta_0^{(0)}(s)=
\frac{s^{1/2}}{2k}\frac{M_\pi^2}{s-\frac{1}{2}z_0^2}\times\\
&&\left\{\frac{z_0^2}{M_\pi\sqrt{s}}+B_0+B_1w(s)+B_2w(s)^2+B_3w(s)^3\right\},
\nonumber\\
&&w(s)=\frac{\sqrt{s}-\sqrt{s_0-s}}{\sqrt{s}+\sqrt{s_0-s}}, \qquad s_0=4M_K^2.
\end{eqnarray}
Above that energy, and up to 1.42 GeV, we use the KPY06 polynomial
parametrization for the phase shift, but with one more term in the expansion. 
For definiteness, we provide here the polynomial parametrization once
it has been matched to Eq.~\eqref{eq:AppendixS0lowparam} above, by imposing continuity and
a continuous derivative at $s=s_M$, namely
\begin{widetext}
\begin{equation}
  \label{eq:Appendixnewparam}
  \delta_0^{(0)}(s)=\left\{ 
\begin{array}{ll}
\displaystyle{d_0\left(1-\frac{\vert k_2 \vert}{k_M}\right)^2+\delta_M \, \frac{\vert k_2 \vert}{k_M}\left(2-\frac{\vert k_2 \vert}{k_M}\right)+
 \vert k_2 \vert (k_M-\vert k_2 \vert)
\left(8\delta_M'+c \frac{(k_M-\vert k_2 \vert)}{M_K^3}\right)
},& (0.85 \,{\rm GeV})^2<s<4 M_K^2, \\
 \\
\displaystyle{d_0+B\frac{k_2^2}{M_K^2}+C\frac{k_2^4}{M_K^4}}+D\,\theta(s-4M_\eta^2)\frac{k_3^2}{M_\eta^2},&
4 M_K^2<s<(1.42 \gev)^2,
\end{array}\right.
\end{equation}
\end{widetext}
where $k_2=\sqrt{s/4-M_K^2}$. Note that we have defined $\delta_M=\delta(s_M)$ and $\delta_M'=d\delta(s_M)/ds$, 
which are obtained from 
Eq.~\eqref{eq:AppendixS0lowparam}, and $k_M=\vert k_2(s_M) \vert$.

Finally, we assume an elastic S0 wave, $\eta_0^{(0)}=1$, up to the two-kaon threshold,
whereas above that energy, we use:
\begin{eqnarray}
\eta_0^{(0)}(s)=\exp\bigg[\frac{-k_2(s)}{s^{1/2}}\left(\tilde\epsilon_1+
\tilde\epsilon_2\,
\frac{k_2}{s^{1/2}}+\tilde\epsilon_3\,\frac{k_2^2}{s}\right)^2\\\nonumber
-\tilde\epsilon_4 \theta(s-4M_\eta^2)\frac{k_3(s)}{s^{1/2}}\bigg].
\label{eq:AppendixinelasticityS0}
\end{eqnarray}

We have collected the values of the parameters for the UFD and CFD set in 
Table~\ref{tab:S0parameters}.

\begin{table}[h]
  \centering
  \begin{tabular}{ccc}
S0 wave   & UFD &CFD\\
\hline
$ B_0$ &$7.26\pm0.23$ &$7.14\pm0.23$ \\
   $B_1$& $-25.3\pm0.5$&$-25.3\pm0.5$\\
 $B_{2}$&$-33.1\pm1.2$&$-33.2\pm1.2$\\
 $B_{3}$&$-26.6\pm2.3$&$-26.2\pm2.3$\\
 $z_0$ & $M_\pi$& $M_\pi$\\
\hline
$d_0$ & $(227.1\pm1.3)^\circ$& $(226.5\pm1.3)^\circ$\\
$c$ & $(-660\pm  290)^\circ$& $(-81\pm  294)^\circ$\\
$B$ &  $(94.0\pm 2.3)^\circ$& $(93.3\pm 2.3)^\circ$ \\
$C$ & $(40.4\pm 2.9)^\circ$& $(48.7\pm 2.9)^\circ$ \\
$D$ & $(-86.9\pm 4.0)^\circ$& $(-88.3\pm 4.0)^\circ$ \\
$\tilde\epsilon_1$ & $4.7\pm0.2$  & $4.9\pm0.2$  \\
$\tilde\epsilon_2$ & $-15.0\pm0.8$  & $-15.1\pm0.8$  \\
$\tilde\epsilon_3$ & $4.7\pm2.6$ & $4.7\pm2.6$ \\
$\tilde\epsilon_4$ & $0.38\pm0.34$ & $0.32\pm0.34$ \\
\hline
  \end{tabular}
  \caption{S0 wave parameters for the UFD and CFD sets.
The first four lines correspond to the low energy parametrization, 
$\sqrt{s}\leq 0.85\,\gev$, and the last nine to the parametrization up to 
$\sqrt{s}=1.42\,\gev$.}
  \label{tab:S0parameters}
\end{table}

\subsection{S2 wave}

As we have already done with the S0 wave, we have also set the
 matching point between intermediate and low energy parametrizations for this wave at
$s_M^{1/2}= 850\, \mev$.
Thus, at energies $s^{1/2}\leq s_M^{1/2}$ we use:
\begin{eqnarray}
&&  \cot\delta^{(2)}_0(s)=\frac{s^{1/2} }{2k}\frac{M_\pi^2}{s-2z_2^2}\left\{
B_0+B_1 w_l(s)\right\},
\nonumber\\
&&w_l(s)=\frac{\sqrt{s}-\sqrt{s_l-s}}{\sqrt{s}+\sqrt{s_l-s}},\quad s_l^{1/2}=1.05 \,\gev,
  \label{eq:S2lowparam}
\end{eqnarray}
whereas at intermediate energies, $850\, \mev\leq s^{1/2}\leq 1420 \,\mev$, we use:
\begin{eqnarray}
&&  \cot\delta^{(2)}_0(s)=\frac{s^{1/2} }{2k}
\frac{M_\pi^2}{s-2z_2^2}\times
\nonumber\\
&&\left\{
B_{h0}+B_{h1}[ w_h(s)-w_h(s_M)]+B_{h2}[ w_h(s)-w_h(s_M)]^2\right\},
\nonumber
\end{eqnarray}
where
\begin{eqnarray}
&&w_h(s)=\frac{\sqrt{s}-\sqrt{s_h-s}}{\sqrt{s}+\sqrt{s_h-s}},
\quad s_h^{1/2}=1.42 \,\gev,
\nonumber\\
&& B_{h0}=B_0+B_1 w_l(s_M),\quad
B_{h1}=B_1\left.\frac{\partial w_l(s)}{\partial w_h(s)}\right\vert_{s=s_M}\nonumber\\
&&B_{h1}=B_1\frac{s_l}{s_h}\frac{\sqrt{s_h-s_M}}{\sqrt{s_l-s_M}}
\left( \frac{\sqrt{s_M}+\sqrt{s_h-s_M}}{\sqrt{s_M}+\sqrt{s_l-s_M}}
\right)^2
\label{eq:S2highparam}
\end{eqnarray}
Note that, with these definitions,
both the parametrization and its derivative are continuous at the matching point.

Note that we have explicitly factorized the Adler zero at
$s_A=2 z_2^2$. For the unconstrained fit, $z_2$ is fixed to the pion
mass. As explained in the main text in Sect.~\ref{sec:ADlerS2}, 
we then calculate the Adler
zero position using Roy and GKPY equations, and feed the weighted average
into the constrained fit. This change 
is very small in
terms of the total values and uncertainties of other quantities, but
it is relevant in the differences when calculating the fulfillment of GKPY equations. 

For the inelasticity, we set $\eta_0^{(2)}(s)=1$ for $s$ below
$\hat s=(1.05\, \gev)^2$ and above that energy we use the empirical fit:

$$\eta_0^{(2)}(s)=1-\epsilon(1-\hat s/s)^{3/2}.$$

% \begin{equation}
%   \eta_0^{(2)}(s)=\left\{ 
% \begin{array}{ll}
%  \displaystyle{1},&
%  s>\hat s=(1.05\, \gev)^2,\\
%  \\
% \displaystyle{1-\epsilon(1-\hat s/s)^{3/2}
% },& s<\hat s=(1.05 \,\gev)^2,
% \nonumber
% \end{array}\right.
% \end{equation}

The S2 wave  parameters for UFD and CFD sets are given in Table~\ref{tab:S2parameters}.

\begin{table}[h]
  \centering
  \begin{tabular}{ccc}
S2 wave    & UFD &CFD\\
\hline
$ B_0$ &$-80.4\pm2.8$ &$-79.4\pm2.8$\\
$B_1$& $-73.6\pm10.5$&$-63.0\pm10.5$\\
 $z_2$ & $M_\pi$&$143.5\pm3.2\,\mev$\\
\hline
$B_{h2}$&$112\pm38$&$32\pm38$\\
$\epsilon$&$0.28\pm0.12$&$0.28\pm0.12$\\
\hline
  \end{tabular}
  \caption{S2 wave parameters for the UFD and CFD sets.}
  \label{tab:S2parameters}
\end{table}

\subsection{P wave}
For this wave we have set the
 matching point between low and intermediate energy parametrizations at
$s_M^{1/2}= 2 M_K$. Thus, at low energies $s^{1/2}\leq 2M_K$, we use:
\begin{eqnarray}
&&\cot\delta_1(s)=\frac{s^{1/2} }{2k^3}(M_\rho^2-s)\left\{
\frac{2M_\pi^3}{M_\rho^2\sqrt{s}}+B_0+B_1 w(s)\right\},
\nonumber\\
&&w(s)=\frac{\sqrt{s}-\sqrt{s_0-s}}{\sqrt{s}+\sqrt{s_0-s}},\quad
s_0^{1/2}=1.05 \,\gev,
  \label{eq:Plowparam} 
\end{eqnarray}
where the $\rho$ mass is fixed to $M_\rho=773.6\pm0.9 \,\mev$.
At intermediate energies, $2M_K\leq s^{1/2}\leq 1420 \,\mev$, we use
a purely phenomenological parametrization:
\begin{eqnarray}
&&\delta_1(s)=\lambda_0+\lambda_1\left(\sqrt{s}/2M_K-1\right)+
\lambda_2\left(\sqrt{s}/2M_K-1\right)^2,
\nonumber\\
&&\eta_1(s)=1-\epsilon_1\sqrt{1-4M_K^2/s}-\epsilon_2(1-4M_K^2/s),
  \label{eq:Phighparam}
\end{eqnarray}
where $\lambda_0$ is fixed from the value of $\delta_1(4 M_K^2)$
obtained from the low energy parametrization, 
so that the phase shift is continuous. Note the possible presence
of a discontinuity in the derivative,  allowed by
the presence of the $K\bar{K}$ threshold. 
The values of the UFD and CFD parameters are given in Table~\ref{tab:Pparameters}.

\begin{table}[h]
  \centering
  \begin{tabular}{ccc}
P wave    & UFD &CFD\\
\hline
$ B_0$ &$1.055\pm0.011$ &$1.043\pm0.011$\\
   $B_1$&$0.15\pm0.05$&$0.19\pm0.05$\\
\hline
 $\lambda_1$ & $1.57\pm0.18$ &$1.39\pm0.18$\\
 $\lambda_2$ & $-1.96\pm0.49$ &$-1.70\pm0.49$\\
$\epsilon_1$&$0.10\pm0.06$&$0.00\pm0.06$\\
$\epsilon_2$&$0.11\pm0.11$&$0.07\pm0.11$\\
\hline
  \end{tabular}
  \caption{P wave parameters for the UFD and CFD sets.}
  \label{tab:Pparameters}
\end{table}

\subsection{The D0 wave}

As it was the case for the P wave, the matching energy between low and intermediate energies 
is now taken at $s_M^{1/2}=2 M_K$. At low energies, $s^{1/2}\leq2M_K$, 
we parametrize this wave by:
\begin{eqnarray}
&&\cot\delta_2^{(0)}(s)=\frac{s^{1/2} }{2k^5}(M_{f_2}^2-s)M_\pi^2\left\{
B_0+B_1 w(s)\right\},
\nonumber\\
&&w(s)=\frac{\sqrt{s}-\sqrt{s_0-s}}{\sqrt{s}+\sqrt{s_0-s}},\quad
s_0^{1/2}=1.05 \,\gev,
  \label{eq:D0lowparam} 
\end{eqnarray}
where the mass of the $f_2(1270)$ resonance is fixed 
at $M_{f_2}=1275.4 \,\mev$.
In the intermediate region, $2 M_K\leq s^{1/2}\leq 1420 \,\mev$,
we use a rather similar parametrization:
\begin{eqnarray}
&&\cot\delta_2^{(0)}(s)=\frac{s^{1/2} }{2k^5}(M_{f_2}^2-s)M_\pi^2\left\{
B_{0h}+B_{1h} w_h(s)\right\},
\nonumber\\
&&w_h(s)=\frac{\sqrt{s}-\sqrt{s_h-s}}{\sqrt{s}+\sqrt{s_h-s}},\quad
s_h^{1/2}=1.45 \,\gev.
  \label{eq:D0highparam} 
\end{eqnarray}
Imposing continuity at the matching point
fixes $B_{h0}$ from the
value of $\delta_2^{(0)}(4M_K^2)$ obtained from the low energy parametrization.
We take the inelasticity to be different from 1 only for $s>4M_K^2$, in which case we write:
\begin{equation}
\eta_2^{(0)}=
1-\epsilon\left( \frac{1-4M_K^2/s}{1-4M_K^2/M_{f_2}^2} \right)^{5/2}
\left[1+r\left(1-\frac{k_2(s)}{k_2(M_{f_2}^2)}\right)\right].
\end{equation}
The parameters of the D0 wave are given in Table~\ref{tab:D0parameters}.

\begin{table}[h]
  \centering
  \begin{tabular}{ccc}
D0 wave    & UFD &CFD\\
\hline
$ B_0$ &$12.47\pm0.12$ &$12.40\pm0.12$\\
   $B_1$&$10.12\pm0.16$&$10.06\pm0.16$\\
\hline
 $B_{h1}$ & $43.7\pm1.8$ &$43.2\pm1.8$\\
$\epsilon$&$0.284\pm0.030$&$0.254\pm0.030$\\
$r$&$2.54\pm0.31$&$2.29\pm0.31$\\
\hline
  \end{tabular}
  \caption{D0 wave parameters for the UFD and CFD sets.}
  \label{tab:D0parameters}
\end{table}

\subsection{The D2 wave}

We use the following parametrization from threshold up to $1420\, \mev$:
\begin{eqnarray}
&&\cot\delta_2^{(2)}(s)=\frac{s^{1/2} }{2k^5}\times\hspace{4cm}
\nonumber\\
&&\qquad\times\frac{M_\pi^4\,s}{4(M_\pi^2+\Delta^2)-s}
\left\{B_0+B_1 w(s)+B_2w(s)^2\right\},
\nonumber\\
&&w(s)=\frac{\sqrt{s}-\sqrt{s_0-s}}{\sqrt{s}+\sqrt{s_0-s}},\quad
s_0^{1/2}=1.45 \,\gev,
  \label{eq:D2lowparam} 
\end{eqnarray}
and we consider that the inelasticity differs from 1 for
$s^{1/2}>1.05\,\gev$, as follows:
\begin{equation}
\eta_2^{(2)}(s)=1-\epsilon(1-\hat s/s)^3,\quad\hat s^{1/2}=1.05\,\gev,
\end{equation}
which is almost negligible up to 1.25 \gev. The values of the parameters 
for the UFD and CFD sets are given in 
Table~\ref{tab:D2parameters}.

\begin{table}[h]
  \centering
  \begin{tabular}{ccc}
D2 wave    & UFD &CFD\\
\hline
$ B_0$ &$(2.4\pm0.5)\,10^3$ &$(4.1\pm0.5)\,10^3$\\
   $B_1$&$(7.8\pm1.0)\,10^3$&$(8.6\pm1.0)\,10^3$\\
   $B_2$&$(23.7\pm4.2)\,10^3$&$(25.5\pm4.2)\,10^3$\\
 $\Delta$ & $196\pm25\,\mev$ & $233\pm25\,\mev$\\
$\epsilon$&$0.2\pm0.2$&$0.0\pm0.2$\\
\hline
  \end{tabular}
  \caption{D2 wave parameters for the UFD and CFD sets.}
  \label{tab:D2parameters}
\end{table}

\subsection{The F wave}
We neglect the inelasticity up to $1420\,\mev$ and simply
use the following parametrization from threshold:
\begin{eqnarray}
&&\cot\delta_3(s)=\frac{s^{1/2} }{2k^7}M_\pi^6
\left\{\frac{2\lambda M_\pi}{\sqrt{s}}+B_0+B_1 w(s)\right\},
\nonumber\\
&&w(s)=\frac{\sqrt{s}-\sqrt{s_0-s}}{\sqrt{s}+\sqrt{s_0-s}},\quad
s_0^{1/2}=1.45 \,\gev.
  \label{eq:Flowparam} 
\end{eqnarray}
The parameters for the UFD and CFD sets are given in
Table~\ref{tab:Fparameters}. Note that they do not change at all from one set to another.

\begin{table}[h]
\centering
  \begin{tabular}{ccc}
F wave    & UFD &CFD\\
\hline
$ B_0$ &$(1.09\pm0.03)\,10^5$ &$(1.09\pm0.03)\,10^5$\\
   $B_1$&$(1.41\pm0.04)\,10^5$&$(1.41\pm0.04)\,10^5$\\
$\lambda$&$0.051\times10^5$&$0.051\times10^5$\\
\hline
  \end{tabular}
  \caption{F wave parameters for the UFD and CFD sets.}
  \label{tab:Fparameters}
\end{table}

\subsection{The G waves}
The contribution of the G0 and G2  waves was shown to be
completely negligible for
the calculations. The details can be found in the Appendix of KPY08
\cite{Kaminski:2006qe}.

\subsection{Regge Parametrizations}
\label{sec:Regge}

Next we show the Regge parametrizations that we use in the high energy region,
i.e. above $1420 \,\mev$. The forward $(t=0)$ Regge parametrizations
were obtained from fits 
to high energy data \cite{Pelaez:2003ky}. 
 For the $t\neq0$ 
behavior we \cite{Kaminski:2006qe} 
simply covered the uncertainties
between the different fits in \cite{Rarita:1968zz}.
These Regge fits are expected to represent {\it experimental} data 
when $1.42\, \gev\leq s^{1/2}\leq\, 20 \,\gev$ and $4 M_\pi^2\geq t\geq-0.4\, \gev^2$, 
somewhat less reliably for the most negative $t$ values.
This is enough
to describe the region of interest, that reaches $t=-0.42\, \gev^2$.
In particular, for the $\rho$ Regge trajectory, we use the 
following expression for the imaginary part, which is all we need in the dispersive integrals:
\begin{eqnarray}
  \label{eq:rhoregge}
&&  {\rm Im}\, F^{(I_t=1)}(s,t)=\beta_\rho\frac{1+\alpha_\rho(t)}
{1+\alpha_\rho(0)}
\varphi(t){\rm e}^{bt}\left(\frac{s}{\hat s}\right)^{\alpha_\rho(t)},
\nonumber\\
&&\alpha_\rho(t)=\alpha_\rho(0)+t\,\alpha_\rho'+\frac{1}{2}t^2\,\alpha_\rho'',
\nonumber\\
&&\varphi(t)=1+d_\rho t +e_\rho t^2,
\end{eqnarray}
where we fix:
\begin{eqnarray}
&&\hat s=1\, \gev^2,\quad  b=2.4\pm0.2 \,\gev^{-2}, \\
&&\alpha'_\rho=0.90\,\gev^{-2},\alpha''_\rho=-0.3\,\gev^{-4}, \nonumber\\
&&d_\rho=2.4\pm0.5\,\gev^{-2},\nonumber
\label{eq:fixrhoregeeparams}
\end{eqnarray}
whereas the rest of the parameters are allowed to vary in the fits.

\begin{table}
  \centering
  \begin{tabular}{ccc}
    Regge param.  & UFD&CFD\\
\hline
$\beta_\rho$& $1.22\pm0.14 $&$1.48\pm0.14$\\
$\alpha_\rho(0)$& $0.46\pm0.02$&  $0.53\pm0.02$\\
$\beta_P$& $2.54\pm0.04 $& $2.50\pm0.04 $ \\
$c_P$&$0.0\pm1.0\,\gev^{-2}$&$0.6\pm1.0\,\gev^{-2}$\\
$c_{P'}$&$-0.4\pm0.4\,\gev^{-2}$&$-0.38\pm0.4\,\gev^{-2}$\\
$\beta_{P'}$& $0.83\pm0.05 $& $0.80\pm0.05 $\\
$\alpha_{P'}(0)$& $0.54\pm0.02$& $0.53\pm0.02$\\
$\beta_2$& $0.2\pm0.2 $& $0.08\pm0.2 $\\
$e_\rho$ & $0\pm2.5\,\gev^{-4}$& $2.7\pm2.5\,\gev^{-4}$\\
\hline
  \end{tabular}
  \caption{UFD and CFD  parameters for the $\rho$, Pomeron and $I=2$ Regge
contributions to $\pi\pi$ scattering amplitudes.}
  \label{tab:Reggeparameters}
\end{table}

For both the Pomeron $P$ and the $P'$ pole, we have used for $s^{1/2}=1420 \,\mev$:
\begin{eqnarray}
  \label{eq:Pomeron}
&&  {\rm Im} \,F^{(I_t=0)}(s,t)=P(s,t)+P'(s,t),\nonumber\\
&&\nonumber\\
&&P(s,t)=\beta_P\Psi_P(t)\alpha_P(t)\frac{1+\alpha_P(t)}{2}{\rm e}^{bt}
\left(\frac{s}{\hat s}\right)^{\alpha_P(t)},\nonumber\\
&&\alpha_P(t)=1+t\alpha'_P,\qquad\qquad\Psi_P(t)=1+c_Pt,\nonumber\\
&&\nonumber\\
&&P'(s,t)=\beta_{P'}\Psi_{P'}(t)
\frac{\alpha_{P'}(t)[1+\alpha_{P'}(t)]}{\alpha_{P'}(0)[1+\alpha_{P'}(0)]}{\rm e}^{bt}
\left(\frac{s}{\hat s}\right)^{\alpha_{P'}(t)},\nonumber\\
&&\alpha_{P'}(t)=\alpha_{P'}(0)+t\alpha'_{P'},\quad\Psi_{P'}(t)=1+c_{P'}t,\nonumber\\
\end{eqnarray}
where, once again, we fix:
\begin{eqnarray}
&&\hat s=1\, \gev^2,\quad  b=2.4\pm0.2 \,\gev^{-2}, \\
&&\alpha'_P=0.20\pm0.10 \,\gev^{-2},\alpha'_{P'}=0.90\,\gev^{-2}, \nonumber\\
&&c_P=0.0\pm1.0\,\gev^{-2},c_{P'}=-0.4\pm0.4\,\gev^{-2},\nonumber
\label{eq:fixpomeronparams}
\end{eqnarray}
and allow the rest of the parameters to vary in the fits.

Finally, the Regge exchange of isospin two is parametrized as:
\begin{eqnarray}
  \label{eq:ReggeI2}
&&  {\rm Im}\,F^{(I_t=2)}=\beta_2\,{\rm e}^{bt}
\left(\frac{s}{\hat s}\right)^{\alpha_\rho(t)+\alpha_\rho(0)-1}.
\end{eqnarray}

In Table \ref{tab:Reggeparameters} we show the values of the Regge parameters 
obtained from the direct fit to high energy data (UFD) and how they
are modified when imposing the dispersive constraints in the fits (CFD).

\section{Derivation of the once subtracted dispersion relations}
\label{app:gkpy}
A once subtracted dispersion relation for a scattering amplitude
of definite isospin $I$ has the following expression:
\begin{multline}
  \label{eq:gkpy:1s-dr}
  F^{(I)}(s,t)=F^{(I)}(s_0,t)
  + \frac{s-s_0}{\pi} \int_{4\mps}^\infty ds'
  \frac{\im F^{(I)}(s',t)}{(s'-s_0)(s'-s)} \\
  + \frac{s-s_0}{\pi} \int_{-t}^{-\infty} ds'
  \frac{\im F^{(I)}(s',t)}{(s'-s_0)(s'-s)},
\end{multline}
with $s_0$ the subtraction point. This expression assumes that the
point $s$ is regular. However, we are especially interested in what
happens for $s$ in the physical region, that is, on the cuts of the
function $F(s,t)$. The usual prescription is to define the amplitude
for physical values of $s$ as:
\begin{equation}
  \nonumber
  F_{phys}(s,t) = \lim_{\epsilon\ra 0^+} F(s+i\epsilon,t).
\end{equation}
With this prescription, we have:
\begin{multline}
  \nonumber
  F_{phys}^{(I)}(s,t)=\lim_{\epsilon\ra0^+} F^{(I)}(s+i\epsilon,t) = \\
  F^{(I)}(s_0,t)
  + \frac{s-s_0+i\epsilon}{\pi} \int_{4\mps}^\infty ds'
  \frac{\im F^{(I)}(s',t)}{(s'-s_0)(s'-s-i\epsilon)} \\
  + \frac{s-s_0+i\epsilon}{\pi} \int_{-t}^{-\infty} ds'
  \frac{\im F^{(I)}(s',t)}{(s'-s_0)(s'-s-i\epsilon)}.
\end{multline}
To obtain the physical amplitude, we must take the limit $\epsilon\ra0^+$
in this expression. Suppose $s$ is on the right hand cut (RHC), $4\mps<s<\infty$.
Since
\begin{equation}
  \nonumber
  \inv{x \pm i\epsilon} = \pepe\left[\inv{x}\right] \mp i\pi\delta(x),
  \qquad \epsilon \ra 0^+,
\end{equation}
we can write the RHC integral as:
\begin{equation}
  \nonumber
  \frac{s-s_0}{\pi} \,\pepe \int_{4\mps}^\infty ds'
  \frac{\im F^{(I)}(s',t)}{(s'-s_0)(s'-s)}
  + i \,\im F^{(I)}(s,t),
\end{equation}
whereas the left hand cut (LHC) integral presents no problems when
$\epsilon$ vanishes. Then we have
\begin{multline}
  \nonumber
  F^{(I)}_{phys}(s,t) = F^{(I)}(s_0,t) + i \,\im F^{(I)}(s,t) \\
  + \frac{s-s_0}{\pi}\,\pepe \int_{4\mps}^\infty ds'
  \frac{\im F^{(I)}(s',t)}{(s'-s_0)(s'-s)} \\
  + \frac{s-s_0}{\pi} \int_{-t}^{-\infty} ds'
  \frac{\im F^{(I)}(s',t)}{(s'-s_0)(s'-s)},
\end{multline}
thus the dispersive integrals only reconstruct the {\em real part}
of the amplitude, instead of the total amplitude. Had we chosen
$s$ to be on the LHC, the reasoning would be analogous, but the
principal value should be taken on the LHC integral, instead of
on the RHC one. We finally obtain:
\begin{multline}
  \nonumber
  \re F_{phys}^{(I)}(s,t) = \re F^{(I)}(s_0,t) \\
  + \frac{s-s_0}{\pi} \,\pepe \int_{4\mps}^\infty ds'
  \frac{\im F^{(I)}(s',t)}{(s'-s_0)(s'-s)} \\
  + \frac{s-s_0}{\pi} \,\pepe \int_{-t}^{-\infty} ds'
  \frac{\im F^{(I)}(s',t)}{(s'-s_0)(s'-s)},
\end{multline}
with the principal value taken on the cut on which $s$ lies.
This is valid for any $s$ on the cuts of $F^{(I)}(s,t)$, i.~e.,
for physical $s$.
We can now recast the LHC integral on the $s$-channel
in Eq.~\eqref{eq:gkpy:1s-dr} in terms of the $u$-channel RHC
by renaming the dummy variable $s'$
as $u'$ in the LHC integral and performing the substitution
\begin{equation}
  \nonumber
  u' \ra 4\mps - s' - t.
\end{equation}
Taking both integrands under the same integral sign,
and choosing $s_0=0$---in analogy with Roy's derivation---we obtain:
\begin{multline}
  \nonumber
  % \label{eq:gkpy:1s-dr}
  \re F^{(I)}(s,t) = \re F^{(I)}(0,t) \\
  + \frac{s}{\pi} \int_{4\mps}^\infty ds'
  \left[
    \frac{\im F^{(I)}(s',t)}{s'(s'-s)}
    - \frac{\im F^{(I)}(u',t)}{u'(u'-s)}
  \right].
\end{multline}
Each of these integrals is potentially divergent if taken
by themselves due to the Pomeron contribution coming from
the $I_t=0$ channel, which grows like $\im F^{(I_t=0)}(s,t) \sim s$
for large $s$. We now show that this is not the case
when taken together.

%\subparagraph{Proof}
Bose statistics require that the
$I_t=0$ amplitude be symmetric under $s-u$
exchange,
\begin{equation}
  \nonumber
  F^{(I_t=0)}(s,t) = F^{(I_t=0)}(u,t).
\end{equation}
Since the amplitudes with well-defined isospin in the $s$-
and $t$-channels are related via the usual crossing matrices,
\begin{equation}
  \nonumber
  C_{st} =
  \begin{pmatrix}
    1/3 & 1 & {5}/{3} \\
    1/3 & 1/2 & -{5}/{6} \\
    1/3 & -1/2 & 1/{6}
  \end{pmatrix},
  \quad
  C_{su} =
  \begin{pmatrix}
    1/{3} & -1 & {5}/{3} \\
    -1/{3} & 1/{2} & {5}/{6} \\
    1/{3} & 1/{2} & 1/{6}
  \end{pmatrix},
\nonumber
\end{equation}
we know that each amplitude with well-defined isospin in the
$s$-channel has a contribution from each of the amplitudes
with well-defined isospin in the $t$-channel. In particular,
the contribution from the $I_t=0$ channel to the integrand
can be written as:
\begin{multline}
  \nonumber
  \left[
    \inv{s'(s'-s)} - \inv{u'(u'-s)}
  \right] \im F^{(I_t=0)}(s',t) = \\
  \frac{(s+t-4\mps)(2s'+t-4\mps)\im F^{(I_t=0)}(s',t)}
  {s'(s'-s)(s'+t-4\mps)(s'+s+t-4\mps)}.
\end{multline}
The $s'^2$ terms in the numerator cancel out, and the integrand
decays as $1/s'^2$ when $s'\ra\infty$, so that the integral converges.
This is in contrast with the expected $1/s'$ asymptotic behavior,
which would spoil convergence.
The contributions from the other $t$-channel isospin
contributions $I_t=1,2$ are not problematic, since they
grow as $(s')^\alpha$ with $\alpha<1$, and are convergent even
if taking the integrals separately.
Note that this cancellation does not depend on the explicit
parametrizations we use for the Pomeron,
but rather, on very general asymptotic
properties of the amplitudes. %$\square$

In order to rewrite the RHC contribution from the $u$ channel
in terms of amplitudes on the RHC $s$-channel,
we take into account the crossing symmetry relation:
\begin{equation}
  \label{eq:gkpy:s-u-crossing}
  F^{(I)}(4\mps-s'-t,t)=\sum_{I'} C_{su}^{II'} F^{(I')}(s',t),
\end{equation}
with $C_{su}$ the crossing matrix defined above. Also,
\begin{equation}
  \label{eq:gkpy:s-t-crossing}
  F^{(I)}(0,t) = \sum_{I''} C_{st}^{II''} F^{(I'')}(t,0),
\end{equation}
and we now write a dispersion relation for $F^{(I'')}(t,0)$:
%\begin{widetext}
  \begin{multline}
    \label{eq:gkpy:t-dr}
    F^{(I'')}(t,0) = F^{(I'')}(t_0,0) \\ +
    \frac{t-t_0}{\pi} \int_{4\mps}^\infty ds'
    \left[
      \frac{\im F^{(I'')}(s',0)}{(s'-t)(s'-t_0)} \qquad \right. \\ \left. -
      \frac{\sum_{I'''}C_{su}^{I''I'''}\im F^{(I''')}(s',0)}
      {(4\mps-t-s')(4\mps-s'-t_0)}
    \right].
  \end{multline}
%\end{widetext}
Again, in analogy with Roy, we take $t_0=4\mps$. Thus:
\begin{widetext}
  \begin{multline}
    \label{eq:gkpy:final-dr}
    \nonumber \re F^{(I)}(s,t) =
    \sum_{I'} C_{st}^{II'} F^{(I')}(4\mps,0)
    + \frac{s}{\pi} \,\pepe \int_{4\mps}^\infty ds' \left[ \frac{\im
        F^{(I)}(s',t)}{s'(s'-s)} - \frac{\sum_{I'} C_{su}^{II'} \im
        F^{(I')}(s',t)} {(s'+t-4\mps)(s'+s+t-4\mps)} \right] \\ % &
    + \frac{t-4\mps}{\pi} \,\pepe \int_{4\mps}^\infty ds' \sum_{I''}
    C_{st}^{II''} \left[ \frac{\im F^{(I'')}(s',0)}{(s'-t)(s'-4\mps)} -
      \frac{\sum_{I'''}C_{su}^{I''I'''}\im
        F^{(I''')}(s',0)} {s'(s'+t-4\mps)} \right].
  \end{multline}
\end{widetext}
Now, to project into partial waves, we define first the following kernels:
\begin{eqnarray}
K_{\ell\ell'}(s,s')&\hspace{-.3cm}=&\hspace{-.3cm}\frac{s}{\pi s'(s-s')}
\int_{0}^{1}\hspace{-.2cm}dx P_\ell(x)P_{\ell'}\left(y\right), \label{KernelK}\\ \nonumber
L_{\ell\ell'}(s,s')&\hspace{-.2cm}=&\hspace{-.2cm}\frac{s}{\pi}
\int_{0}^{1}dx P_\ell(x)
\frac{P_{\ell'}\left(y\right)}
{u'\left(u'-s\right)}, \label{KernelL} \\\nonumber
M_\ell(s,s')&\hspace{-.2cm}=&\hspace{-.2cm}\frac{1}{\pi(s'-4M_{\pi}^2)}\int_{0}^{1}dx P_\ell(x)
\frac{t-4M_{\pi}^2}{s'-t}, \label{KernelM} \\\nonumber
N_\ell(s,s')&\hspace{-.2cm}=&\hspace{-.2cm}\frac{1}{\pi s'}\int_{0}^{1}dx P_\ell(x)
\frac{4M_{\pi}^2-t}{u'} \label{KernelN},
\end{eqnarray}
where $P_\ell(x)$ and $P_{\ell'}(y)$ are Legendre polynomials, and
\begin{eqnarray}
  \nonumber
&&t = \frac{(s-4M_{\pi}^2)(x-1)}{2}, \\
  \nonumber
% \end{eqnarray}
% and 
% \begin{equation}
&&u' = 4M_{\pi}^2 - s' - t, \,\,\,\,\,\, y = \frac{u'-t}{u'+t}.
%\end{equation}
\end{eqnarray}
Note we have taken advantage of the symmetry 
of the integrands to change the integration 
limits from $(-1,1)$ to $(0,1)$.

With the normalization chosen in Section~\ref{sec:notation},
and recalling that $a_0^{(1)}=0$, we find:
\begin{widetext}
  \begin{eqnarray}
    \label{eq:gkpy:eqs} \nonumber
    \re t_\ell^{(I)}(s) &=& \xi_\ell\,\sum_{I''} C_{st}^{II''}a_0^{(I'')}
    + \sum_{\ell'} (2\ell'+1)
    \int_{4\mps}^{\infty} ds' \Bigg\{K_{\ell\ell'}(s,s')\im t^{(I)}_{\ell'}(s')
    -L_{\ell\ell'}(s,s')\sum_{I'}C^{su}_{II'}\im t^{(I')}_{\ell'}(s') \\ \nonumber
    && +  \sum_{I''}C^{st}_{II''}\left[M_\ell(s,s')\im t^{(I'')}_{\ell'}(s')-
      N_\ell(s,s')\sum_{I'''}C^{su}_{I''I'''}\im t^{(I''')}_{\ell'}(s')\right]\Bigg\}.
  \end{eqnarray}
\end{widetext}
In order to simplify the previous expression, we define
  \begin{eqnarray}
    \overline{K}_{\ell\ell'}^{I I'}(s,s') & = & (2\ell'+1) \\
   & \times& [ K_{\ell\ell'}(s,s') \delta^{I I'}
    - L_{\ell\ell'}(s,s') (C_{su})^{I I'}\nonumber\\
   & +&  M_\ell(s,s') (C_{st}^{I I'}) - N_\ell(s,s') (C_{st}C_{su})^{I I'}].\nonumber
  \end{eqnarray}
We thus arrive at the final result used in Eq.~\eqref{1SEquations}:
  \begin{eqnarray}
  \nonumber
    \re t_\ell^{(I)}(s) &=&\overline{ST}_\ell^{I} + \overline{DT}^{I}_\ell(s) \\
    &+&\sum_{I'=0}^2 \sum_{\ell'=0}^{1}
    \mbox{P.P.}
    \int_{4\mps}^{s_{max}} ds'  \overline{K}_{\ell\ell'}^{I I'}(s,s') \im t_{\ell'}^{I'}(s'),
\nonumber
  \end{eqnarray}
%\end{eqnarray}
where, for simplicity, the high energy part of the integrals ($s'>s_{max}$) and
the higher partial waves ($\ell'>1$) are grouped in the so-called
driving terms, $\overline{DT}_\ell^I(s)$. The subtraction 
terms $\overline{ST}_\ell^I$, which are now constant, are:
\begin{equation}
  \nonumber
  \overline{ST}_\ell^I = \xi_\ell\,\sum_{I''} C_{st}^{II''}a_0^{I''},
\end{equation}
with the $\xi_\ell$ coefficients defined in Eq.~\eqref{eq:xi:l}.
For our purposes we will only need  $\xi_0=1$
and  $\xi_1=1/2$.
Note that the subtraction term $\overline{ST}_\ell^I$ is a {\em constant}, and
does not depend on $s$. This is a relevant feature of GKPY equations
versus Roy Equations, as explained in Section~\ref{sec:decomposition}.

\begin{widetext}
  \begin{equation}
%  \nonumber
    \label{eq:xi:l}
    \xi_\ell = \int_0^1 dx P_\ell(x) = \frac{\sqrt{\pi}}{2 \Gamma(1-\frac{\ell}{2}) \Gamma(\frac{3+\ell}{2})}
    = \left\{
      \begin{array}{ll}
        1, & \ell=0 \\
        0, & \ell=2m,\,m>0 \\
        \frac{(-1)^m}{2^{m+1}(m+1)!} \prod_{k=0}^{m-1} \left[ 2m - (2k+1) \right], & \ell=2m+1
      \end{array}
    \right. .
  \end{equation}
\end{widetext}

\section{Integral kernels in GKPY equations}
\label{app:kernels}

All kernels in Eqs.~(\ref{KernelK})-(\ref{KernelN}) can be calculated analytically. 
One has to note, however, that the $L_{\ell\ell'}(s,s')$ and $N_\ell(s,s')$ kernels
% , as
% well as the piece
% $$\int_0^1 dx P_\ell(x)\frac{ \im F^{I_t}\left(s',t\right)}{u'\left(u+s'\right)}$$
% in
are singular at $u' = 0$, namely $x = -(2s'-s-4M_{\pi}^2)/(s-4M_{\pi}^2)$,
where a principal value over the integral is understood.

In this work we need 18 $\overline{K}^{II'}_{\ell\ell'}(s,s')$ kernels,
since we are considering the dispersion relation for the S0, P and S2
waves, but using S0, P, S2, D0, D2 and F waves as input.
However, following \cite{Wanders:2000mn}, we know that, since
the $K$, $L$, $M$ and $N$ kernels in Eqs.~(\ref{KernelK})-(\ref{KernelN})
do not depend on isospin, the $\overline{K}^{II'}_{\ell\ell'}(s,s')$ are not
all independent and can be expressed in terms of four of the $K_{\ell\ell'}$
above, and eight combinations of the other kernels, which we
call $I_{\ell\ell'}(s,s')$. Namely:

\begin{align}
\nonumber
&\overline{K}^{00}_{00}  =  K_{00} - I_{00}/3, \quad 
&&\overline{K}^{02}_{00}  =  -\frac{5}{3} I_{00},  \\ \nonumber        
&\overline{K}^{01}_{01}  =  3 I_{01},  \quad               
&&\overline{K}^{00}_{02}  = 5(K_{02} - \frac{1}{3}I_{02}),  \\ \nonumber 
&\overline{K}^{02}_{02}  =  -\frac{25}{3} I_{02},   \quad        
&&\overline{K}^{01}_{03}  =  7 I_{03}, 
\end {align}

\begin{align} 
\nonumber                                                                      
&\overline{K}^{10}_{10}  =  I_{10}/3, \quad 
&&\overline{K}^{12}_{10}  =  -\frac{5}{6} I_{10},\\ \nonumber           
&\overline{K}^{11}_{11}  =  3(K_{11}- \frac{1}{2} I_{11}), \quad 
&&\overline{K}^{10}_{12}  =  \frac{5}{3} I_{12},  \\ \nonumber          
&\overline{K}^{12}_{12}  =  -\frac{25}{6} I_{12}, \quad 
&&\overline{K}^{11}_{13}  =  7(K_{13}- \frac{1}{2}I_{13}), 
\end{align}

\begin{align}       
\nonumber                                                                     
&\overline{K}^{20}_{00}  =  -I_{00}/3, \quad 
&&\overline{K}^{22}_{00}  =  K_{00}- I_{00}/6,  \\ \nonumber         
&\overline{K}^{21}_{01}  =  -\frac{3}{2} I_{01},  \quad 
&&\overline{K}^{20}_{02}  =  -\frac{5}{3} I_{02}, \\ \nonumber          
&\overline{K}^{22}_{02}  =  5(K_{02} - \frac{1}{6}I_{02}),  \quad 
&&\overline{K}^{21}_{03}  =   -\frac{7}{2} I_{03},
\end{align}
where
\begin{eqnarray}                                            
I_{00} =  L_{00}- M_0 + N_0,\quad
I_{01} =  L_{01}+ M_0 - N_0,\\ \nonumber
I_{10} =  L_{10}+ M_1 + N_1,\quad
I_{11} =  L_{11}- M_1 - N_1,\\ \nonumber
I_{02} =  L_{02}- M_0 + N_0,\quad
I_{03} =  L_{03}+ M_0 - N_0,\\ \nonumber
I_{12} =  L_{12}+ M_1 + N_1,\quad
I_{13} =  L_{13}- M_1 - N_1.
\end{eqnarray}                                                                               
The analytic expressions for the $K_{ll^\prime}$ kernels are:
\begin{eqnarray}   
\nonumber            
K_{00}  &= & -\frac{s}{\pi s' (s-s')}, \\ \nonumber
K_{02}  &= & -\frac{s(4M_{\pi}^2+s-2s')}{2\pi s' (s'-4M_{\pi}^2)^2}, \\ \nonumber
K_{11}  &= & \frac{s(8M_{\pi}^2+s-3s')}{6\pi s' (s-s')(s'-4M_{\pi}^2)}, \\ \nonumber
K_{13}  &= & \frac{s(4M_{\pi}^2+s-2s')^2}{8\pi s' (s'-4M_{\pi}^2)^3},\\ 
%\nonumber
%K_{20}  &= & 0, 
%\\
%K_{22}  &= & \frac{(s-4M_{\pi}^2) s (7 s-15 s'+32M_{\pi}^2)}{40 \pi  (s-s') (s'-4M_{\pi}^2)^2s'}.
\end{eqnarray}  
The diagonal kernels $K_{00}(s,s')$ and $K_{11}(s,s')$ 
%and $K_{22}(s,s')$ 
contain a singularity at $s = s'$,
which is the only type of singularity in the GKPY equations.

By defining the following $a_i$ functions:
\begin{eqnarray}
a_1 &=&  \frac{s'}{s+s'-4M_{\pi}^2},\\\nonumber
% \end{eqnarray}
% \begin{eqnarray}
a_2 &=&  \frac{(s+2 s'-4M_{\pi}^2)^2}{4 (s+s'-4M_{\pi}^2)^2},\\ \nonumber
a_3 &=& -\frac{s^2-4 (s'-2M_{\pi}^2)^2}{4 (s'-4M_{\pi}^2) (s+s'-4M_{\pi}^2)},\\ \nonumber
a_4 &=& -\frac{(s-2 s'+4M_{\pi}^2) (s+s'-4M_{\pi}^2)}{(s'-4M_{\pi}^2) (s+2 s'-4M_{\pi}^2)},\\ \nonumber
a_5 &=& \frac{s' (-s+2 s'-4M_{\pi}^2)}{(s'-4M_{\pi}^2) (s+2 s'-4M_{\pi}^2)}, \\ \nonumber
a_6 &=& -\frac{(s-2 s'+4M_{\pi}^2) (s+2 s'-4M_{\pi}^2)}{4 (s'-4M_{\pi}^2) s'},\nonumber
\end{eqnarray}
the analytical expressions for the $I_{\ell\ell'}(s,s')$ can be recast as:
\begin{eqnarray} 
I_{00}(s,s') &=&
2\frac{(s-4M_{\pi}^2) (s'-2M_{\pi}^2)/(s'-4M_{\pi}^2)+ s'
\log \left(a_1\right)}{\pi s'  (s-4M_{\pi}^2)},\\
I_{01}(s,s') &=& 
-\frac{2(s'-2M_{\pi}^2)}{\pi (s'-4M_{\pi}^2)s'}\;\;\;\;\;\;\;\;\;\;\;\;\;\;\;\;\;\;\;\;\;\;\;\;\;\;\;\;\;\;\;\;\;\;\;\;\;\;\;\;\;\;\;\;\;\;\;\;\;\;\;\;\;\;\;\;\;\;\;\;\\
&&-2\frac{(s'-4M_{\pi}^2) s' \log \left(a_1\right)+s s' \log \left(a_2\right)}{\pi(s-4M_{\pi}^2) (s'-4M_{\pi}^2) s'}, \nonumber
\end{eqnarray}
\begin{widetext}
\begin{eqnarray} 
I_{02}(s,s') &=& 
\frac{1}{\pi}\left(\frac{6 s}{(s'-4M_{\pi}^2)^2}+\frac{1}{s'-4M_{\pi}^2}+\frac{1}{s'}\right)
+\frac{1}{\pi  (s-4M_{\pi}^2)}\left(2 \log \left(a_1\right)+\frac{6 s (s+s'-4M_{\pi}^2) 
\log \left(a_2\right)}{(s'-4M_{\pi}^2)^2}\right), \\\nonumber
I_{03}(s,s') &=& -\frac{1}{\pi  (s-4M_{\pi}^2)}
\Bigg\{\frac{(s-4M_{\pi}^2) \left(2 s'^3+10 (s-2M_{\pi}^2) s'^2+\left(25
s^2-60M_{\pi}^2 \
s+64M_{\pi}^4\right) s'-64M_{\pi}^6\right)}{(s'-4M_{\pi}^2)^3 s'} \\ 
&+& 2 \log \left(a_1\right)+\frac{2 s \left(10 s^2+15 \
(s'-4M_{\pi}^2) s+6 (s'-4M_{\pi}^2)^2\right) \log \left(a_2\right)}{(s'-4M_{\pi}^2)^3}\Bigg\},
\end{eqnarray}

\begin{eqnarray} \nonumber
I_{10}(s,s') &=& 
-\frac{2}{\pi  (s-4M_{\pi}^2)^2 (s'-4M_{\pi}^2) s'} 
\Bigg\{s^2M_{\pi}^2+2 s'^2 s - 8 s' sM_{\pi}^2-8 sM_{\pi}^4-8 s'^2M_{\pi}^2+32
s'M_{\pi}^4+16M_{\pi}^6  \\ 
&+&(s'-4M_{\pi}^2) s'\log \left(a_1\right) s 
+ 2 s' \left(s'^2-6 s'M_{\pi}^2+8M_{\pi}^4\right) \log \left(a_1\right)\Bigg\},\\\nonumber
I_{11}(s,s') &=& \frac{2}{\pi  (s-4M_{\pi}^2)^2} 
\Bigg\{\frac{(2 s'+M_{\pi}^2) s^2+2 \left(s'^2-8 \
s'M_{\pi}^2-4M_{\pi}^4\right) s-8 \left(s'^2-4M_{\pi}^2
s'-2M_{\pi}^4\right)M_{\pi}^2}{(s'-4M_{\pi}^2)s'}\\ \nonumber
&+&
\frac{1}{s'-4M_{\pi}^2}
\Big[s (s+3 s'-8M_{\pi}^2) \log \left(a_3\right)-\left(s^2+2 \
(s'-2M_{\pi}^2) s+2 \left(s'^2-6 s'M_{\pi}^2+8M_{\pi}^4\right)\right) 
\log \left(a_4\right)\Big] \\
&+& 2 (s'-2M_{\pi}^2) 
\log \left(a_5\right)-s 
\log \left(a_6\right)\Bigg\} ,
\end{eqnarray}

\begin{eqnarray} \nonumber
I_{12}(s,s') &=& \frac{1}{2 \pi  (s-4M_{\pi}^2)^2}
\Bigg\{-\frac{2 (s-4M_{\pi}^2) \left(9 s' s^2+2 \left(6 s'^2-17 \
s'M_{\pi}^2-4M_{\pi}^4\right) s+4 \left(s'^3-8 s'^2M_{\pi}^2+14M_{\pi}^4 \
s'+8M_{\pi}^6\right)\right)}{(s'-4M_{\pi}^2)^2 s'} \\\nonumber
&+&
\frac{4}{(s'-4M_{\pi}^2)^2}
\Big[ s \
\left(3 s^2+3 (3 s'-8M_{\pi}^2) s+7 s'^2-44M_{\pi}^2 s'+64M_{\pi}^4\right) 
\log \left(a_3^{-1}\right) \\ \nonumber
&+&\left(3 s^3+3 (3 s'-8M_{\pi}^2) s^2+6 \
\left(s'^2-6 s'M_{\pi}^2+8M_{\pi}^4\right) s+2 (s'-4M_{\pi}^2)^2 \
(s'-2M_{\pi}^2)\right)  
\log \left(a_4\right)\Big]\\
&-& 8 (s'-2M_{\pi}^2) 
\log \left(a_5\right)+4 s 
\log \left(a_6\right)
\Bigg\},
\end{eqnarray}

\begin{eqnarray} \nonumber
I_{13}(s,s') &=& \frac{1}{2 \pi  (s-4M_{\pi}^2)^2}
\Bigg\{\frac{2 (s-4M_{\pi}^2) \left(85 s' s^3+5 s' (33 s'-100M_{\pi}^2) \
s^2+\left(72 s'^3-510 s'^2M_{\pi}^2+784M_{\pi}^4 s'+96M_{\pi}^6\right) s\right)}{3 (s'-4M_{\pi}^2)^3 s'} \\ \nonumber
&-&
\frac{4}{(s'-4M_{\pi}^2)^3}
\Bigg[s \bigg(10 s^3+5 (7 s'-20M_{\pi}^2) \
s^2+12 \left(3 s'^2-19 s'M_{\pi}^2+28M_{\pi}^4\right) s \\ \nonumber
&+& (s'-4M_{\pi}^2)^2 (13 s'-28M_{\pi}^2)\bigg)
\log \left(a_3^{-1}\right) \\ \nonumber
&+&
\bigg(10 s^4+5 (7 \
s'-20M_{\pi}^2) s^3+12 \left(3 s'^2-19 s'M_{\pi}^2+28M_{\pi}^4\right) s^2
+12 (s'-4M_{\pi}^2)^2 (s'-2M_{\pi}^2) s\\ \nonumber
&+&2 (s'-4M_{\pi}^2)^3 (s'-2M_{\pi}^2)\bigg)\log \left(a_4\right)
\Bigg] \\
&+& \frac{8 (s-4M_{\pi}^2) \left(s'^2-4M_{\pi}^2 s'-2M_{\pi}^4\right)}{(s'-4M_{\pi}^2) s'} +
8 (s'-2M_{\pi}^2) 
\log \left(a_5\right)-4 s 
\log \left(a_6\right)\Bigg\}.
\end{eqnarray}
\end{widetext}

The behavior around threshold is also interesting  when considering the
expansions of the kernels around $s-4M_{\pi}^2$.
In particular, the threshold expansions of $\overline{K}^{II'}_{\ell\ell'}(s,s')$ 
around $s=4M_{\pi}^2$ 
behave like $a + b(s-4M_{\pi}^2)+\ldots$

\vspace*{.5cm}
\section{ROY-GKPY weighted phases}
\label{sec:ponderatedphases}

\begin{table}
\begin{tabular}{c|c|c|c|}
$\sqrt{s}$ (MeV)&
$\delta^{0}_{0}(^\circ)$&$\delta^{1}_{1}(^\circ)$&$\delta^{2}_{0}(^\circ)$\\\hline
310&7.1$\pm$0.3 &0.2$\pm$0.1&-1.5$\pm$0.1\\
340&11.7$\pm$0.5&0.6$\pm$0.1&-2.5$\pm$0.1\\
370&16.5$\pm$0.7&1.2$\pm$0.1&-3.5$\pm$0.1\\
400&21.5$\pm$1.0&1.9$\pm$0.2&-4.6$\pm$0.2\\
430&26.6$\pm$1.3&2.8$\pm$0.2&-5.7$\pm$0.2\\
460&31.9$\pm$1.8&3.9$\pm$0.2&-6.7$\pm$0.3\\
490&36.9$\pm$3.0&5.3$\pm$0.2&-7.8$\pm$0.3\\
520&40.7$\pm$7.5&7.0$\pm$0.2&-8.9$\pm$0.3\\
550&50.5$\pm$5.4&9.1$\pm$0.2&-9.9$\pm$0.4\\
580&54.7$\pm$3.2&12.0$\pm$0.2&-11.0$\pm$0.4\\
610&59.3$\pm$2.5&15.9$\pm$0.3&-12.0$\pm$0.5\\
640&63.8$\pm$2.1&20.7$\pm$0.5&-13.1$\pm$0.6\\
670&68.1$\pm$1.8&28.7$\pm$0.5&-14.1$\pm$0.6\\
700&72.2$\pm$1.7&40.6$\pm$2.6&-15.1$\pm$0.7\\
730&76.2$\pm$1.6&56.1$\pm$1.1&-16.2$\pm$0.8\\
760&80.3$\pm$1.6&79.0$\pm$0.8&-17.2$\pm$0.9\\
790&84.3$\pm$1.6&101.8$\pm$0.8&-18.2$\pm$1.0\\
820&88.6$\pm$1.7&118.8$\pm$0.9&-19.2$\pm$1.1\\
850&93.5$\pm$1.8&128.3$\pm$1.9&-20.2$\pm$1.2\\
880&99.7$\pm$2.2&142.0$\pm$2.0&-21.2$\pm$1.3\\
910&108.8$\pm$3.4&147.0$\pm$1.3&-22.1$\pm$1.4\\
940&122.7$\pm$7.0&150.5$\pm$1.2&-22.9$\pm$1.5\\
970&152.0$\pm$6.3&153.3$\pm$1.2&-23.9$\pm$1.7\\
\end{tabular}
\caption{ Phases from the dispersive data analysis. Central values are
obtained as a weighted average between the output of Roy and GKPY equations,
using the CFD fit as input. We do not weight the uncertainty
but take the smallest of the two, 
since both results come from the same data.
}
\label{tab:ponderated phases}
\end{table}

In  Table~\ref{tab:ponderated phases}  we give
the central values of the phase 
in the elastic regions,  as the weighted averaged 
{\it obtained from the output of Roy and GKPY equations}, when using the CFD set as input.
These results could be understood as a traditional ``energy dependent data
analysis''. We do not weight the uncertanty but take the smallest of the two
outputs, since both results come from the same data.

\vspace*{5cm}


\end{document}